UNIVERSITÀ
DEGLI STUDI
DE L'AQUILA

*Department of Information Engineering, Computer Science and Mathematics*

# IoT-based Emergency Evacuation Systems

*PhD in*

**Information and Communication Technologies**

**XXXII Doctoral Cycle**

*PhD Candidate*: **Mahyar Tourchi Moghaddam**

*Supervisor*: **Prof. Henry Muccini**

*Co-Supervisor*: **Prof. Claudio Arbib**

La borsa di dottorato è stata cofinanziata con risorse del
Programma Operativo Nazionale Ricerca e Innovazione 2014-2020 (CCI 2014IT16M2OP005),
Fondo Sociale Europeo, Azione I.1 "Dottorati Innovativi con caratterizzazione Industriale"

UNIONE EUROPEA
Fondo Sociale Europeo

Ministero dell'Istruzione, dell'Università e della Ricerca

PON
RICERCA
E INNOVAZIONE
2014 - 2020

**DECLARATION**

I here by declare that the report entitled "IoT-based Emergency Evacuation Systems" submitted by me, as my PhD at the University of L'Aquila is a record of bonafide work carried out by me under the supervision of Professor Henry Muccini, co-supervision of Professor Claudio Arbib and collaboration with my other colleagues and co-authors. Thus, as the personal pronouns used in this thesis, *we* is preferred to *I*.

I further declare that this work has not been submitted and will not be submitted, either in part or in full, for the award of any other degree or diploma in this institute or any other institute or university.

The PhD is co-financed by the PON-RI 2014-2020 program, using European funds dedicated to innovative industrial PhD.

Place: L'Aquila

Date: 05/01/2020 **Mahyar Tourchi Moghaddam**

# DEDICATION

*Tomy beloved parents an family, who have been my source of inspiration…*

*To my love, Mina, who encourage me to pursue my dreams…*

# ACKNOWLEDGMENTS

I would like to express my deepest appreciation to my dear friend, my supervisor, Prof. Henry Muccini, without whom I would never have been able to learn how to be a good researcher. He was always patient, supportive, and helpful in both my academic career and personal life.

I would like to acknowledge the efforts given by my Co-Supervisor, Prof. Claudio Arbib. He always encouraged me to come up with new ideas and implement those ideas within my research.

I would like to thank Prof. Julie Dugdale for her guidance in completing this dissertation and her great support in my research period in France.

I would like to thank my colleagues in Sweden: Prof. Paul Davidsson, Prof. Romina Spalazzese, and Prof. Ivica Crnkovic.

Finally, I would like to thank all my colleagues and friends at DISIM.

# Contents

# List of Figures

# List of Tables

# INTRODUCTION

The increasing topological changes in urban environment have caused the human civilization to be subjected to an increasing risk of emergencies. Fires, earthquakes, floods, hurricanes, overcrowding or and even pandemic viruses endanger human lives. Hence, designing infrastructures to handle possible emergencies has become an ever increasing need. The safe evacuation of occupants from the building takes precedence when dealing the necessary mitigation and disaster risk management. Nowadays, evacuation plans appear as static maps (e.g. Figure 1, left), designed by civil protection operators, that provide some pre-selected routes through which pedestrians should move in case of emergency. The static models may work in low congested spacious areas. However, the situation may barely be imagined static in case of a disaster.

The static emergency map expose several limitations such as: *i)* ignoring abrupt congestion, obstacles or dangerous routes and areas; *ii)* leading all pedestrians to the same route and making specific areas highly crowded; *iii)* ignoring the individual movement behavior of people and special categories (e.g. elderly, children, disabled); *iv)* lack of providing proper trainings for security operators in various scenarios; *v)* lack of providing a comprehensive situational awareness for evacuation managers.

By simply tracking people in an indoor area, possible congestions can be detected, best evacuation paths can be periodically re-calculated, and minimum evacuation time under ever-changing emergency conditions can be evaluated. Using a well-designed internet of things (IoT) infrastructure can provide various solutions in both design-time and real-time.

At design-time, a building architecture can be assessed regarding safety conditions, even before its (re-) construction. Simulations are among feasible solutions to assess the evacuability of buildings and feasibility of evacuation plans. At design-time, an IoT-based evacuation system provides: *i)* Safety considerations for building architecture in early (re-) construction phase; *ii)* Finding out the building dimensions that lead to an optimum evacuation performance; *iii)* Bottleneck discovery that is tied with the building characteristics; *iv)* Comparing various routing optimization models to pick the best match one as a base of real-time evacuation system; *v)* Visualizing dynamic evacuation executions to demonstrate a variety of scenarios to security operators and train them.

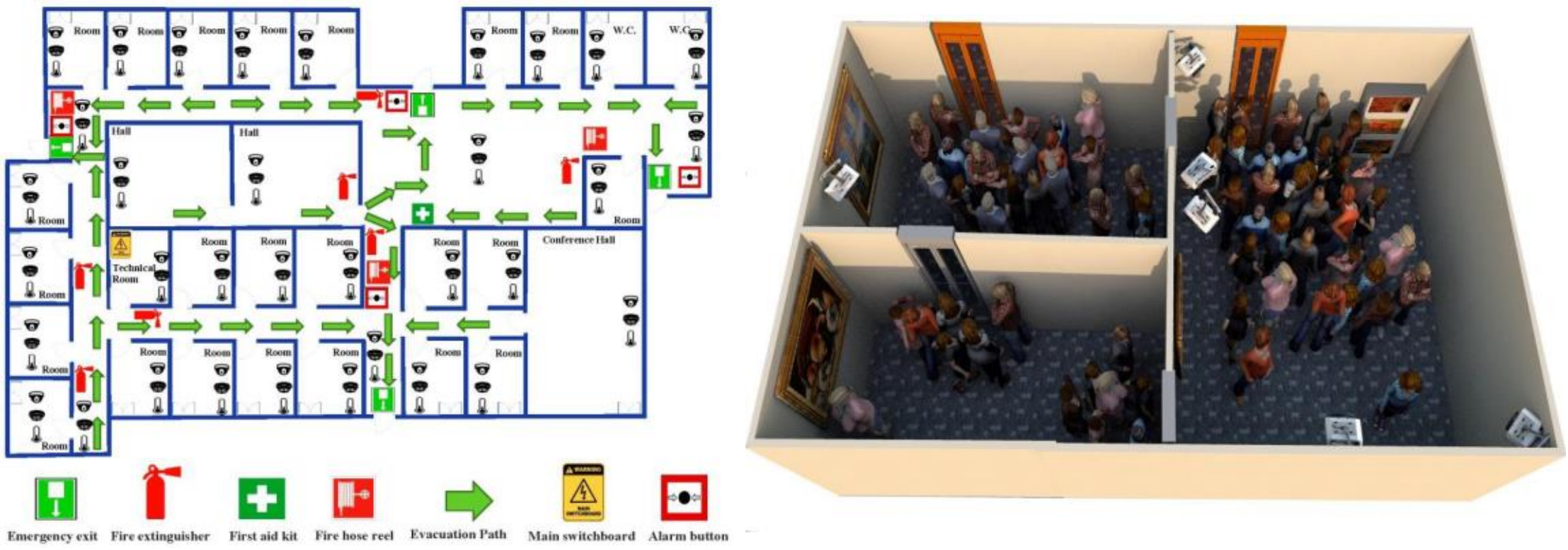

**Figure 1** Traditional evacuation maps (left) VS IoT-based systems (right).

At real-time, an IoT architecture supports the gathering of data that will be used for dynamic monitoring and evacuation planning. At real-time, an IoT-based evacuation system provides: *i)* Optimal solutions that can be continuously updated, so evacuation guidelines can be adjusted according to visitors position that evolve overtime; *ii)* Paths that become suddenly unfeasible can automatically be discarded by the system; *iii)* The model can be incorporated into a mobile app supporting emergency units to evacuate closed or open spaces.

Since the evacuation time of people from a scene of an emergency (e.g. building) is crucial, IoT-based evacuation infrastructures need to have an optimization algorithm as their core. In order to reduce the time taken for evacuation, better and more robust exit strategy are developed. Some algorithms are used to model participating agents for their exit patterns and strategies and in order to evaluate their movement behaviour based on performance, efficiency and practicality attributes. The algorithms normally provide a way to evacuate the occupants as quickly as possible.

While this research and all associated experiences are carried out in Italy, we see the problem from an international viewpoint. Within this thesis, we carried out the following research and experiments to analyze and develop an IoT-based emergency evacuation system:

The first two chapters present systematic mapping studies to review the state of the art and help designing high quality IoT architectures. More specifically, *chapter one* investigates on IoT software architectural styles and *chapter two* assesses the architectural fault-tolerance. *Chapter three* proposes some adaptive architectural styles and their associated quality of energy consumption. After taking the preliminary design decisions about the architecture, in *chapter four* we propose a core computational component to be in charge of minimizing the time necessary to evacuate people from a building. We developed a network flow algorithm that decomposes the building space and time into finite elements: unit cells and time slots. In *chapter five*, we assessed the effectiveness of the IoT system in providing good real-time and design-time solutions. *Chapter six* focuses on real-time performance and minimize computational and evacuation delays to a minimum, by using queuingnetwork.

During our research, we designed and implemented a hardware and software IoT infrastructure. We installed sensors throughout the selected building, whose data constantly feed into the algorithm to show the best evacuation routes to the occupants. We further realized that, such a system may lack the accuracy since: *i)* a purely optimization approach can lack realism as building occupants may not evacuate immediately; *ii)* occupants may not always follow the recommended optimal paths due to various behavioural and organizational issues; *iii)* the physical space may prevent an effective emergency evacuation. Therefore, in *chapter seven* we introduced a simulation-optimization approach. The approach allows us to test more realistic evacuation scenarios and compare them with an optimal approach. We simulated the optimized netflow algorithm under different realistic behavioral agent-based modeling (ABM) constraints, such as social attachment and improved IoT system accordingly.

This thesis makes the following main contributions:

**Contributions on new and legitimate IoT architectures:**

- Addressing to an up to date state of the art class for IoT architectural styles and patterns.
- Proposing a set of self-adaptive IoT patterns and assessing their specific quality attributes (fault-tolerance, energy consumption and performance).
- Designing an IoT infrastructure and testing its performance in both real-time and design-time applications.

**Algorithmic contribution:**

- Developing a network flow algorithm that facilitates minimizing the time necessary to evacuate people from a scene of disaster.

**Evaluation / experimentation environment contributions:**

- Modeling various social agents and their interactions during an emergency to improve the IoT system accordingly.
- Evaluating the system by using empirical and real case studies.

# Part I

# Internet of Things Architecture

## Introduction to Part I

This part is written based on the following peer reviewed articles:

- **IoT Architectural Styles**, *Published in: European Conference on Software Architecture, 2018.*
  *DOI:* `https://doi.org/10.1007/978-3-030-00761-4_5`
- **Fault-tolerant IoT**, *Published in: International Workshop on Software Engineering for Resilient Systems, 2019.*
  *DOI:* `https://doi.org/10.1007/978-3-030-30856-8_5`
- **Self-adaptive IoT Architectures**, *Published in: European Conference on Software Architecture: Companion. 2018.*
  *DOI:* `https://doi.org/10.1145/3241403.3241424`

**Abstract.** IoT components are becoming more and more ubiquitous. Thus, the necessity of architecting IoT applications is bringing a substantial attention towards software engineering community. On this occasion, different styles and patterns can facilitate shaping the IoT architectural characteristics. This part aims at classifying state of the art to design a class of IoT architectural styles and patterns. The architecture is generally exposed to various quality issues, such as fault-tolerance, energy efficiency and performance.

Regarding *IoT architectural styles and patterns*, the part followed a systematic mapping study (SMS) procedure picking up 63 studies among over 2,300 candidate studies. The first chapter of this part goes through cloud-based distributed collaborative and hybrid architectures and discusses their pros and cons. It further provides a set of abstract IoT reference architectures beneficial for academic and industrial applications.

On the subject of *Fault-tolerance*, the part again followed an SMS process to design a reference IoT architecture and assess its various layers based on fault-tolerant standards. The motivation is that within an IoT system, sensor and actuator nodes can be missed, network links can be down, and processing and storage components can fail to perform properly. The study reveled that the IoT components distribution, collaboration and intelligent elements location (which are discussed in first chapter) impact the system resiliency. We further noticed that in addition to FT-IoT in cloud level, several studies extend the application to fog and edge computing. There is also foreseen a growing scientific interest on using the microservices architecture to address FT in IoT systems. The chapter gives a foundation to classify the existing and future approaches for fault tolerant IoT, by classifying a set of methods, techniques and architectures that are potentially capable to reduce IoT systems failure.

In another chapter, we critically analyzed a set of IoT distribution and self-adaptation patterns coming from previous chapters and identify their suitable architectural combinations. Further, we used our IoT modeling framework (CAPS) to model an emergency handling system. Based on these, we design two quality driven architectures to be used for a forest monitoring and evacuation example and evaluate and compare them w.r.t their quality of *energy efficiency*.

**Keywords.** *IoT, Software architecture, Styles, Patterns, Systematic mapping study, Fault-tolerance, Energy efficiency, Performance*

**Overview.** It is foreseen that 26 billion devices by 2020 and 500 billion devices by 2030 will be connected to the Internet (37) and business to business spending on IoT technologies, apps and solutions will reach 267 billion dollars by 2020 (28). Another estimation says that the IoT has a potential economic impact of 11 trillion dollars per year by 2025, which would be equivalent to about 11 % of the world economy (64). Such predictions are a matter of encouragement for companies to invest on IoT based applications and to build their pillars on IoT in order to achieve their desired value creation and sustained competitive advantage. Along with a suitable degree of maturity regarding technologies and solutions applied on the identification, connectivity and computation of IoT components, a slope up over architectural concerns is further apparent. Hence, a role of the academic community might be providing a set of standard architectures to assure the efficiency and quality of IoT hardware and software components in practice.

Our attention goes to one specific pillar of software architecture, that is, architectural styles and patterns for engineering IoT applications. Such a focus is driven by a concrete need: since our team is involved in the design and implementation of IoT-based urban security systems, we have been looking for architectural styles and patterns driving the way IoT components shall be combined. Various IoT elements (such as identification, sensing, communication, computing, service and semantic) handle different tasks to build an IoT system. However, these elements need to adapt themselves to changes in their own situation and execution environment. Feedback control loops have been identified as crucial elements in realizing self-adaptation of software sys-tems (108) (74). Among various control loops such as OODA (observe, orient, decide, act), MAPE-K (monitoring, analysis, plan, executing, knowledge), and cognitive cycle (sensing, analysis, decision, action) (71), MAPE-K loop is more often used to perform self-adaptation.

Another challenge is to assess the combined *self-adaptive IoT architectures* based on their quality attributes satisfaction level. Muccini et al (77) specified the most recognized quality attributes to be satisfied with a proper IoT architecture, which are for example, scalability, security, interoperability and performance. Considering the case of indoor/outdoor emergency handling (the domain of this thesis), an IoT-based system should provide a strong degree of fault-tolerance (FT), energy efficiency and performance. A dependable IoT system should provide reliable and fault-free services. A fault is a defect within the hardware or software systems that impacts the correct functionality.

It is particularly difficult to establish a pattern for FT in IoT, since the IoT devices are heterogeneous, highly distributed, powered on battery, relied upon wireless communication and affected by scalability. The distribution of IoT devices cause the system to suffer from, e.g., server crash, server omission, incorrect response and arbitrary failure. The wireless and battery dependency makes the IoT devices barely recoverable. Furthermore, being exposed to new devices and services impacts the system performance.

In safety critical systems such as evacuation, performance becomes a critical issue since the time from gathering data from IoT sensors, running evacuation algorithms and performing the actuation should be such minimized to be complaint with the real-time nature of system. Energy efficiency also becomes an important factor if no power source is available (anymore) and the IoT devices should sill support handling the emergency situation. Hence, the software architecture should help to keep the energy consumption in a minimum level.

The audience of this study are both research and industry communities interested to improve their knowledge and select a suitable architectural style for their IoT system.

## Chapter 1

# IoT Architectural Styles

Conforming a systematic mapping study (SMS) selection procedure, we picked out 63 papers among over 2,300 candidate studies. To this end, we applied a rigorous classification and extraction framework to select and analyze the most influential domain-related information. This chapter is structured as follows. Section 1.1 motivates the need for this study. Section 1.2 reveals the design of this systematic study. Section 1.3 presents a taxonomy on IoT architectures and provides background. Sections 1.4, 1.5, 1.6 and 1.7 elaborate on the obtained results whilst Sect. 1.8 runs a number of horizontal analysis over the results and discusses the obtained results. Section 1.9 analyses threats to validity and Sect. 1.10 closes the chapter and discusses future works.

## 1.1 Motivation

This section discusses the motivation that this research arose from and argues the potential scientific value of it. Thus, an extensive search has been performed in Subsect. 2.1 to discover the related existed systematic reviews. Subsection 2.2 gives a concise reasoning upon the necessity for a systematic mapping study on IoT Architectural styles.

### 1.1.1 Existing Mapping Studies Related to IoT Architectures

Toward learning the already conducted systematic studies (literature review (SLR) and SMS) related to this research topic, we performed a manual search using the following search string:

> ("systematic mapping study" OR SMS OR "systematic literature rev" OR LR) AND (IoT OR "Internet of Things" OR "Internet-ofthings" OR "Internet of Everything" OR "Internet-of-everything") AND ("software architecture" OR "system architecture" OR architecture)

Subsequently, in order to best organize the search, following inclusion and exclusion criteria are determined.

*Inclusion Criteria*: *i)* Studies performed a systematic literature review or mapping study on architectural solutions, methods, styles, patterns or languages specific for IoT and IoE; *ii)* Studies written in English language and available in full-text; *iii)* Studies subject to peer review.

*Exclusion Criteria*: *i)* Studies that are focusing only on architecture or only on IoT (and IOE); *ii)* Studies that are NOT secondary (systematic literature reviews and mapping studies); *iii)* Studies in the form of tutorial papers, editorials, etc.

Further, a multi-stage search and selection process has been performed based on three authentic databases: the ACM Digital Library, ISI Web of Science, and Wiley Inter Science. We initially found a total number of 317 papers and after impurity removal, merge and duplication removal, the selection process applied on 214 remaining studies. After all, we did not find any systematic study on the topic. However, a slightly related study with different objective and scope has been chosen to be compared with our research. The search and selection procedure can be find at the following link: `https://www.dropbox.com/s/bxri9gv91sv5ttu/DE.ECSA-IoT.Style.xlsx?dl=0`.

The research (23) conducted a systematic survey that purposed on categorizing the challenges arise from cloud-based software systems architecture. Strengths: The paper is well-structured and follow a clear methodology and research questions, concluded by a framework for future researches. Why it is different from our work: the paper (23) tries to discover the related literature on software architecture of cloud-based systems; it is merely a review and they do not conclude it with proposing any architecture pattern; it is not specifically related to IoT. Our objective is instead to propose different styles and patterns for IoT architecture, applicable on all IoT domain solutions, whether based on cloud or not.

### 1.1.2 The Need for a SMS on IoT Architectural Styles

This research complements the existing studies regarding the IoT architectures with introducing a literature-based classification of its styles and patterns. So far, a large body of knowledge has been proposed in both IoT systems and software architecture styles, however, a lack of harmonizing and integrating them together is undeniable.

Although the IoT has been introduced more than one decade ago, the research and industry communities are still trying to define its different aspects effectively. Trying to discover the impact of existing literature on proposing a new set of IoT architectures, we identify, describe, and classify different styles to help the community to choose the best architecture for their IoT models.

## 1.2 Research Method

The goal of this research is formulated based on the Goal-Question-Metric perspectives (21) as follow:

*Purpose*—to propose a class of IoT architectures
*Issue*—with identifying, describing, and classifying different styles and patterns
*Object*—based on existing IoT architecture approaches
*Viewpoint*—from the research and industry viewpoints.

### 1.2.1 Search Strategy

To achieve the aforementioned goal, we arranged for a set of questions along with their rationale:

- **RQ1.** *What sort of architectural styles can be used in order to model an IoT system?* This research question aims at categorizing different types of IoT architecture styles in detail.

- **RQ2.** *How IoT architectures can be categorized based on their distribution level?* This research question aims at classifying the IoT architectures based on their intelligent edge and element collaboration.

- **RQ3.** *How scientific publications on IoT architectural styles evolved over time? What strategy they used to structure their research?* This research question aims at identifying and classifying the interest of researchers in IoT architectural styles and their various characteristics over time.

- **RQ4.** *What type of evidence (evaluation or assurance) is provided by existing literature on IoT architectural styles?* This question reviews whether the primary studies guaranteed their functionality through a kind of validation or not.

Furthermore, an optimum search strategy is expected to provide effective solutions to the following questions: which, where, what, and when (113).

*Which Approaches?* The search strategy consists of two phases: *i)* an automatic search on academic database; and *ii)* a snowballing. The first step has been performed using a search string (see below) followed by the selection criteria applied on the set of results. Snowballing refers to using the reference list of a paper (backward snowballing) or the citations to the paper (forward snowballing) to identify additional papers. The start set for the snowballing procedure is composed by the selected papers retrieved by automatic search, namely the primary studies, which are selected by applying inclusion/exclusion criteria to the automatic search results. In any case,inclusion/exclusion criteria will be applied to each paper and if a paper considered to be included, snowballing will be applied iteratively, and the procedure ends when no new papers can be found.

*Where to Search?* The electronic databases that we used for the automatic search (ACM, IEEE, ELSEVIER, SPRINGER, ISI Web of Science, and WILEY Inter Science) are known as the main source of literature for potentially relevant studies on software engineering (24).

*What to Search?* Following some test executions and refinements, the search string has been finalized as shown in the listing. We tried to codify the string in a way to be best adapted to specific syntax and criteria of each selected electronic data source.

*(*IoT OR "Internet of Things" OR "Internet-ofthings" OR "Internet of Everything" OR "Internet-of-everything") AND (architecture OR "software architecture") AND (patterns OR styles)

Further, we combined all studies into a single dataset, after removal of impurities and duplicates.

*When and What Time Span to Search?* We did not consider publication year as a criterion for the search and selection steps. Thus, all studies coming from the selection steps, until February 2018, were included regardless of their publication time.

### 1.2.2 Selection Strategy

A multi-stage selection process (Fig. 1.1) has been designed to give a full control on the number and characteristics of the studies coming from different stages. As it is shown in Fig. 1.1, we are not mentioning Science Direct since we did not achieve any result on that. Furthermore, we used "Software Engineering" as a refinement criterion for Springer engine as it led to over 183,000 results that were potentially outside of our intended research area. We did not use Google Scholar since it may generate many irrelevant results and have considerable overlap with ACM and IEEE; nevertheless, we used Google Scholar in the forward snowballing procedure. Hence, we considered all the selected studies and filtered them according to a set of well-defined inclusion and exclusion criteria (Table 1.1).

**Data Extraction.** This step is aimed at identifying, collecting, and classifying data from the selected primary studies (the list is available on online data extraction file) to answer the research questions. To this end, a detailed classification framework has been designed to structure the extracted data. Indeed, designing an effective classification framework needs a comprehensive analysis of the primary studies' content. Further-more, the IoT standards and formal software architecture classifications supported us through categorizing the data extraction. The systematic process that we followed for this phase consists of collecting and clustering the keywords of primary studies.

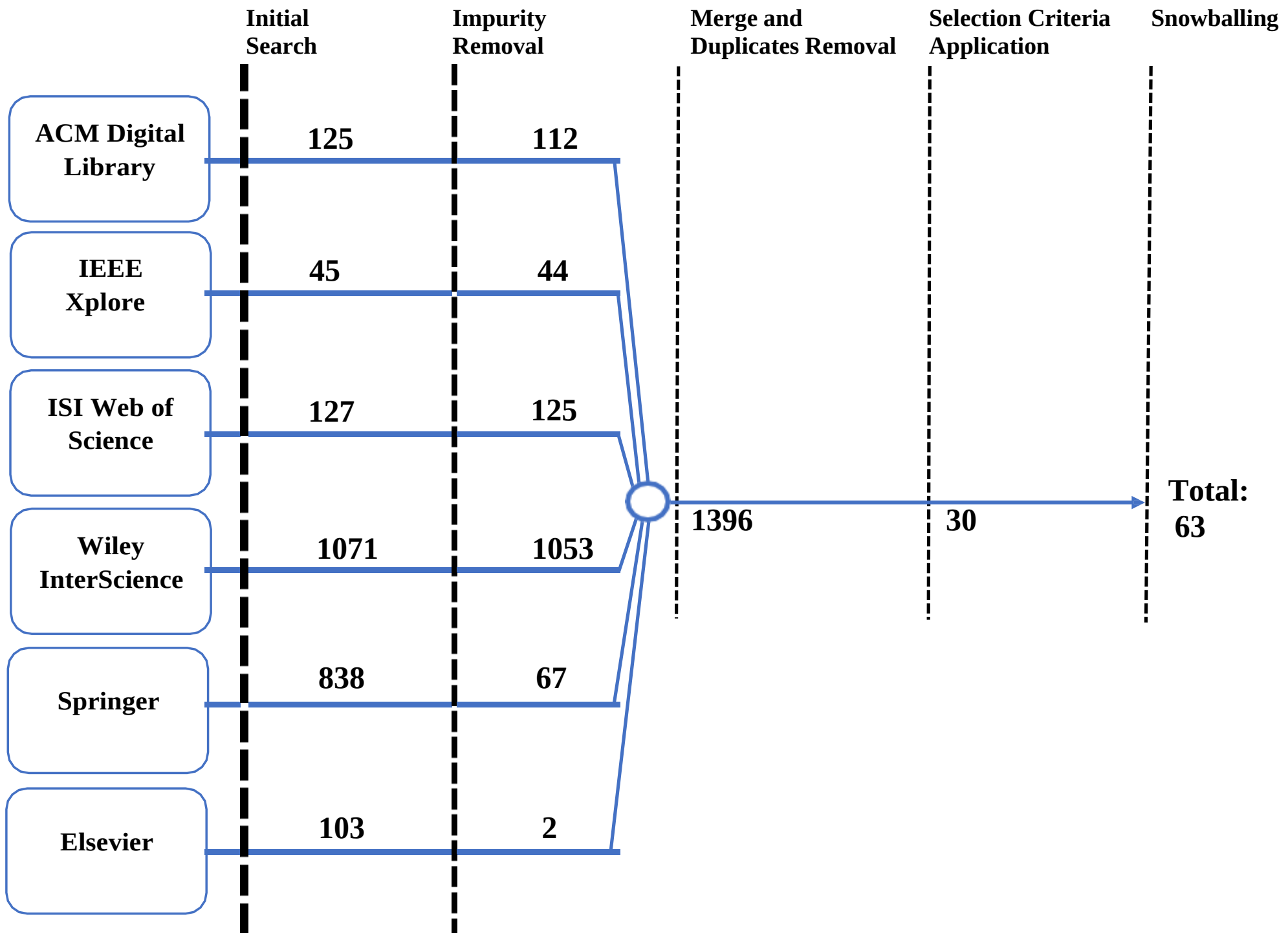

**Figure 1.1** Search and selection process.

**Table 1.1** Inclusion and exclusion criteria.

| Inclusion criteria | Exclusion criteria |
|---|---|
| Studies proposing, leveraging, or analysing architectural solutions, methods, techniques, or styles and patterns, specific for IoT and IoE | Studies that, while focusing on IoT, do not explicitly deal with their architecture (e.g., studies focussing only on technological aspects, inner details of IoT) |
| studies subject to peer review (e.g., journal papers, papers published as part of conference proceedings, workshop papers, and book chapters) | Secondary or tertiary studies (e.g., systematic literature reviews, surveys, etc.) |
| Studies written in English language and available in full-text | Studies in the form of tutorial papers, editorials, etc. because they do not provide enough information |

**Data Synthesis.** The data synthesis activity involves collating and summarizing the data extracted from the primary studies (53) with the main goal of understanding, analysing, and classifying current research on IoT architectures. The data synthesis has been structured of following two phases. *Vertical analysis*: *i)* analysis of extracted data individually to track the trends and collect information of each study with respect to the research questions; *ii)* analysis the discrete extracted data as a whole to reason about potential patterns and trends. *Horizontal analysis: i)* analysis of extracted data to explore possible relations across different dimensions and facets of the research. *ii)* using contingency tables analysis to cross-tabulate and group the data and made comparisons between two or more concepts of the classification framework.

**Study Replicability.** A replication package is provided to tackle the page limits of a the chapter: (`https://www.dropbox.com/s/bxri9gv91sv5ttu/DE.ECSA-IoT.Style.xlsx?dl=0`). The package is available as an excel file with different sheets that include all necessary information such as primary studies, data extraction, keywording and clustering, snowballing, primary studies distribution, validity examination and etc.

## 1.3 Background and Taxonomy

### 1.3.1 Reference Definition of IoT

This section provides some various definitions of IoT mostly derived from our primary studies, then suggest a reference definition for the purpose of this work.

According to P5 (99), the Internet of Things comprises large numbers of smart devices at the network edge that may have to collaborate and interact with each other in real time. P54 (104) defines IoT as an environment in which objects (devices) are given unique identifiers and the ability to transfer data over a network without having humanto- human or human-to-computer interaction. From another view (P32) (90), IoT could be specified as a worldwide network of interconnected entities. As stated in P21 (78) IoT is an ecosystem that interconnects physical objects with telecommunication networks, joining the real world with the cyberspace and enabling the development of new kinds of services and applications.

All aforementioned definitions have their focus on the networking aspect of IoT, whilst the following two definitions emphasize on its computational environment too. IoT is a construction paradigm of computational systems where the objects around us will be in the network in order to extend the capabilities of the environment (P16) (65). The Internet of Things is a technological revolution that represents the future of computing and communications (P34) (110).

IoT can be considered as the future evaluation of the Internet that realizes machineto- machine (M2M) learning. Thus, IoT provides connectivity for everyone and everything (P48) (55). P9 (91) focuses on IoT objectives that are: Convergence, Communication, Connectivity, Content, Computing, and Collections. In the Cluster of European Research Projects report, IoT is defined as an integrated part of the future Internet, which ensures that 'things' with identities can communicate with each other.

From our point of view, IoT is: *the internal/external communication of intelligent components via internet in order to improve the environment through proving smarter services.*

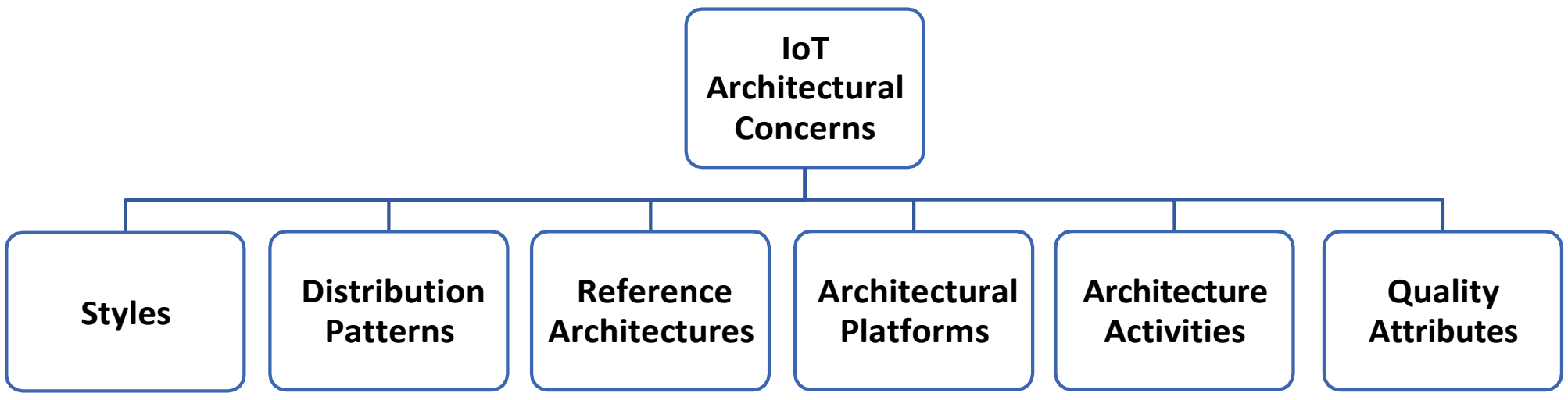

**Figure 1.2** IoT architectures taxonomy.

### 1.3.2 Taxonomy

By analyzing the primary studies under such a dimension, a set of representative concepts have been identified as shown in Fig. 1.2. The taxonomy shows various architectural concerns on IoT systems. The focus of this study goes to the architectural styles and distribution patterns in the following sections, hence, the remaining features are briefly addressed here.

**Reference Architectures.** An IoT reference architecture shall provide a uniform basis to understand, compare and evaluate different IoT solutions. Among our primary studies, (15/63) papers try to develop a kind of IoT reference architecture. For instance, P61 (44) introduces an abstract IoT reference architecture with an abstract view on the components of IoT and their possible connections, in order to ensure a broad applicability. However, a number of more extendable, scalable and flexible IoT reference architectures are presented as architectural platforms.

**Architectural Platforms.** Most of IoT platforms are cloud- based and open-source. Amazon web service IoT platform (AWS) dominates the consumer cloud market. AWS provides multiple data processing services (Amazon S3, Amazon DynamoDB, AWS Lambda, Amazon Kinesis, Amazon SNS, Amazon SQS). However, the core logic of the platform is located within the Message Broker, Thing Registry, Thing Shadows, Rules Engine, and the Security and Identity component, and hence, they are encompassed by the IoT Integration Middleware (44). Microsoft Azure IoT Hub is another example. Its reference architecture is composed of core platform services and application-level components to facilitate the processing needs across three major areas of a typical IoT solution: *i)* device connectivity, *ii)* data processing, analytics, and management *iii)* presentation and business connectivity (29). There are other platforms such as OpenMTC, FIWARE, and SiteWhere, that can be find over selected primary studies.

**Architecture Activities.** The architecture activities variables have been extracted from Li et al. (62) paper. Most discussed activities in architectural level are analysis (32/63) and understanding (30/63) a kind of IoT architecture. This denotes that each study tries to define its own IoT architecture to address a specific problem. However, (19/63) studies reused a special style of architecture that was mostly layered archi-tecture. Evaluation (22/63), description (18/63), synthesis (14/63) are among the superlative used activities but impact analysis (11/63), implementation (10/63), recovery (9/63), and maintenance (8/63) are rarely discoursed.

**Quality Attributes.** The standard used to categorize quality attributes comes from ISO 25010 tied with some specific IoT attributes derived from the primary studies keywording. The architectural style of an IoT system can have effect on quality attributes but does not guarantee all of them. The most recognized quality attributes that are supposed to be satisfied with a proper IoT architecture are scalability (45/63), security (43/63), interoperability (38/63), and performance (37/63). Scalability is an essential attribute as IoT should be capable to perform at an acceptable level with this scale of devices. Furthermore, security gains a high concern in an IoT system, in which different components and entities are connected to each other through a network. Interoperability helps heterogenous components of IoT to work together efficiently. Privacy (32/63), availability (28/63), mobility (26/63), reliability (24/63), resiliency (12/63), and evolvability (9/63) are positioned in the lower degree of concern. Resiliency, that is effective handling the failures and is a critical aspect, is not addressed vastly through primary studies but has a huge capacity to be studied in future researches.

## 1.4 Architectural Styles (RQ1)

The primary studies used one or more overlaid style(s) to design their software architecture. However, among the various IoT architectural styles, layered architecture (34/63) was the clear winner as reported in Table 1.2. In the Layered View the system is viewed as a complex heterogeneous entity that can be decomposed into interacting parts (11). The primary studies designed their layered architecture in different ways, ranged from 3 to 6 layers. As shown in Fig. 1.3, a three-layer IoT architecture is composed of the perception layer, processing and storage layer, and application layer.

The *perception* layer consists of the physical objects and sensor devices (P48) [16] in order to identify and collect environmental information and bring them to the virtual space. The *Processing and storage* layer is in charge of analysing and storing the data gathered by sensors. Various techniques such as cloud computing, ubiquitous computing, database software and intelligent processing are being used to best handle the collected information. The *application* layer provides the service requested by customers (P63) (2) ranging from agriculture to smart healthcare.

Four-layer IoT architecture has one more substrate on the top, that is called *business* layer. This layer is responsible for the handling of entire IoT system. By creating the business models according to dynamic value propositions, this layer designs the roadmap of IoT system. To build a five-layer IoT architecture, a *network* layer can be added to transfer information from perception to processing layer. The transmission medium can be wired or wireless and technology can be 3G, UMTS, Wi-Fi, Bluetooth, infrared, ZigBee, etc. depending upon the sensor devices (P48) (55). A number of studies brought an *adaptation* layer into the IoT architecture to make it six-layer. This layer is positioned between perception and network layers. This layer is an adapter that facilitate interoperability of IoT heterogenous devices.

**Table 1.2** Architectural styles.

| Architecture style | #studies | Studies |
|---|---|---|
| Layered | 34 | P1, P3, P4, P7, P12, P17, P18, P20, P21, P25, P26, P27, P33, P34, P35, P39, P41, P42, P43, P44, P45, P48, P49, P50, P52, P53, P54, P55, P57, P58, P59, P61, P62, P63 |
| Cloud based | 32 | P1, P2, P5, P6, P8, P9, P10, P11, P15, P16, P20, P21, P24, P26, P28, P29, P32, P33, P40, P44, P45, P48, P51, P52, P55, P56, P57, P58, P60, P61, P62, P63 |
| Service oriented | 15 | P3, P9, P13, P14, P16, P19, P22, P23, P26, P28, P37, P38, P51, P55, P63 |
| Microservices | 6 | P6, P13, P16, P19, P46, P47 |
| Restful | 5 | P22, P29, P30, P37, P43 |
| Publish/subscribe | 3 | P10, P27, P31 |
| Information Centric Networking | 2 | P14, P18 |

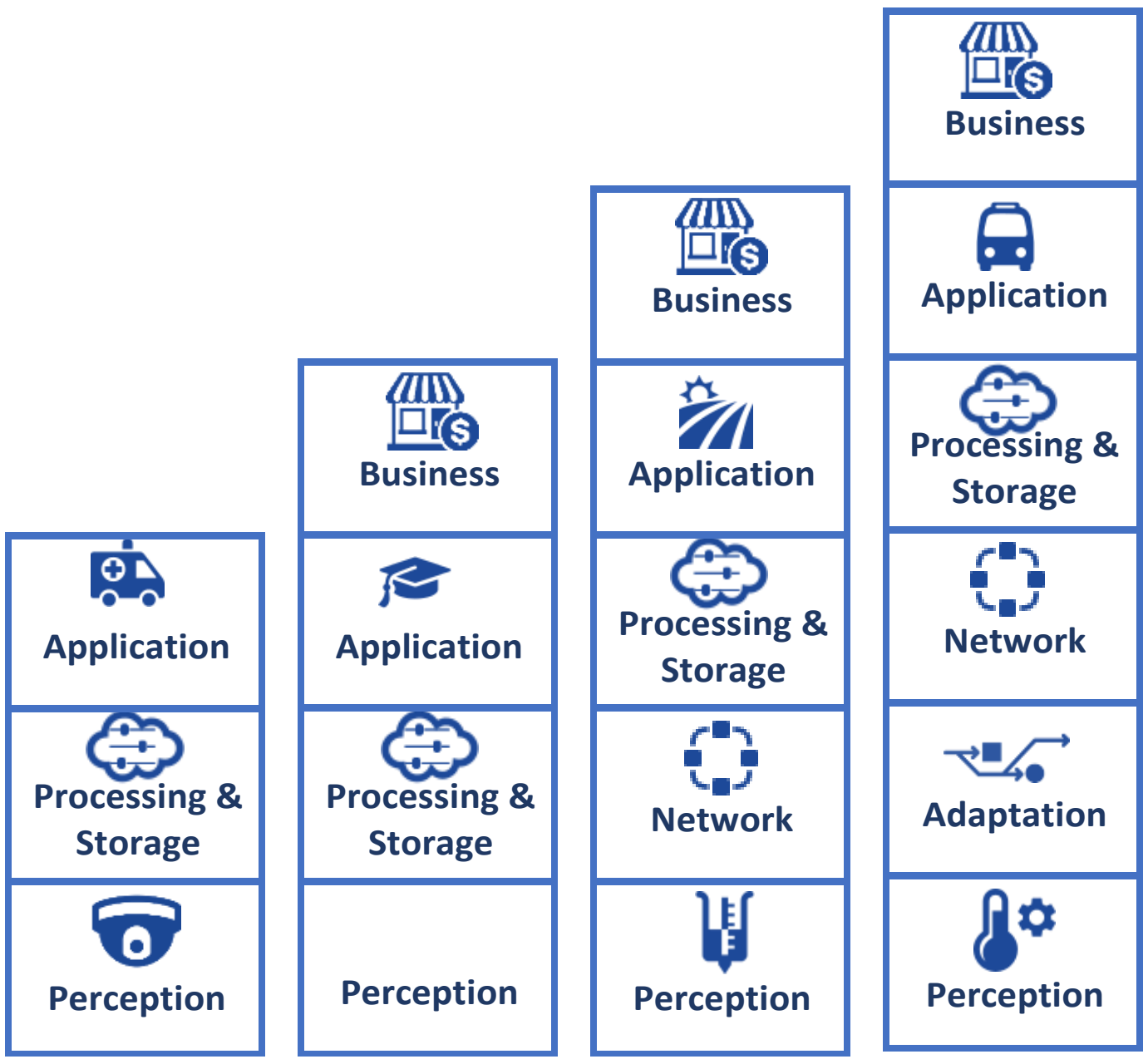

**Figure 1.3** Layered IoT architecture.

Cloud-based architecture (32/63) that has a cloud as the core of their computational part has the second position. Capability of processing and storing big amount of data and providing contextual information, is making cloud computing as an inseparable part of IoT. Fog Computing is a significant extension to cloud environment. Few studies (7/63) addressed fog, as it is a new cloud computing paradigm. Fog brings virtualized cloud services to the edge of the network to control the devices in the IoT (P5) (99).

Cloud architecture is characterized by its various services towards providing an IoT system. As mentioned in P1 (81), Infrastructure as a Service (IaaS), provides virtualized computing resources. The physical machines and virtual machines are stored in the IaaS, and the task of the engines in the IaaS is to mine the data. Data Storage as a Service (DSaaS) provides data storage and information retrieval by a database manager. Platform as a Service (PaaS) provides the tools to work with the machines in the cloud. Software as a Service (SaaS) provides resources to the users for interpretation and visualization of data in the cloud. Fog is positioned between cloud and IoT devices and facilitates the devices to communicate with cloud and provides them processing, storage, and networking services.

Service oriented architectures (SoA) (15/63) put the service at the centre of their IoT service design. In fact, the core application component makes the service available for other IoT components over a network. SOA consists of following three elements. A service provider that is the primary engine underlying the services. A service broker that describes the location of the service and ensures its availability. A service consumer or client that asks the service broker to locate a service and determine how to communicate with that service (92).

Microservices (6/63) and the SOA approach in the IoT have the same goal, that is building one or multiple applications from a set of different services (P19) (19). A microservice is a small application which can be deployed independently, scaled independently, tested independently and which has a single responsibility (100). Literally, the microservice architecture approach utilizes the SoA together with knowledge of software virtualization to overtake the architecture quality limitations like scalability. In this style, an application is built by the composition of several microservices.

Restful (5/63) is underlying architecture organization style of the Web and provides a decoupled architecture, and light weight communication between service producer and service consumers, that is suitable for cloud-based APIs. Restful has its essence on creating loosely coupled services on the Web so that it can be easily reused. It further has advantages for a decentralized and massive-scale service system align well the field of pervasive computing (43).

In Publish/subscribe architectural style (3/63) publisher sends a message on a specific topic, regardless of receiver, and a subscriber can subscribe and receive the same topic asynchronously. The system is generally mediated by a number of brokers which receive published messages from publishers and send them to subscribers.

Information Centric Networking (ICN) (2/63) instead, makes the information as a base of the device communication. ICN matches the application pattern of IoT systems and provides an efficient and intelligent communication paradigm for IoT (45).

## 1.5 Distribution Patterns (RQ2)

On the other hand, IoT distribution patterns classify the architectures according to edge intelligence and elements collaboration (P32) (37). The IoT architecture patterns are classified as: centralized, collaborative, connected intranets, and distributed based on a layered architectural style (Fig. 1.4).

**Centralized.** In this pattern, the perception layer provides data for the central processing and storage component to be provided as services in the next layer. Connecting to this central component is mandatory to use the IoT service. The central component can be a server, cloud, or a fog network connected to cloud.

**Collaborative.** Here a network of central intelligent components can communicate in order to form and empower their services.

**Connected Intranets.** In this pattern, sensors provide data within a local intranet to be used locally, remotely, and centrally. The advantage is that if the central component fails, local service is still in access. The disadvantage is that there is no fully distributed framework to facilitate the communication among components.

**Distributed.** Here all components are fully interconnected and capable to retrieve, process, combine, and provide information and services to other components towards the common goals.

Table 1.3 shows the distribution patterns that are used by the primary studies. Most of studies used centralized pattern (51/63) followed by collaborative (10/63), fully distributed (4/63) and connected intranets (2/63) patterns. Distributed patterns are not widely discussed for IoT architecture, however, there is foreseen a grow specially for industrial applications.

Towards our objectives, we present a three and four layered architecture that are composed of the following layers (Fig. 1.4). Perception: represents the physical sensors and actuators of the IoT that aim to collect information. Processing and Storage: is the central IoT component that stores and analyses the data gathered by perception components to be in access of other entities for their application purposes. Application:

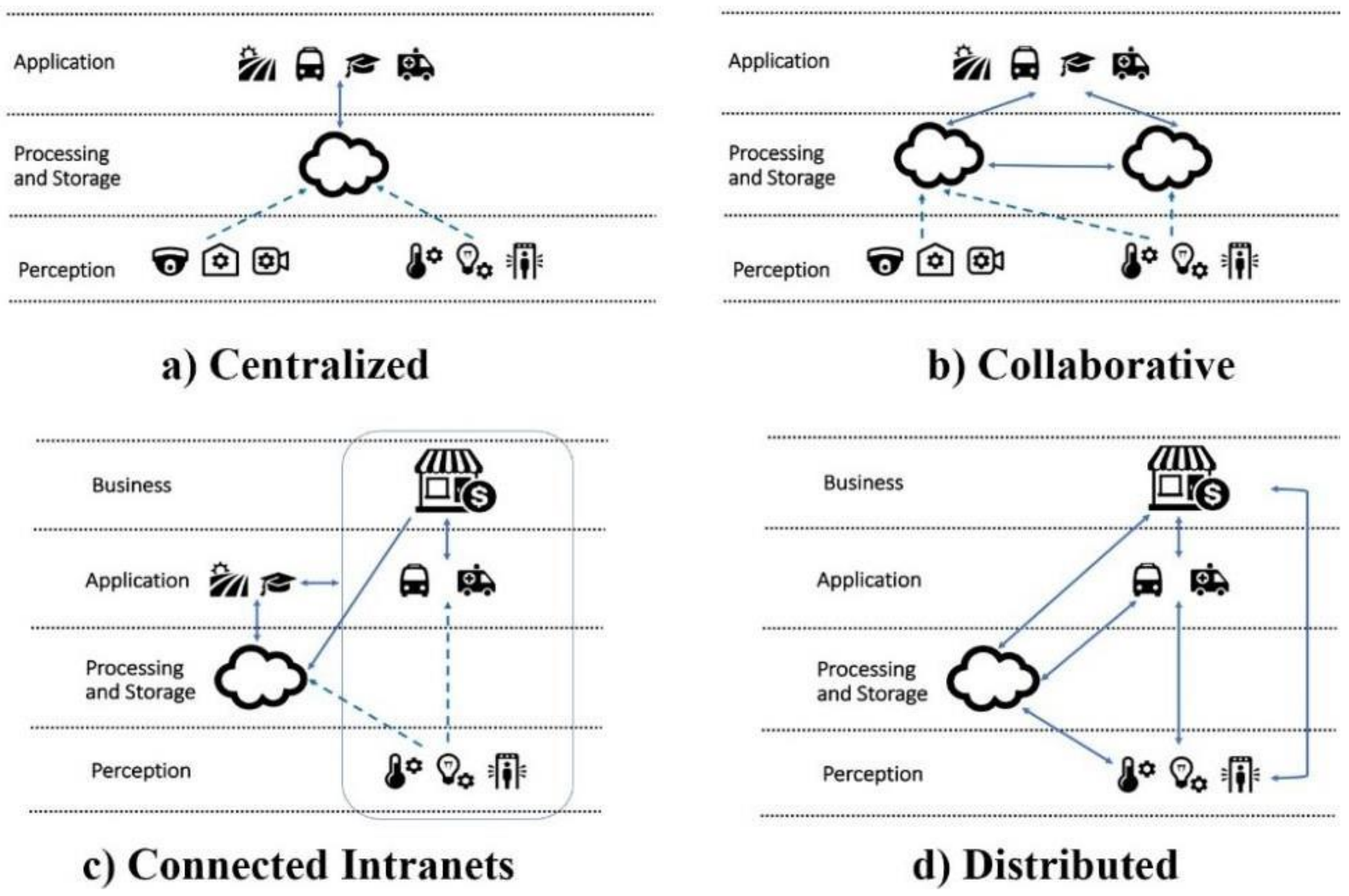

**Figure 1.4** IoT architectural patterns.

determines the class of services provided by IoT. Business: manages the IoT system for its specific goal, by creating business models derived from the information of application layer. The styles are described as follow:

**Table 1.3** IoT distribution patterns.

| Distribution patterns | #studies | Studies |
|---|---|---|
| Centralized | 51 | P1, P2, P5, P6, P7, P9, P10, P11, P12, P13, P14, P16, P17, P18, P19, P20, P21, P22, P23, P24, P27, P28, P29, P31, P32, P33, P34, P35, P37, P38, P40, P41, P42, P43, P44, P46, P47, P48, P49, P50, P51, P53, P54, P55, P56, P57, P59, P60, P61, P62, P63 |
| Collaborative | 10 | P3, P8, P15, P25, P26, P32, P36, P45, P51, P58 |
| Connected intranets | 4 | P4, P32, P39, P58 |
| Distributed | 2 | P32, P52 |

## 1.6 Publication Trend (RQ3)

In this section the publication evolution on IoT architectural styles are presented. To this end, publication year, venue, type and strategy are extracted and discussed below.

**Publication Year.** Figure 1.5 shows the distribution of IoT architectural styles literature. It noticeably indicates that the number of papers grows by time and there are few papers published before 2014. This result confirms the scientific interest and research necessity on IoT architecture issues in the last few years.

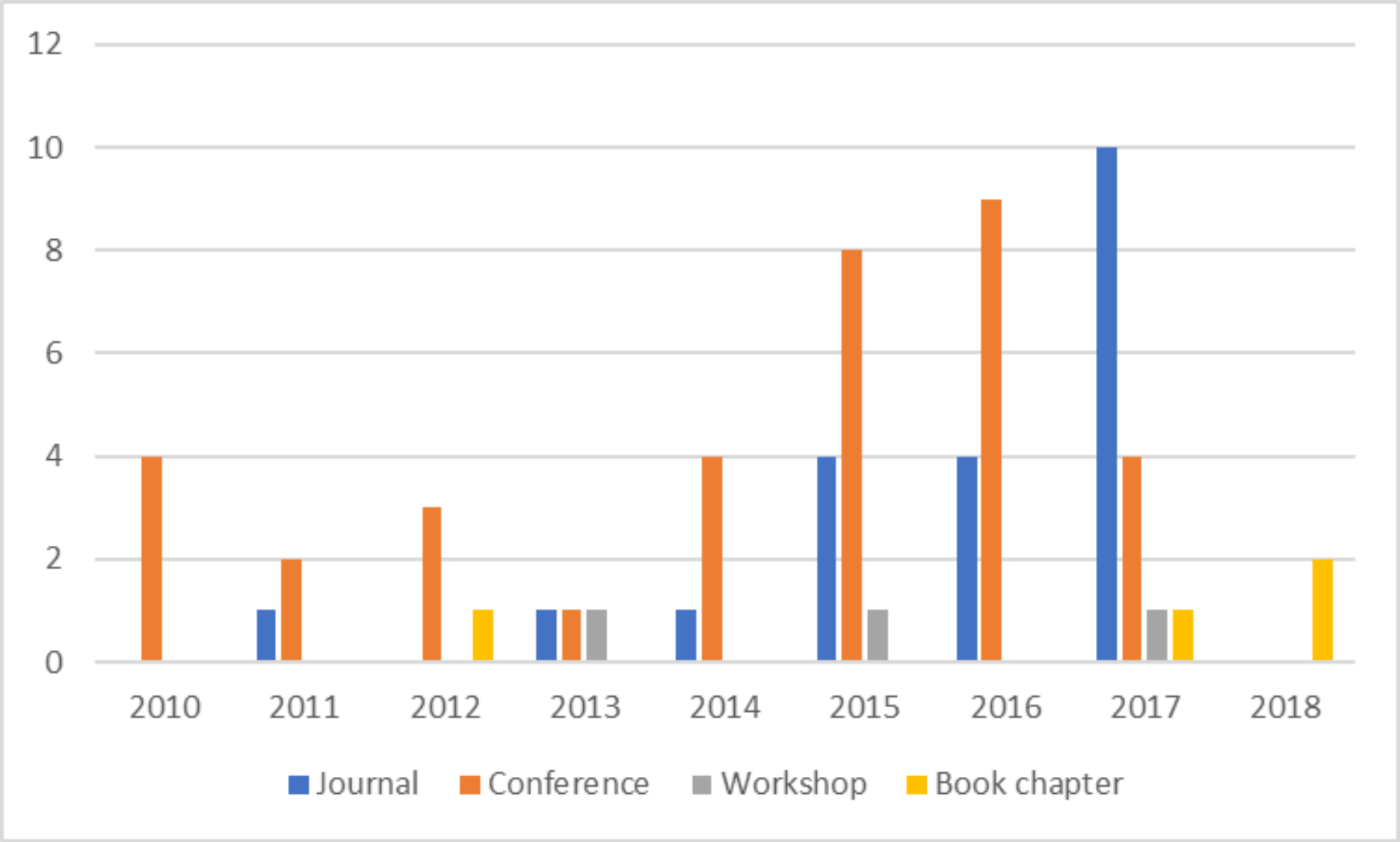

**Figure 1.5** Distribution of primary studies by type of publication.

**Publication Type.** The most common publication type is conference paper (35/63), followed by journal (21/63), book chapter (4/63) and workshop paper (3/63). Such a high number of conference and journal papers may point out that architecting IoT is maturing as a research topic despite its still relatively young.

**Publication Venues.** From the extracted data we can notice that research on IoT architecture is spread across many venues, spanning different research areas such as telecommunications, software engineering, cloud computing, security, etc. This can be figured out as an indication, which IoT architectural styles area is perceived today as orthogonal with respect to many other research areas, rather than a specific research topic.

**Research Strategies.** To learn the dispersion of research strategy across primary studies, we take advantage of well-known research approaches proposed by Wieringa et al. in (109). Solution proposal (39/63) is the most common strategy, followed by philosophical papers (17/63). Considering the IoT as a novel concept, it is justifiable that most of studies try to provide their own solution for architecting it. Evaluation research (16/63) is the third most common strategy highlighting the efforts through industrializing the conducted studies. Validation (10/63) comes afterward, to show the degree of evidence provided by researches. Experience (2/63) and opinion (1/63) research strategies are also used butrarely.

## 1.7 Provided Evidence (RQ4)

**Empirical Method.** Lots of primary studies did not provide any type of evaluation to validate their work (26/63). However, the other empirical methods are used as follows: Experiments (13/63), illustrative examples for evaluation (13/63), case studies (12/63), and prototype (10/63).

**Assurance.** Concerning assurances, (15/63) studies provide some level of evidence for claims using experimental results and (10/63) of the studies use simulation. Few studies used emulation (3/63), formal method (3/63) and consistency checking (2/63) to assure their study functionality. However, in most of studies (41/63), no assurance is provided at all.

These results confirm that the evidence provided by studies is often obtained from experiments, and application of the researches results to toy examples.

## 1.8 Horizontal Analysis and Discussion

This section reports the results orthogonal to the vertical analysis presented in the previous sections. For the purpose of this section, we cross-tabulated and grouped the data, we made comparisons between pairs of concepts of our classification framework and identified perspectives of interest.

### 1.8.1 Architectural Styles VS IoT Distribution Patterns

Here the question is, "Which architectural style is more often used for different IoT distribution patterns?" As shown in Table 1.4, (26/63) studies used the centralized layered architecture and again (26/63) based the centralized architecture on a cloud component. 4 over 11 studies that used collaborative pattern, presented their architecture in a layered style, whilst (7/11) made it based on cloud. The attention on cloud confirms the close relation between IoT and DevOps culture, and the necessity of developing a software computational core for such a system.

**Table 1.4** Styles vs distribution patterns.

| IoT distr. patterns | IoT styles | | | | | | |
|---|---|---|---|---|---|---|---|
| | Layered | Cloud based | SOA | Micro service | Restful | Pub/sub | ICN |
| Centralized (#: 58) | 26 | 26 | 12 | 6 | 4 | 3 | 2 |
| Collaborative (#: 11) | 4 | 7 | 3 | - | - | - | - |
| Connected intranets (#: 4) | 3 | 2 | - | - | - | - | - |
| Distributed (#: 2) | 1 | 2 | - | - | - | - | - |

However, there is a clear research shortcoming on IoT distributed patterns development. The level of distribution has a direct impact on quality attributes satisfaction.

### 1.8.2 IoT Elements vs Quality Attributes

"*What quality attributes need to be best satisfied for each main element of IoT?*" Previous paragraph investigated on deciphering the best software architectural style for IoT. A software architectural style over another, exposes a set of specific quality attributes for the IoT system. Moreover, the wisdom of various IoT elements over the architecture is crucial to design a quality-oriented system. Six main elements of IoT (2)along with their relevant primary studies count are: communication (55/63), sensing (55/63), computing (39/63), service (30/63), identification (27/63), and semantics (22/63). However, we made this horizontal analysis to learn what quality attributes should be focused on for each IoT element. Scalability is the most respected feature for identification element. To improve the scalability of this element, a certain design choice of identification devices can be made. Security is also in the center of attention for IoT elements, despite, interoperability is strongly tied with security and privacy in IoT.

### 1.8.3 Distribution Patterns vs Quality Attributes

"*Which IoT quality attributes should particularly be assured to design an appropriate IoT pattern?*" To answer, the horizontal analysis shows that other than security, scalability, and interoperability that are most respected; IoT distribution patterns are strongly addressing the IoT system's performance. Regarding the rapid development and extension of devices in the edge of the network, performance of IoT should be maintained in an appropriate level. Performance highly depends on the data storage and application logic distribution among edge and central servers. Fog computing is introduced to improve performance level tied with the response time.

## 1.9 Threats to Validity

According to Petersen et al. (82), the quality rating for this systematic mapping study assessed and scored as 73%. This value is the ratio of the number of actions taken in comparison to the total number of actions reported in the quality checklist. The quality score of our study is far beyond the scores obtained by existing systematic mapping studies in the literature, which have a distribution with a median of 33% and 48% as absolute maximum value. However, the threats to validity is unavoidable. Below we shortly define the main threats to validity of our study and the way we mitigated them.

**External Validity**: In our study, the most severe threat related to external validity may consist of having a set of primary studies that is not representative of the whole research on IoT architectural styles. We mitigated this potential threat by *i)* following a search strategy including both automatic search and backward-forward snowballing of selected studies; *ii)* defining a set of inclusion and exclusion criteria. Along the same lines, gray and non-English literature are not included in our research as we want to focus exclusively on the state of the art presented in high-quality scientific studies in English.

**Internal Validity**: It refers to the level of influence that extraneous variables may have on the design of the study. We mitigated this potential threat to validity by (i) rigorously defining and validating the structure of our study, (ii) defining our classification framework by carefully following the keywording process, (iii) and conducting both the vertical and horizontal analysis.

**Construct Validity**: It concerns the validity of extracted data with respect to the research questions. We mitigated this potential source of threats in different ways. (i) performing automatic search on multiple electronic databases to avoid potential biases; (ii) having a strong and tested search string; (iii) Complementing the automatic by the snowballing activity; (iv) rigorously screen the studies according to inclusion and exclusion criteria.

**Conclusion Validity**: It concerns the relationship between the extracted data and the obtained results. We mitigated potential threats to conclusion validity by applying well accepted systematic methods and processes throughout our study and documenting all of them in the excel package.

## 1.10 Conclusion

In this chapter we present a systematic mapping study with the goal of classifying and identifying the domain state-of-the-art and redesign a class of IoT architectural styles respecting the philosophy and granularity of architectural patterns. Starting from over 2,300 potentially relevant studies, we applied a rigorous selection procedure resulting in 63 primary studies. The results of this study are both research and industry oriented and are intended to make a framework for future research in IoT architectural styles field. *This chapter helped us understanding various architectural patterns and style to be used for designing IoT-based emergency evacuation system.*

**Primary Studies**

**Chapter 2**

# Fault-tolerant IoT

This study aims at identifying and classifying the existing FT mechanisms that can tolerate the IoT systems failure. In line with a systematic mapping study selection procedure, we picked out 60 papers among over 2300 candidate studies. To this end, we applied a rigorous classification and extraction framework to select and analyze the most influential domain-related information. The chapter is organized as follows. Section 2.1 reveals the design of this systematic study. Section 2.2 presents a reference IoT architecture and analyzes its associated FT aspects. Sections 2.3, 2.4, 2.5, 2.6 and 2.7 elaborate on the obtained results while Section 2.8 analyses threats to validity. Section 2.9 closes the chapter and discusses future work.

## 2.1 Research Method

The goal of this research is formulated based on the Goal-Question-Metric perspectives (57; 54) as follow:

*Purpose*: to provide a deep understanding on Fault-tolerant IoT systems

*Issue*: by identifying, classifying and analyzing different methods, techniques and architectures

*Object*: based on existing IoT systems approaches

*Viewpoint*: from both research and industry viewpoints.

### 2.1.1 Search Strategy

To achieve the aforementioned goal, we arranged for a set of questions:

- ***RQ1:*** *What IoT architectural styles and patterns are able to make the system prone to fault?*

- ***RQ2:*** *What traditional and novel techniques and methods can protect IoT systems against failure?*
- ***RQ3:*** *What are the quality attributes associated with Fault-tolerance in IoT systems?*
- ***RQ4:*** *What are the trends and evolution that can be deduced from the scientific publications on FT-IoT?*

Furthermore, a good search strategy should provide effective solutions to the following questions (114):

***Which approaches?*** The search strategy consists of two phases: *i)* an automatic search on academic database; and *ii)* a snowballing. The first step has been performed using the search string below. A selection criteria has been subsequently applied on the set of results. Then a snowballing procedure on the included results of the automatic search has been applied to structure the final set of primary studies.

| |
|---|
| *(IoT OR "Internet of Things" OR "Internet-of-Things") AND ("Fault tolerant" OR "Fault-tolerant" OR "Fault tolerance" OR "Fault-tolerance")* |

***Where to search?*** The electronic databases that we used for the automatic search (ACM, IEEE, Elsevier, Springer, ISI Web of Science, and Wiley Inter Science) are known as the main source of literature for potentially relevant studies on software engineering.

***When and what time span to search?*** We did not consider publication year as a criterion for the search and selection steps. Thus, all studies coming from the selection steps, until May 2019, were included regardless of their publication time.

### 2.1.2 Selection Strategy

A multi-stage selection process (Figure 2.1) has been designed to give a full control on the number and characteristics of the studies coming from different stages[1].

Afterwards, we considered all the selected studies, and filtered them according to a set of well-defined inclusion and exclusion criteria (Table 2.1). According to the standards, the definition of inclusion/ exclusion criteria has been guided by two main drivers: *i)* keeping the focus of the selected papers on the scope of the study; and *ii)* avoiding gray or not scientific works. Thus, Inclusion/exclusion criteria shall be aligned with the research questions. We included studies that satisfied all inclusion criteria, and discarded studies that met any exclusion criterion.

[1]It is worth mentioning that we considered "Software Engineering" as the *Search Topic*, since the original search leaded to 193,000 results.

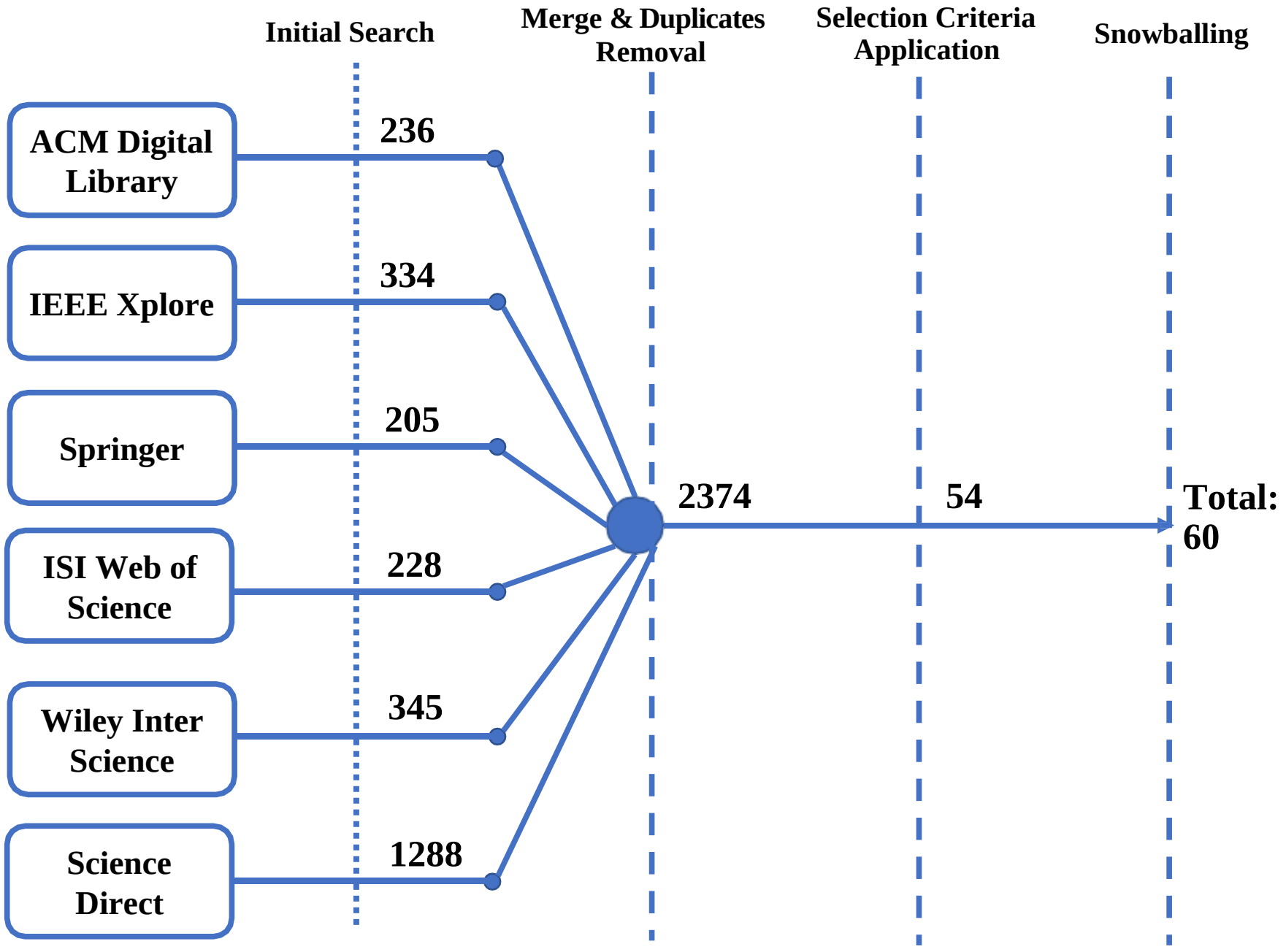

**Figure 2.1** Search and selection process.

On the 2,374 potentially relevant papers, we performed a first manual step applying the selection criteria on title and abstract of the papers. Afterwards, a second manual step of reading the full text of firstly selected papers has been performed and followed by snowballing. The reasons for which we obtained only 60 primary studies over 2,374 potentially relevant papers are that: *i)* our search string was quite inclusive (to avoid ignoring any potentially relevant paper); *ii)* however, selection criteria application has been carefully performed in a way to avoid including the papers that fall out of the scope of the research. In order to minimize bias, the procedure has been performed by the first researcher and the results have been double-checked by the other researcher.

**Table 2.1** Inclusion and exclusion criteria.

| Inclusion criteria | Exclusion criteria |
|---|---|
| Studies that propose, leverage, or analyze software and hardware solutions, methods, techniques and architectures to design fault-tolerant IoT systems. | Studies that, while focusing on IoT, do not focus on its fault-tolerance aspects (e.g., studies focusing only on technological aspects of IoT) or vice versa. |
| Studies subject to peer review (e.g., journal papers, papers published as part of conference proceedings, workshop papers, and book chapters). | Secondary or tertiary studies (e.g., systematic literature reviews, surveys,etc.). |
| Studies written in English language and available in full-text. | Studies in the form of tutorial papers, editorials, etc. because they do not provide enough information. |

After selection of a final set of primary studies, the data has been extracted to answer the research questions.

***Study Replicability.*** A replication package is pro

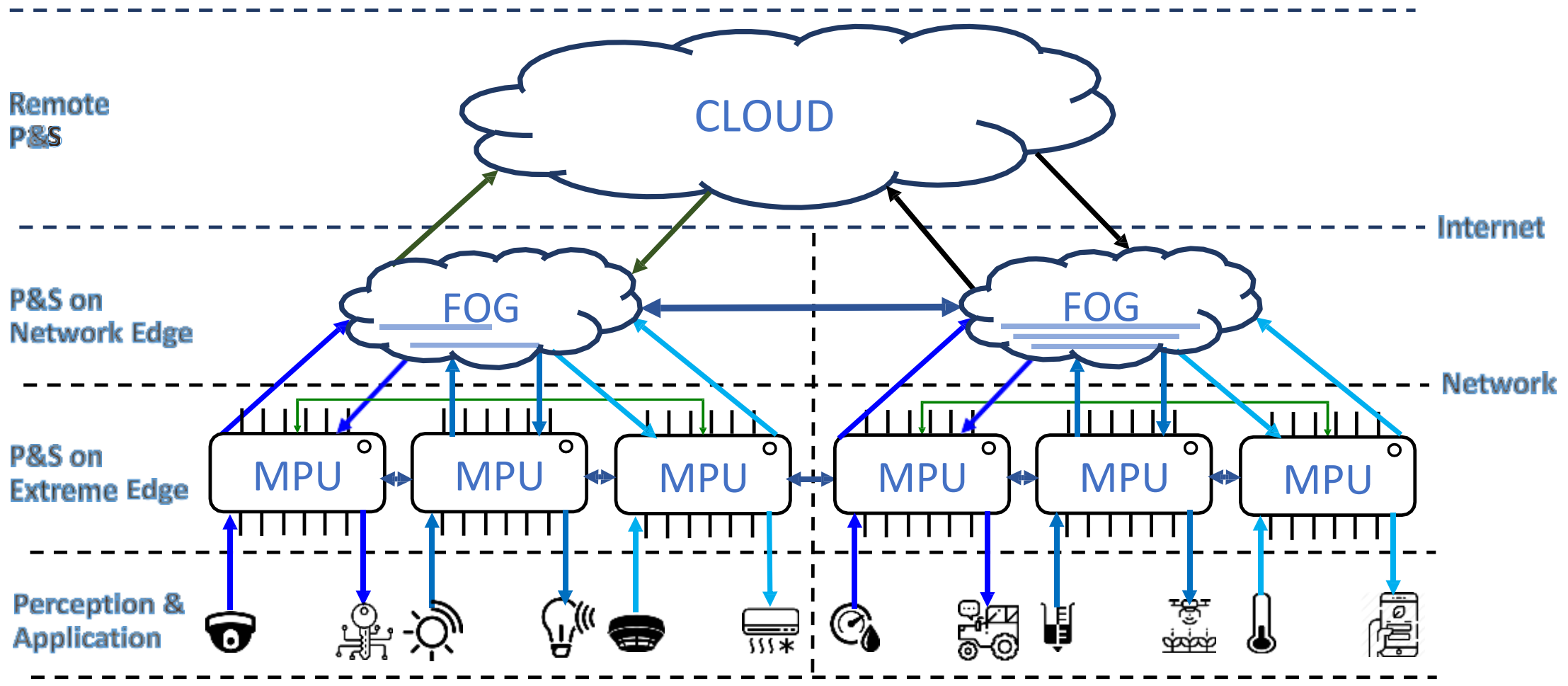

**Figure 2.2** IoT reference architecture (MPU refers to microprocessor unit).

vided to tackle the page limits of the chapter: `https://www.dropbox.com/s/ansb75ncdoqpc9f/DATA-SERENE-2019.xlsx?dl=0`. The package is available as an excel file with different sheets that include all necessary information such as search results, primary studies distribution, data extraction and validity examination.

## 2.2 Background on IoT Architectures

In this section, we present a reference software architecture for the internet of things applications (75; 69; 6). IoT applications typically consist of a set of software components including perception, data processing and storage (PandS) and actuation, which are distributed across network(s). For the purposes of this chapter that has its focus on fault-tolerant data transmission and analysis, we define our architecture based on the following PandS modeling characteristics:

- *Distribution*: this aspect specifies whether data analysis software ought to be deployed on a single node or on several nodes that are distributed across the IoT system. In other words, the distribution is referred to the deployment of the IoT PandS software to hardware. By using a distributed style, the latency will potentially be reduced due to data traffic and bandwidth consumption minimization. Such rapid response time facilitates real-time and fault-tolerant IoT applications. Furthermore, in distributed systems, a faulty PandS will still hold IoT system available since the faulty component can be replaced by another one.

- *Localization*: depending on data size and required analysis complexity, PandS can be executed locally or remotely. Here is the point in which centralized cloud and distributed edge and fog concepts become relevant. The advantage of using a central cloud is that, processing on a cloud component facilitates long-term data analysis for systems that have no constraints on response time. For applications with massive PandS requirement, executing the task on the powerful cloud is the only solution.

Fog nodes are the intermediate PandS, which bring a degree of cloud functionality to the network edge. Fog is not limited to perform on a particular device, so that it can freely be located between device edge and cloud. The analysis capacity of fog is lower than cloud, but it reduces a significant point of failure by shifting towards more than one computational component. However, fog only performs locally so that it does not have a global coverage over a major IoT system. It is worth mentioning that, some IoT devices are able to perform simple PandS by themselves. Performing PandS on IoT device edge, refers to computation capabilities embedded on a smart device to be able to gather and analyze environmental data.

- *Collaboration*: the aforementioned computation components may interact to form and empower IoT services. This collaboration may appear as a level of information sharing, coordinated analysis and/or planning or synchronized actuation. Each IoT sensor network may provide data for many collaborative PandS components, both locally and remotely. Here the advantage is that if the local PandS node fails, local service is still in access.

Considering above definitions, we further design our reference IoT architecture (Figure 2.2). The architecture is composed of a physical layer and several PandS layers. The physical layer is made up of two sub-layers, namely *perception* and *application*. The perception sub-layer hosts a large number of heterogeneous sensors and the application sub-layer consists of various types of actuators. The PandS layers store and analyze data gathered by the perception components to provide the required IoT service.

Looking through primary studies, each of them address the FT for specific layer(s) of the IoT architecture. As shown in Figure 2.3, whilst the faults usually occur in sense (26/60) and actuation (12/60) sub-layers, the primary studies realized the importance of network (38/60) and PandS (33/60) layers for FT-IoT systems. The reason is that, handling FT is under the responsibility of PandS nodes and is based on the transmitted data coming from the physical layer. In Section 5, we discuss various FT strategies and techniques for IoT systems.

## 2.3 Fault-tolerant IoT Architectural Patterns and Styles (RQ1)

This section discusses the specific characteristics of primary studies related to FT-IoT architectural design. The primary studies used one or more overlaid style(s) to design their software system. However, among the various IoT architectural styles, layered architecture (32/60) was the clear winner as reported in Figure 2.4. In the layered view the system is viewed as a complex heterogeneous entity that can be decomposed into interacting parts. The primary studies designed their layered architecture in different ways, ranged from 3 (with a central PandS component only) to 5 (including edge and fog) layers (see Figure 2.2).

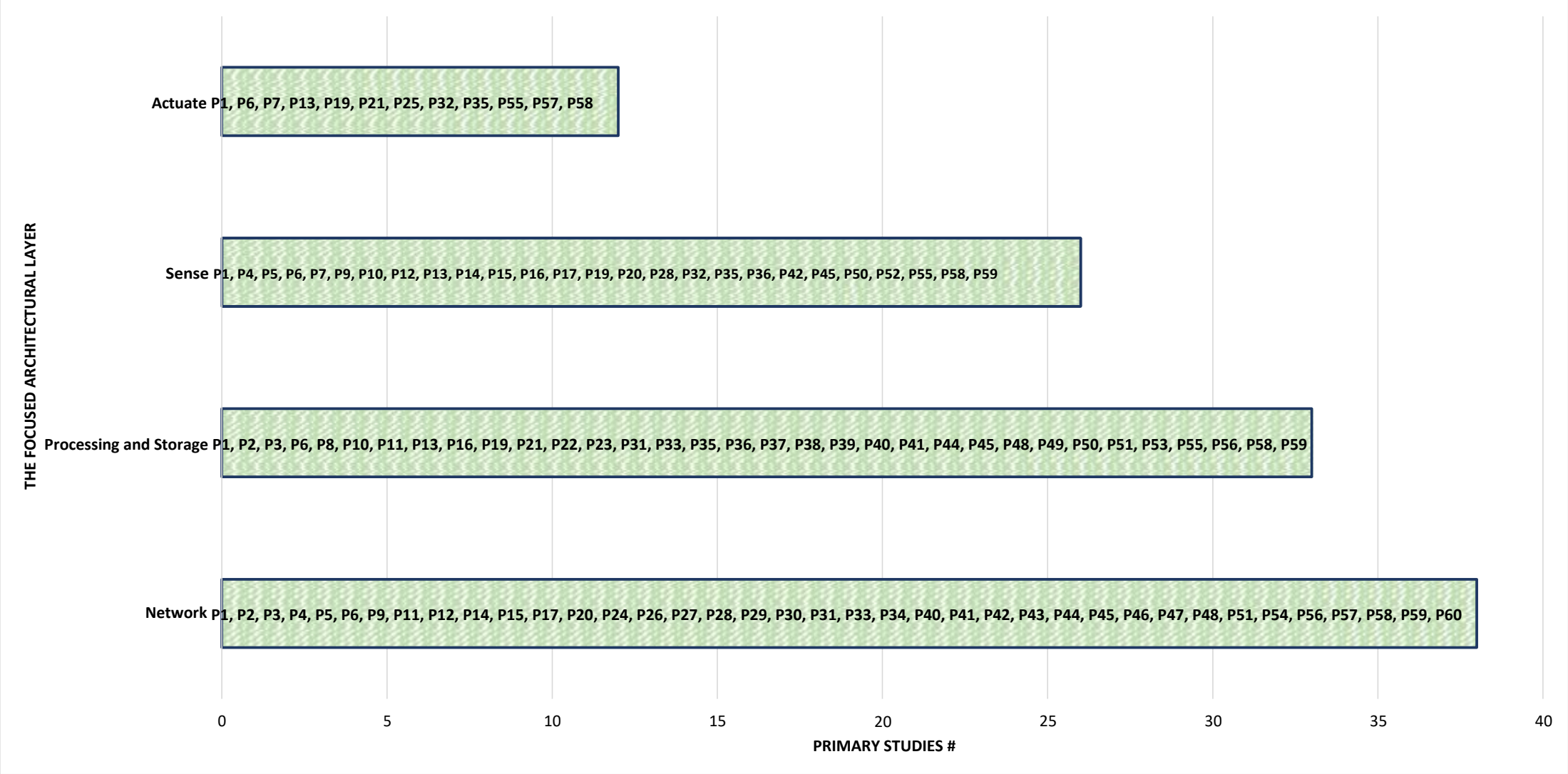

**Figure 2.3** The primary studies focus on each architectural layer.

Cloud-based architecture (28/60) won the second position. Fog that is a significant extension to cloud environment is addressed in 15 studies as well. Few studies (4/60) used the device edge concept to design their FT-IoT architecture. Minimizing the impact of a failed component within an integrated fog-cloud platform needs a common agreement protocol that is able to uniform the system with the minimum rounds of message exchange.

Service oriented architectures (SOA) (9/60) put the service at the centre of their IoT application design. In fact, the core application component makes the service available for other IoT components over a network. Microservices (4/60) and SOA have the same goal in IoT sytems, that is building one or multiple applications from a set of different services. A microservice is a small application with single responsibility, which can be deployed, scaled and tested independently.

P21 proposes a pluggable framework based on a microservices architecture that implements FT support as two complementary microservices: one that uses complex event processing for real-time FT detection, and another that uses online machine learning to detect fault patterns and preemptively mitigate faults before they are activated. P7 propose a system based on container virtualisation that allows IoT clouds to carry out fault-tolerance when a microservice running on an IoT device fails. A reactive microservices architecture and its application in a fog computing case study to investigate FT challenges at the edge of the network is presented in P40 . P56 present a microservices-based mobile cloud platform by exploiting containerization which replaces heavyweight virtual machines to guarantee run-time FT.

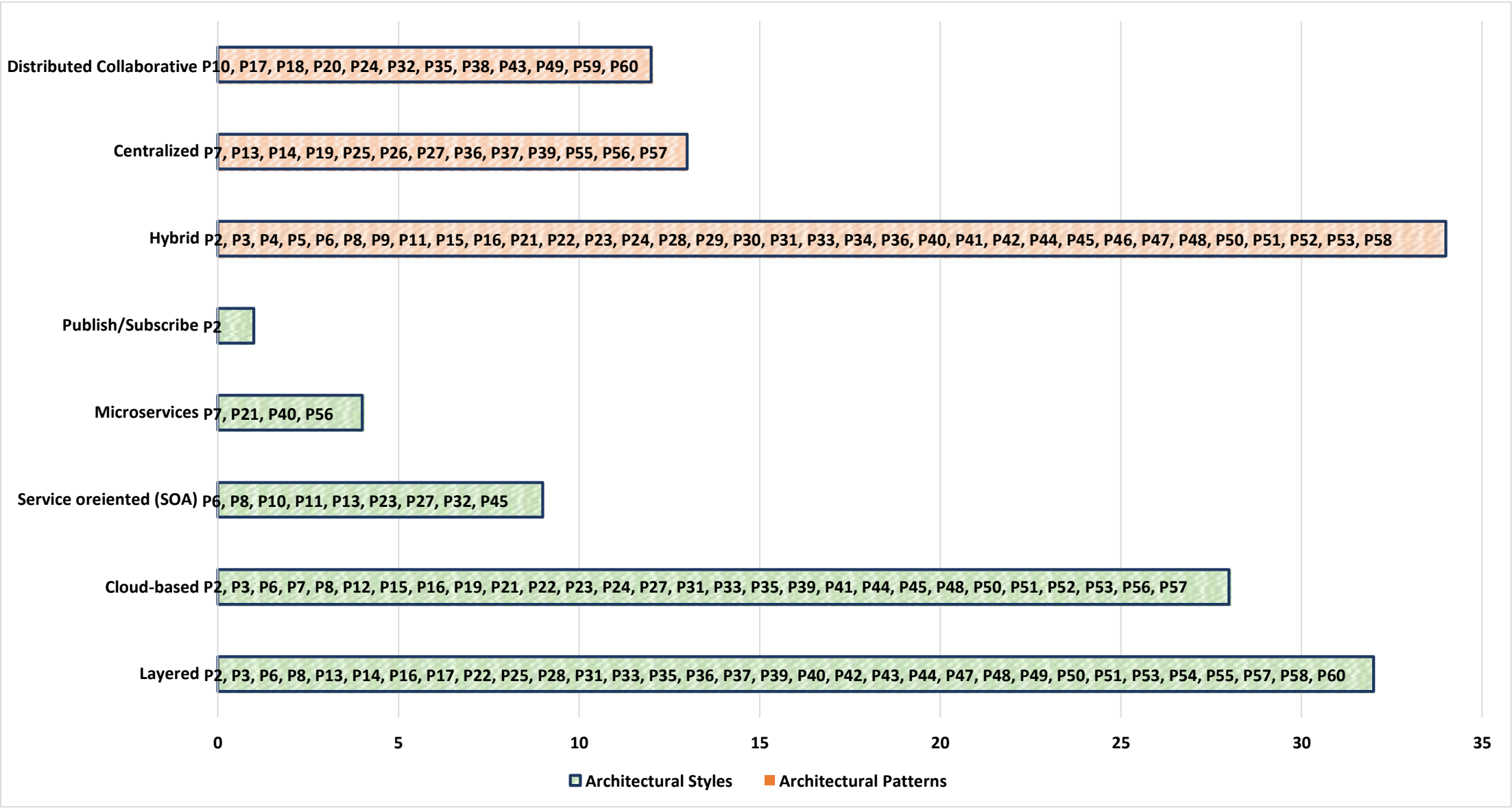

**Figure 2.4** FT-IoT architectural styles and patterns.

On the other hand, as explained in Section 3, IoT distribution patterns classify the architectures according to edge intelligence and elements collaboration. Figure 2.4 shows the distribution patterns that are used by the primary studies. Most of studies used a Hybrid pattern (34/60) followed by the Centralized (13/60) and the Distributed Collaborative (12/60) patterns.

In this section we showed that edge/cloud-based distributed architectures are extensively used by primary studies. The results confirm that: a distributed architecture provides a rapid response time and high availability, and makes the system prone to fault.

## 2.4 Fault-tolerance Techniques for Resilient IoT (RQ2)

As shown in Figure 2.5, the primary studies adopt various techniques to make their IoT system fault-tolerant. These techniques are explained below.

### 2.4.1 Replication

Replication is the process of sharing the data between redundant IoT HW/SW components. Replication guarantees the data consistency, so that failure of a component will not result in system failure. The main replication schemes are known as *active and passive* (39).

In active replication scheme (22/60), processes are replicated in multiple processors to provide fault-tolerance. In IoT context, active replication continuously pushes the group of IoT resources (such as fog or cloud) to execute the same process concurrently. In case of fault, failover can have in very short period to other active resources [P33]. In this way, an extra processing is occurred and redundant and duplicated dataset it sent to endpoint. Despite that active replication takes a lot of processing resources, it is failure transparent and its failure discovery time is deterministic.

In passive replication (24/60), the primary processor performs and the extra IoT components remain idle until a failure occurs. The idle components, however, contact the primary processor in order to be updated and keep consistency. The passive replication scheme imposes additional cost of resources and suffers from slow response to failure.

### 2.4.2 Network Control

In network control scheme (19/60), the IoT network is generally divided into various clusters. A chosen cluster head (CH) periodically makes roll call requests to the other nodes and if it does not receive a reply message, the failure will be confirmed. However, the CH itself makes a single point of failure. Several cluster-based routing protocols have been proposed by the primary studies. Some primary studies took advantage of bio-inspired particle multi-swarm optimization routing algorithm to construct, recover, and select disjoint paths that tolerate the failure while satisfying the quality of service parameters. Some other studies used the virtual CH formation and flow graph modeling to efficiently tolerate the failures of CHs. Multiple traveling salesman is also among the routing algorithms that are addressed by the primary studies.

### 2.4.3 Distributed Recovery Block

In this method (8/60), a single program is concurrently executed on a node pair, from which one is active and the other is inactive. In no-fault situation, the main (active) node performs the task and the other node performs the same task in shadow. Afterwards, both results will be tested and if the test is properly passed, the results associated with the main node will be delivered as the output. If the primary node test fails, the shadow node becomes active and produces the outputs. This method can protect the system only against a single point of failure.

### 2.4.4 Time Redundancy

Time redundancy (1/60) can be performed at both instruction and task levels. At instruction level, the program is duplicated and subsequently the results are compared to discover a potential error. In task level, a software is run twice (or more) to mitigate dynamic faults. Despite that this method does not impose the cost of additional hardware, it increases the time needed to assure redundancy. The method reduces the computing performance and consumes more energy as well.

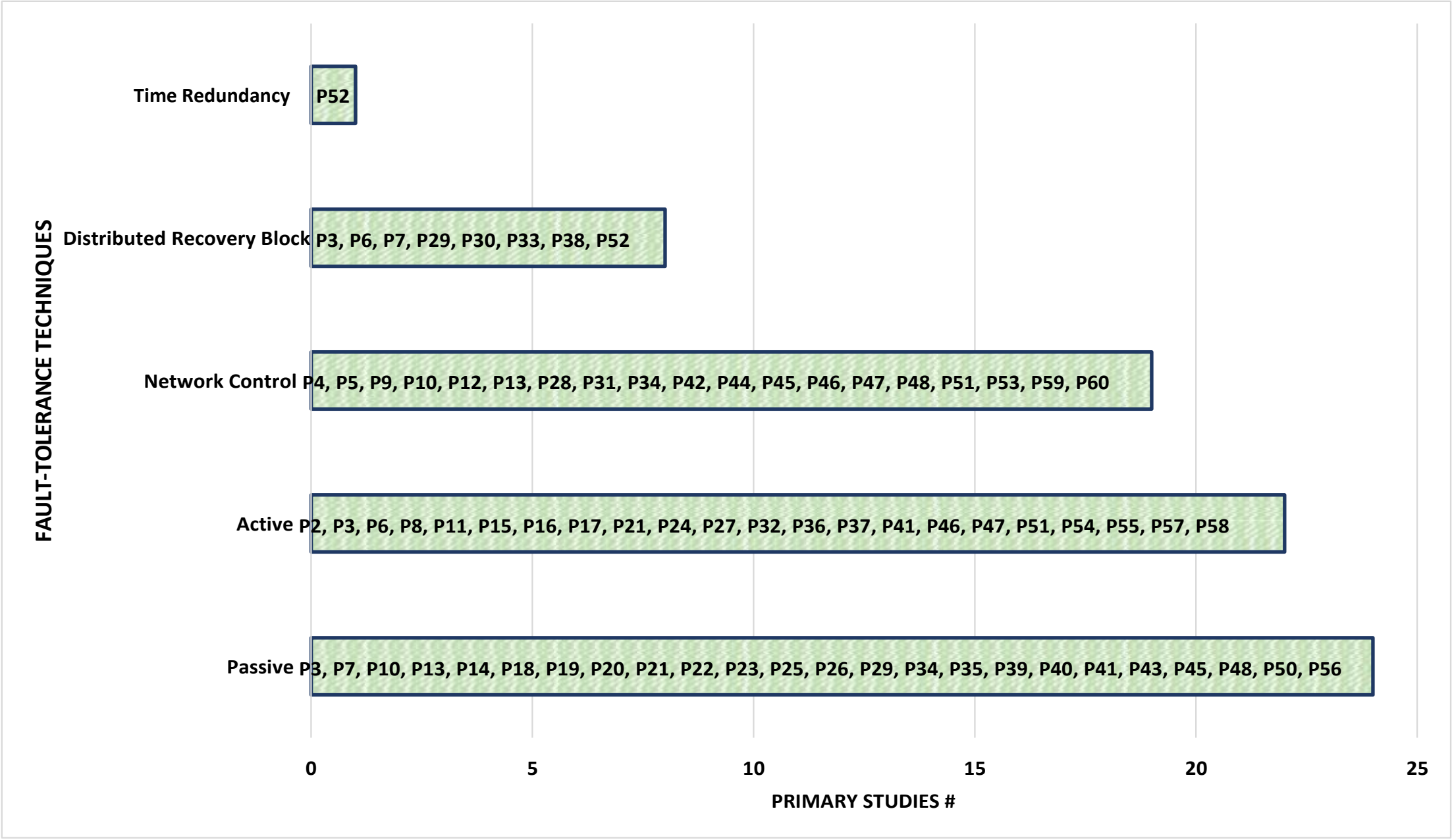

**Figure 2.5** Fault-tolerance techniques.

It is worth mentioning that, the whole IoT system can follow a *Reactive* or *Proactive* strategy. Reactive FT starts to recover the system after the detection of an error (using event processing methods). In proactive FT, the recovery strategy is started even before the detection of an error (using machine learning methods).

## 2.5 Quality of IoT Service Associated with Fault-tolerance (RQ3)

The standard used to categorize quality attributes comes from ISO 25010 and some specific IoT attributes derived from the primary studies keywording.

An IoT system brings many challenges from QoS perspective when takes into account FT. As shown in Figure 2.6, the most recognized quality challenges are related to performance (25/60), availability (20/60), security (20/60) and scalability (16/60), whilst interoperability (8/60) and energy efficiency (2/60) are positioned in a lower degree of concern.

The level of performance depends on how much the processing and storage components are pushed to the edge in a decentralized way. Availability is the ability of a system to be fully or partly operational as and when required. Clearly, FT and availability are not identical since a fault-tolerant system is supposed to maintain the system operational without interruption, but a highly available system may have service interruption. However, A fault-tolerant system should maintain a high level of system availability and performance as well.

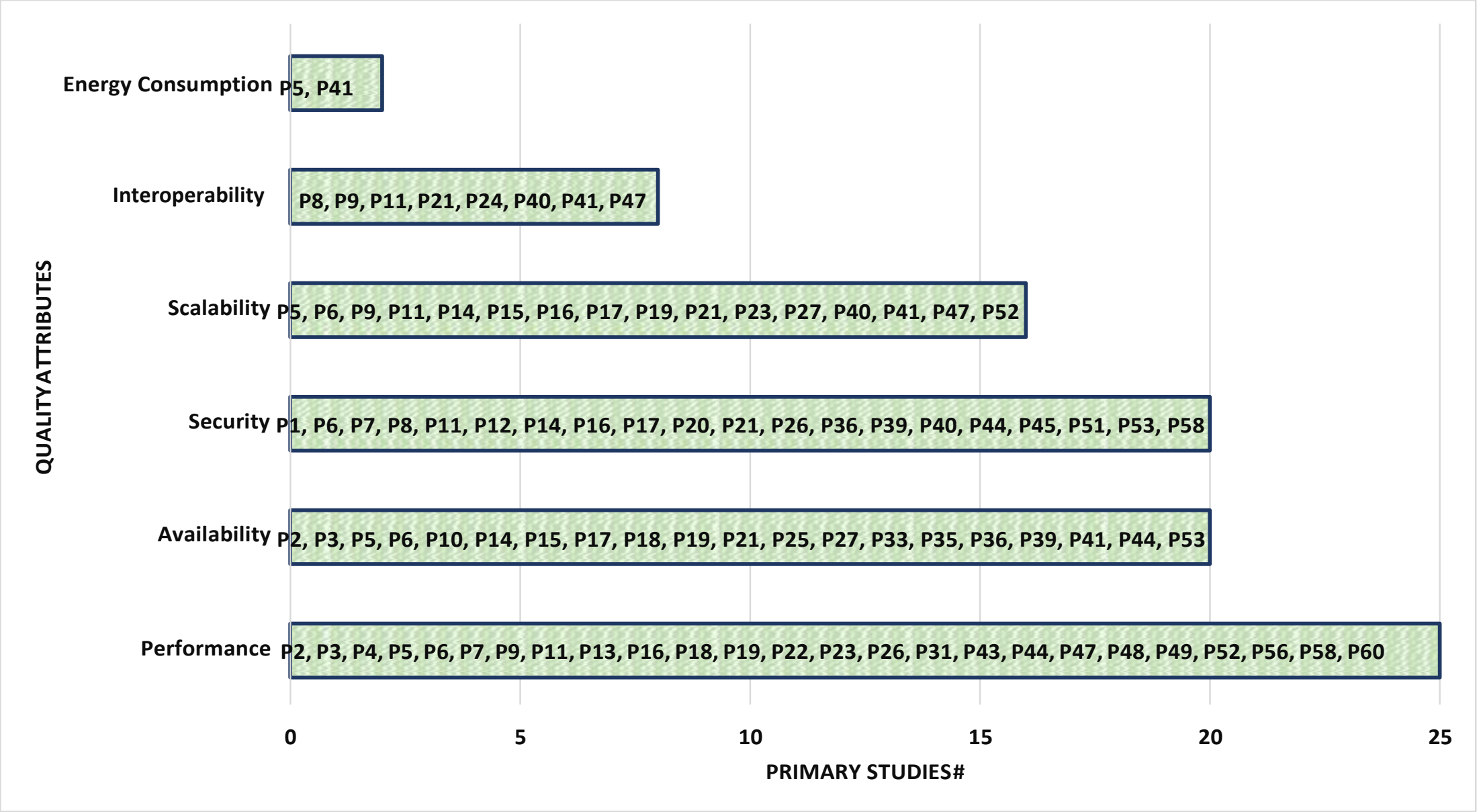

**Figure 2.6** QoS associated with FT-IoT.

In IoT systems that different components and entities are connected to each other through a network, security gains a high concern. Scalability is also an essential attribute as IoT systems should be capable to perform properly considering a huge number of heterogeneous devices. Commenting on scalability of IoT as a whole system is difficult, however, it depends on how new resources can be added on demand. A fault-tolerant system also requires enormous computational efforts to be run in distributed PandS components. Device heterogeneity and PandS elements distribution make the system resistive to scalability.

Interoperability helps IoT heterogeneous components to work together efficiently. It actually depends on how much IoT large-scale heterogeneous devices can communicate directly among each other to gather the required data without having to go through the central/remote components. Since most of IoT devices are battery powered, energy efficiency that is tied to many other quality attributes (such as performance) becomes essential. However, wireless and battery dependency make the IoT devices barely recoverable, flexible to scalability and performant.

## 2.6 Challenges and Emerging Trends (RQ4)

In this section the emerging trends in resilience for FT-IoT are presented. To this end, publication year, type and venue are firstly extracted and an overall discussion is subsequently provided.

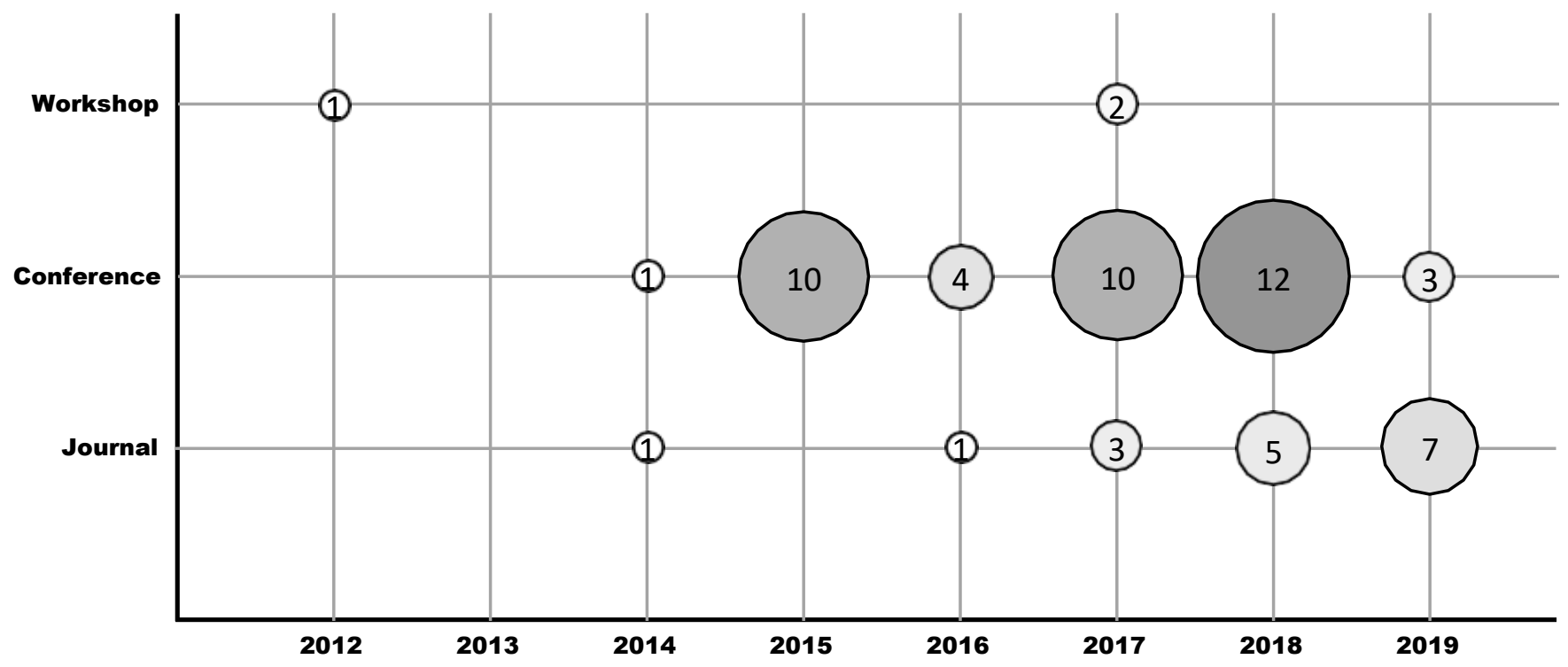

**Figure 2.7** Primary studies distribution by publication type.

### 2.6.1 Publication Year

Figure 2.7 shows the distribution of FT-IoT literature. It noticeably indicates that the number of papers grows by time and there is just one related paper published before 2014. This result confirms the scientific interest and research necessity on FT-IoT issues in the last few years.

### 2.6.2 Publication Type

The most common publication type is conference paper (40/60), followed by journal (17/60), and workshop paper (3/60). Such a high number of journal and conference papers may point out that FT-IoT is maturing as a research topic despite that it is still relatively young.

### 2.6.3 Publication Venues

From the extracted data we can notice that research on FT-IoT is spread across many venues mostly in the span of IoT (e.g. WF-IoT), computing (e.g. ICAC) and networking (e.g. ICOIN) communities. The complete list of venues can be found in the data extraction file. However, the focus on the aforementioned aspects can prove the significance of distributed computing and networking for FT-IoT systems.

### 2.6.4 Emerging Trends in Resilience for FT-IoT

Our study reveals that some of the different Ft-IoT techniques are more rarely covered with respect to others, specifically, distributed recovery block and time redundancy. We clarify that this result by no means implies that there is limited literature or support on such FT techniques, but they appear to have a more limited application on IoT. In architectural level, we observed a significant move toward adopting hybrid architectures, which make the IoT system prone to fault. Furthermore, whilst a growth on using service-oriented and microservices architectures is perceived, their various aspects need to be better investigated regarding FT. The study showed that for FT-IoT architectural layers, the attention especially goes to network and processing and storage components.

What our study reveals is also that performance and availability are tied up with IoT systems fault-tolerance. However, assessing the trade-off between FT and other IoT quality attributes such as scalability, interoperability and energy consumption shall be further investigated. Another result to be further evaluated through a state of the practice analysis, is that only few studies support the interplay between FT techniques and collaborative architectures. The mentioned aspects are to be considered by the domain future work.

## 2.7 Horizontal Analysis

This section reports the results orthogonal to the vertical analysis presented in the previous sections. For the purpose of this section, we cross-tabulated and grouped the data, we made comparisons between pairs of concepts of our classification framework and identified perspectives of interest.

### 2.7.1 FT techniques VS Architectural Patterns

Here the question is, *which architectural pattern is more often used for each FT technique?* As shown in Figure 2.8, (11/60) studies used hybrid pattern to facilitate their passive FT techniques, whilst (15/60) used hybrid for active FT. In contrary, centralized and collaborative architectural patterns are more suitable to address passive FT. Obviously, network control FT technique is better to be addressed by a hybrid architectural pattern. In general, a hybrid architecture guarantees FT-IoT, since if one fog node fails, the IoT system can shift the computation to another fog to avoid the single point of failure.

### 2.7.2 FT techniques VS Quality Attributes

*What quality attributes are satisfied when a specific FT technique is adopted?* As shown in Figure 2.9, passive technique mostly takes into account performance and availability, whilst the active technique gives more weight to security and scalability. Furthermore, network control enhances the performance beside the fault-tolerance. Regarding the rapid development and extension of devices in the edge of the network, performance of IoT should be maintained in an appropriate level. Performance highly depends on the data storage and application logic distribution among edge and central servers. As mentioned before, fog computing can pave the way to improve IoT systems performance level.

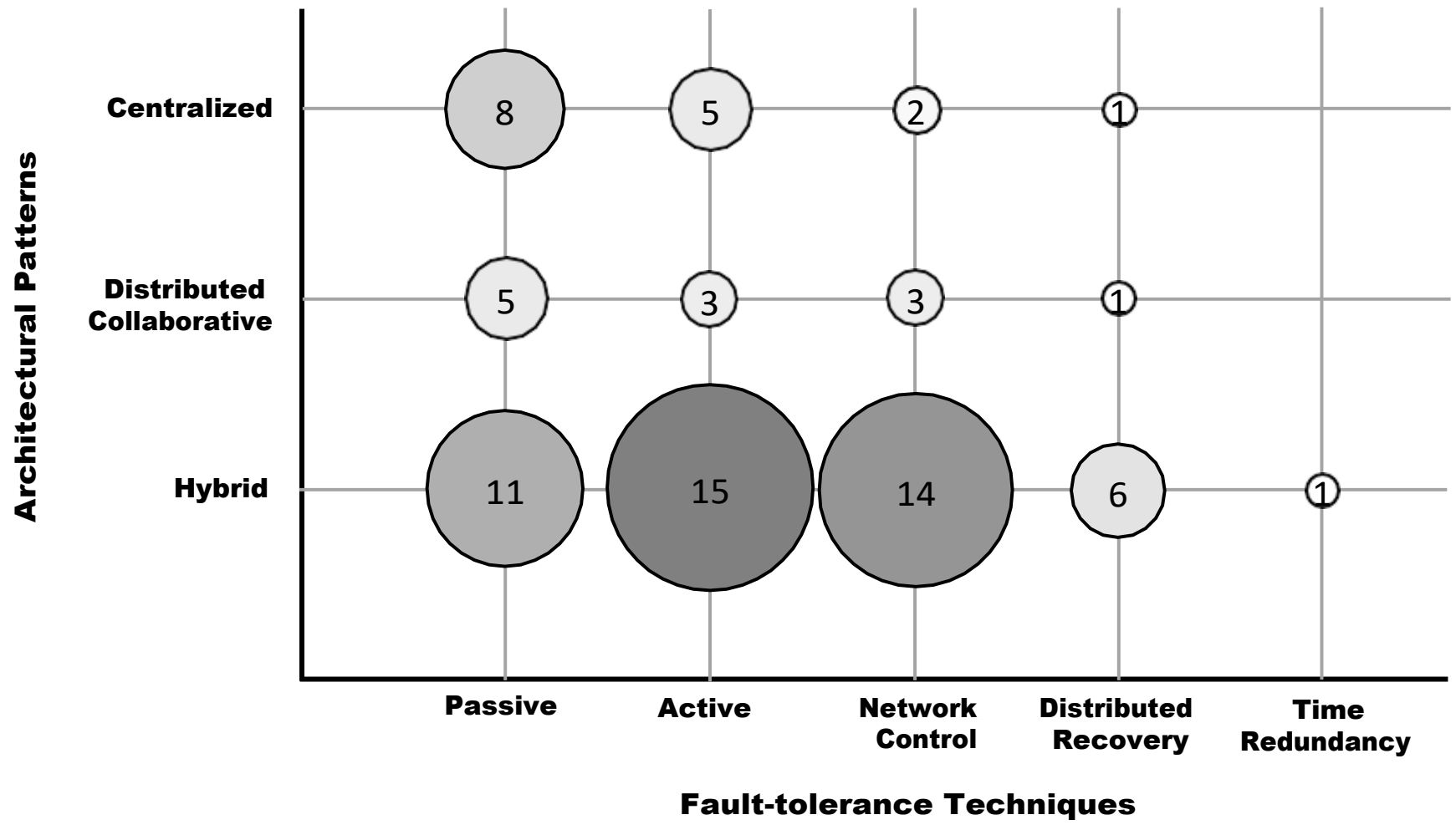

**Figure 2.8** FT techniques VS patterns.

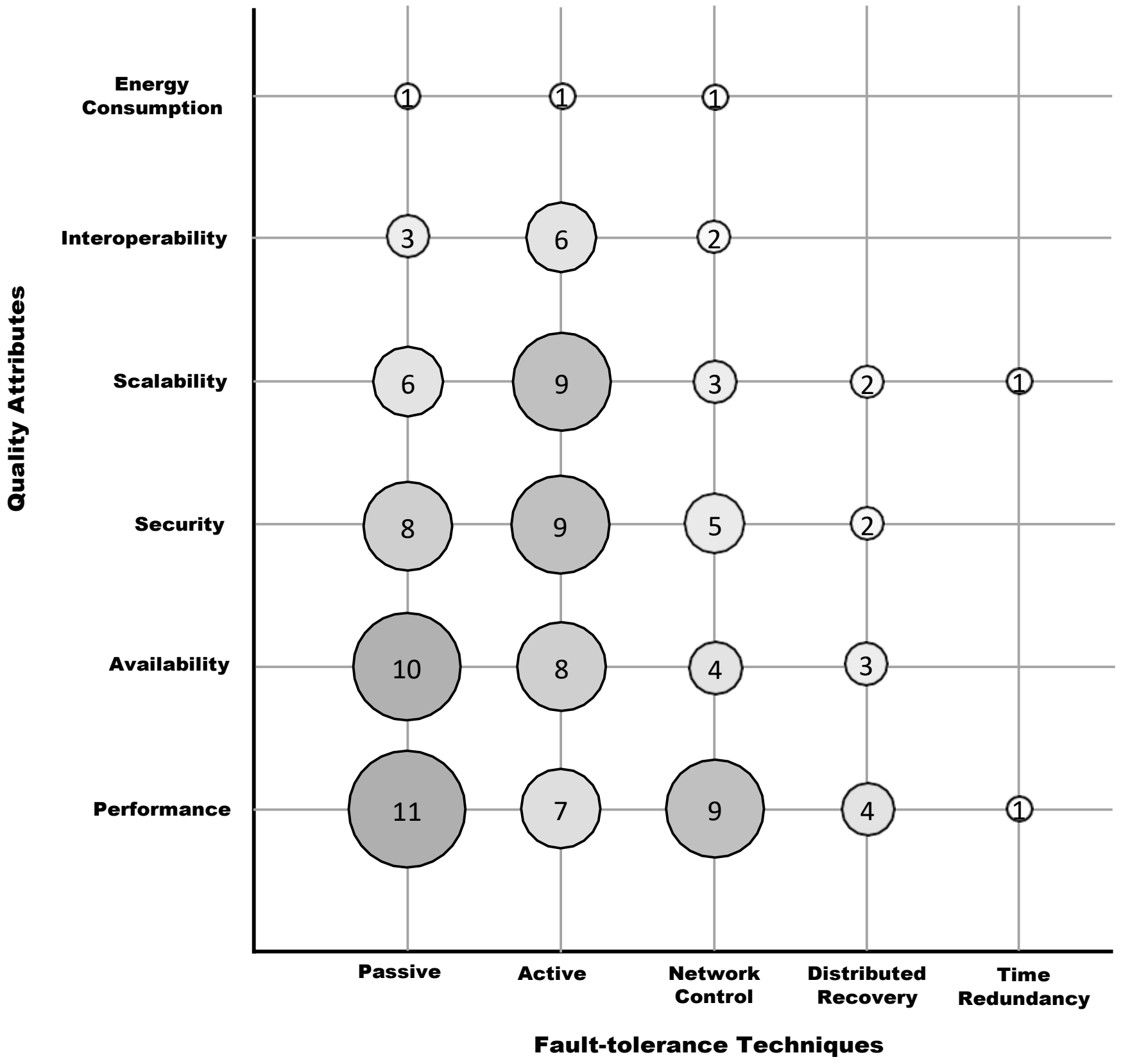

**Figure 2.9** Techniques VS quality attributes.

## 2.8 Threats to Validity

According to Peterson et al (83), the quality rating for this systematic mapping study assessed and scored as 73%. This value is the ratio of the number of actions taken in comparison to the total number of actions reported in the quality checklist. The quality score of our study is far beyond the scores obtained by existing systematic mapping studies in the literature, which have a distribution with a median of 33% and 48% as absolute maximum value. However, the threats to validity are unavoidable. Below we shortly define the main threats to validity of our study and the way we mitigated them.

***External validity:*** in our study, the most severe threat related to external validity may consist of having a set of primary studies that is not representative of the whole research on FT-IoT. We mitigated this potential threat by *i)* following a search strategy including both automatic search and backward-forward snowballing of selected studies; and *ii)* defining a set of inclusion and exclusion criteria. Along the same lines, gray and non-English literature are not included in our research as we want to focus exclusively on the state of the art presented in high-quality scientific studies in English.

***Internal validity:*** it refers to the level of influence that extraneous variables may have on the design of the study. We mitigated this potential threat to validity by *i)* rigorously defining and validating the structure of our study, *ii)* defining our classification framework by carefully following the keywording process, and *iii)* conducting a well-structured vertical analysis. Construct validity: It concerns the validity of extracted data with respect to the research questions. We mitigated this potential source of threats in different ways. *i)* performing automatic search on a couple of databases to avoid potential biases; *ii)* having a strong and tested search string; *iii)* complementing the automatic by the snowballing activity; and *iv)* rigorously screen the studies according to inclusion and exclusion criteria.

***Conclusion validity:*** it concerns the relationship between the extracted data and the obtained results. We mitigated potential threats to conclusion validity by applying well accepted systematic methods and processes throughout our study and documenting all of them in the excel package.

## 2.9 Conclusion

In this chapter we present a systematic mapping study with the goal of classifying and identifying the domain state-of-the-art and extract a set of FT-IoT methods and techniques. Starting from over 2300 potentially relevant studies, we applied a rigorous selection procedure resulting in 60 primary studies. The chapter discusses various aspects of FT-IoT such as techniques, architectural patterns and their associated quality attributes. The results of this study are both research and industry oriented and are in-tended to make a framework for future research in FT-IoT related fields. As a future work, we will assess the potential integration of existing research to an industrial level of IoT.

**Primary Studies**

- **P1**: Toward a New Approach to IoT Fault Tolerance, `https://doi.org/10.1109/MC.2016.238`
- **P2**: CEFIoT: A fault-tolerant IoT architecture for edge and cloud, `https://doi.org/10.1109/WF-IoT.2018.8355149`
- **P3**: Reliable and Fault-Tolerant IoT-Edge Architecture, `https://doi.org/10.1109/ICSENS.2018.8589624`
- **P4**: Efficient Fault-Tolerant Routing in IoT Wireless Sensor Networks Based on Bipartite-Flow Graph Modeling, `https://doi.org/10.1109/ACCESS.2019.2894002`
- **P5**: Optimizing Multipath Routing With Guaranteed Fault Tolerance in Internet of Things, `https://doi.org/10.1109/JSEN.2017.2739188`
- **P6**: Brume - A Horizontally Scalable and Fault Tolerant Building Operating System, `https://doi.org/10.1109/IoTDI.2018.00018`
- **P7**: A Watchdog Service Making Container-Based Micro-services Reliable in IoT Clouds, `https://doi.org/10.1109/FiCloud.2017.57`
- **P8**: Towards Fault Tolerant Fog Computing for IoT-Based Smart City Applications, https://doi.org/10.1109/CCWC.2019.8666447
- **P9**: Device clustering for fault monitoring in Internet of Things systems, `https://doi.org/10.1109/WF-IoT.2015.7389057`
- **P10**: Decentralized fault tolerance mechanism for intelligent IoT/M2M middleware, `https://doi.org/10.1109/WF-IoT.2014.6803115`
- **P11**: Application of Blockchain in Collaborative Internet-of-Things Services, `https://doi.org/10.1109/TCSS.2019.2913165`
- **P12**: A Review of Aggregation Algorithms for the Internet of Things, `https://doi.org/10.1109/ICSEng.2017.43`
- **P13**: Supporting Service Adaptation in Fault Tolerant Internet of Things, `https://doi.org/10.1109/SOCA.2015.38`

- **P14**: Fault tolerant and scalable IoT-based architecture for health monitoring, https://doi.org/10.1109/SAS.2015.7133626
- **P15**: Fault tolerance capability of cloud data center, https://doi.org/10.1109/ICCP.2017.8117053
- **P16**: Reaching Agreement in an Integrated Fog Cloud IoT, https://doi.org/10.1109/ACCESS.2018.2877609
- **P17**: Byzantine Resilient Protocol for the IoT, https://doi.org/10.1109/JIOT.2018.2871157
- **P18**: DRAW: Data Replication for Enhanced Data Availability in IoT-based Sensor Systems, https://doi.org/10.1109/DASC/PiCom/DataCom
- **P19**: Power efficient, bandwidth optimized and fault tolerant sensor management for IOT in Smart Home, https://doi.org/10.1109/IADCC.2015.7154732
- **P20**: Energy efficiency and robustness for IoT: Building a smart home security system, https://doi.org/10.1109/ICCP.2016.7737120
- **P21**: A Microservices Architecture for Reactive and Proactive Fault Tolerance in IoT Systems, https://doi.org/10.1109/WoWMoM.2018.8449789
- **P22**: Management of solar energy in microgrids using IoT-based dependable control, https://doi.org/10.1109/ICEMS.2017.8056441
- **P23**: A hierarchical cloud architecture for integrated mobility, service, and trust management of service-oriented IoT systems, https://doi.org/10.1109/INTECH.2016.7845021
- **P24**: Fault-Tolerant Real-Time Collaborative Network Edge Analytics for Industrial IoT and Cyber Physical Systems with Communication Network Diversity, https://doi.org/10.1109/CIC.2018.00052
- **P25**: Fault-Tolerant mHealth Framework in the Context of IoT-Based Real-Time Wearable Health Data Sensors, https://doi.org/10.1109/ACCESS.2019.2910411
- **P26**:SCONN: Design and Implement Dual-Band Wireless Networking Assisted Fault Tolerant Data Transmission in Intelligent Buildings, https://doi.org/10.1109/VTCFall.2018.8690787

- **P27**: Fault-tolerant application placement in heterogeneous cloud environments, https://doi.org/10.1109/CNSM.2015.7367359
- **P28**: A reliable and energy efficient IoT data transmission scheme for smart cities based on redundant residue based error correction coding, https://doi.org/10.1109/SECONW.2015.7328141
- **P29**: Distributed Continuous-Time Fault Estimation Control for Multiple Devices in IoT Networks, https://doi.org/10.1109/ACCESS.2019.2892905
- **P30**: Trend-adaptive multi-scale PCA for data fault detection in IoT networks, https://doi.org/10.1109/ICOIN.2018.8343217
- **P31**: Adaptive and Fault-tolerant Data Processing in Healthcare IoT Based on Fog Computing, https://doi.org/10.1109/TNSE.2018.2859307
- **P32**: Fault-Recovery and Coherence in Internet of Things Choreographies, https://doi.org/10.1109/WF-IoT.2014.6803224
- **P33**: A Novel Data Reduction Technique with Fault-tolerance for Internet-of-things, https://doi.org/10.1145/3018896.3018971
- **P34**: Performance Comparisons of Fault-Tolerant Rouging Approaches for IoT Wireless Sensor Networks, https://doi.org/10.1145/3195106.3195168
- **P35**: Rivulet: A Fault-tolerant Platform for Smart-home Applications, https://doi.org/10.1145/3135974.3135988
- **P36**: Censorship Resistant Decentralized IoT Management Systems, https://doi.org/10.1145/3286978.3286979
- **P37**: Towards a Foundation for a Collaborative Replicable Smart Cities IoT Architecture, https://doi.org/10.1145/3063386.3063763
- **P38**: Responsible Objects: Towards Self-Healing Internet of Things Applications, https://doi.org/10.1109/ICAC.2015.60
- **P39**: A Multi-agent System Architecture for Self-Healing Cloud Infrastructure, https://doi.org/10.1145/2896387.2896392
- **P40**: Reactive Microservices for the Internet of Things: A Case Study in Fog Computing, https://doi.org/10.1145/3297280.3297402

- **P41**: Fault Tolerance Techniques and Architectures in Cloud Computing - a Comparative Analysis, https://doi.org/10.1109/ICGCIoT.2015.7380625

- **P42**: Energy Efficient Fault-tolerant Clustering Algorithm for Wireless Sensor Networks, https://doi.org/10.1109/ICGCIoT.2015.7380464

- **P43**: Layered Fault Management Scheme for End-to-end Transmission in Internet of Things, https://doi.org/10.1007/s11036-012-0355-5

- **P44**: An Architectural Mechanism for Resilient IoT Services, https://doi.org/10.1145/3137003.3137010

- **P45**: Resilience of Stateful IoT Applications in a Dynamic Fog Environment, https://doi.org/10.1145/3286978.3287007

- **P46**: The Optimal Generalized Byzantine Agreement in Cluster-based Wireless Sensor Networks, https://doi.org/10.1016/j.csi.2014.01.005

- **P47**: A Reliable IoT System for Personal Healthcare Devices, https://doi.org/10.1016/j.future.2017.04.004

- **P48**: Reliable Industrial IoT-based Distributed Automation, https://doi.org/10.1145/3302505.3310072

- **P49**: Low-Cost Memory Fault Tolerance for IoT Devices, https://doi.org/10.1145/3126534

- **P50**: Idea: A System for Efficient Failure Management in Smart IoT Environments, https://doi.org/10.1145/2906388.2906406

- **P51**: Patterns for Things That Fail, https://www.hillside.net/plop/2017/papers/proceedings/papers/07-ramadas.pdf

- **P52**: Fall-curve: A Novel Primitive for IoT Fault Detection and Isolation, https://doi.org/10.1145/3274783.3274853

- **P53**: Multilevel IoT Model for Smart Cities Resilience, https://doi.org/10.1145/3095786.3095793

- **P54**: Energy Efficient Device Discovery for Reliable Communication in 5G-based IoT and BSNs Using Unmanned Aerial Vehicles, https://doi.org/10.1016/j.jnca.2017.08.013

- **P55**: A Programming Framework for Implementing Fault-Tolerant Mechanism in IoT Applications, https://doi.org/10.1007/ 978-3-319-27137-8_56
- **P56**: Transient fault aware application partitioning computational offloading algorithm in microservices based mobile cloudlet networks, https://doi.org/10.1007/s00607-019-00733-4
- **P57**: Channel Dependability of the ATM Communication Network Based on the Multilevel Distributed Cloud Technology, https://doi.org/10.1007/978-3-319-67642-5_49
- **P58**: Design of compressed sensing fault-tolerant encryption scheme for key sharing in IoT Multi-cloudy environment(s), https://doi.org/10.1016/j.jisa.2019.04.004
- **P59**: Fault-Tolerant Temperature Control Algorithm for IoT Networks in Smart Buildings, https://doi.org/10.3390/en11123430
- **P60**: Virtualization in Wireless Sensor Networks: Fault Tolerant Embedding for Internet of Things, https://doi.org/10.1109/JIOT.2017.2717704

## Chapter 3

# Self-adaptive IoT Architectures

In this chapter, we critically analyze a set of IoT distribution and self-adaptation patterns coming from previous chapters and identify their suitable architectural combinations. Further, we use our IoT modeling framework (CAPS) to model an emergency handling system. Based on these, we design two quality driven architectures to be used for a *forest monitoring and evacuation* example and qualitatively evaluate and compare them. The chapter is organized as follows. Literature is briefly discussed in Section 3.1. Section 3.2 defines and categorizes IoT distribution patterns. Self-adaptation control patterns are presented in Section 3.3. Section 3.4 presents the self-adaptive IoT combinational patterns whilst Section 3.5 discusses their quality attributes satisfaction level. The application of the model to a case study is presented in Section 3.6 and conclusions are finally drawn in Section3.7.

## 3.1 Literature Review

So far, a large body of knowledge has been proposed in both IoT architectures and self-adaptation patterns, however, the lack of harmonizing and integrating them together is undeniable.

Regarding IoT architectural styles and patterns, Cavalcante et al (22) introduce two reference architectures for IoT and analyze their characteristics. They realized that, both architectures need to fulfill the essential IoT non-functional requirements such as interoperability, scalability, and security, whilst considering the mandatory requirement of dynamic adaptation for IoT systems. Khan et al (56) present a cloud-based architecture for context-aware services for IoT and smart cities and walk through it using a hypothetical case study. Their architecture is based on cloud and they argue that cloud computing can provide a suitable computing infrastructure for data storage and processing needs of IoT and smart cities applications. Butzin et al (20) investigates on patterns and best practices that are used in the microservices approach and how they can be used in the internet of things.

In self-adaptation domain, Arcaini et al (8) present a conceptual and methodological framework for formal modeling, validating, and verifying distributed self-adaptive systems. They show how MAPE loops for self adaptation can be naturally specified in an abstract stateful language like Abstract State Machines. Weyns et al (108) present a MAPE loop notation and use this notation to describe a number of existing patterns of interacting MAPE loops. They derived these patterns from their use in practice and discussed their ramifications with respect to certain quality attributes. Ribeiro et al (89) designed a management architectural pattern for adaptation system in Internet of Things based on a number of Weyns control loops patterns.

However, few researches considered self-adaptation as a requirement for their IoT architecture. Azimi et al (12) propose a hierarchical computing architecture for IoT-based health monitoring systems. The model benefits from the features of fog and cloud with an adaptive architecture based on MAPE loop that is discussed into a 3-tier IoT-based system. Lee et al (61) propose a self-adaptive software framework for performing runtime verification using the finite state machine-based model checking. For the run-time verification, a self-adaptation process based on a MAPE loop is implemented. Shekhar et al (97) identify the key challenges that inhibit the universal adoption of cloud, especially in the context of IoT applications. They propose a dynamic data driven cloud and edge system, that uses measurement data collected from adaptively instrumenting the cloud and edge resources.

## 3.2 IoT Distribution Patterns

IoT distribution patterns classify the architectures according to edge intelligence and elements collaboration.

As discussed in previous chapter, (77) we classify IoT distribution patterns as: centralized, collaborative, connected intranets, and distributed based on a layered architectural style. Figure 1.4 in chapter 1 shows the four aforementioned distribution patterns. In these layered architectures, *Perception* layer represents the IoT physical sensors and actuators that aim to collect information. *Processing and Storage* layer is the central IoT component that stores and analyses the data gathered by perception components to be in access of other entities for their application purposes. *Application* layer determines the class of services provided by IoT; and *Business* layer manages the IoT system for its specific goal, by creating business models derived from the information of application layer (67).

This layered style makes a better understanding for IoT elements distribution level. In a *centralized* distribution pattern, the perception layer provides data for the central processing and storage component to prepare for a service in the next layer. In order to use the IoT service, one must connect to this central component. The central component can be a server, cloud, or a fog network connected to cloud. In a *collaborative* pattern, a network of central intelligent components can communicate in order to form and empower their services. In a *Connected Intranets* pattern, sensors provide data within a local intranet to be used locally, remotely, and centrally. Here the advantage is that if the central component fails, local service is still in access. The disadvantage is that there is no fully distributed framework to facilitate the communication among components. In a *distributed* pattern, all components are fully interconnected and capable to retrieve, process, combine, and provide information and services to other components towards common goals.

## 3.3 Self-adaptation Control Logic

Weyns et al (108) present six control patterns based on MAPE loop (Monitoring, Analysis, Planning, Execution) that model different types of interacting loops with different level of decentralization. A control loop is a model objected on imposing automatic control on dynamic behavior of a system and has been used in various fields such as software engineering. Figure 3.1 shows the self-adaptation control patterns. In the figure, managed subsystems (MS) comprise the application logic that provides the systems domain functionality. The managing subsystems instead manage the managed subsystems and comprise the adaptation logic.

A centralized self-adaptation pattern performs the adaptation through a central control loop. In a regional planning self-adaptation pattern, a physical space can be divided into different regions and the regions local planners coordinate to find the best adaptation solution for a local or global problem. Coordinated control and information sharing, both are based on a fully decentralized approach, however, with a different level of components coordination. More precisely, in the coordinated pattern, all *MAPE* components coordinate with their corresponding peers, whilst in information sharing, only *M* components communicate with one another. The other three patterns, are based on a hierarchical distribution model.

In master/slave pattern, a hierarchical relationship between one centralized master component *(A and P)* and multiple slave components *(M and E)* is created. Regional planning provides one *P* for each region to supervise the other elements of loop, in a way to interact different regions *P* one to another. The hierarchical control pattern provides a layered separation of concerns to manage the complexity of self-adaptation as a hierarchy of MAPE loops.

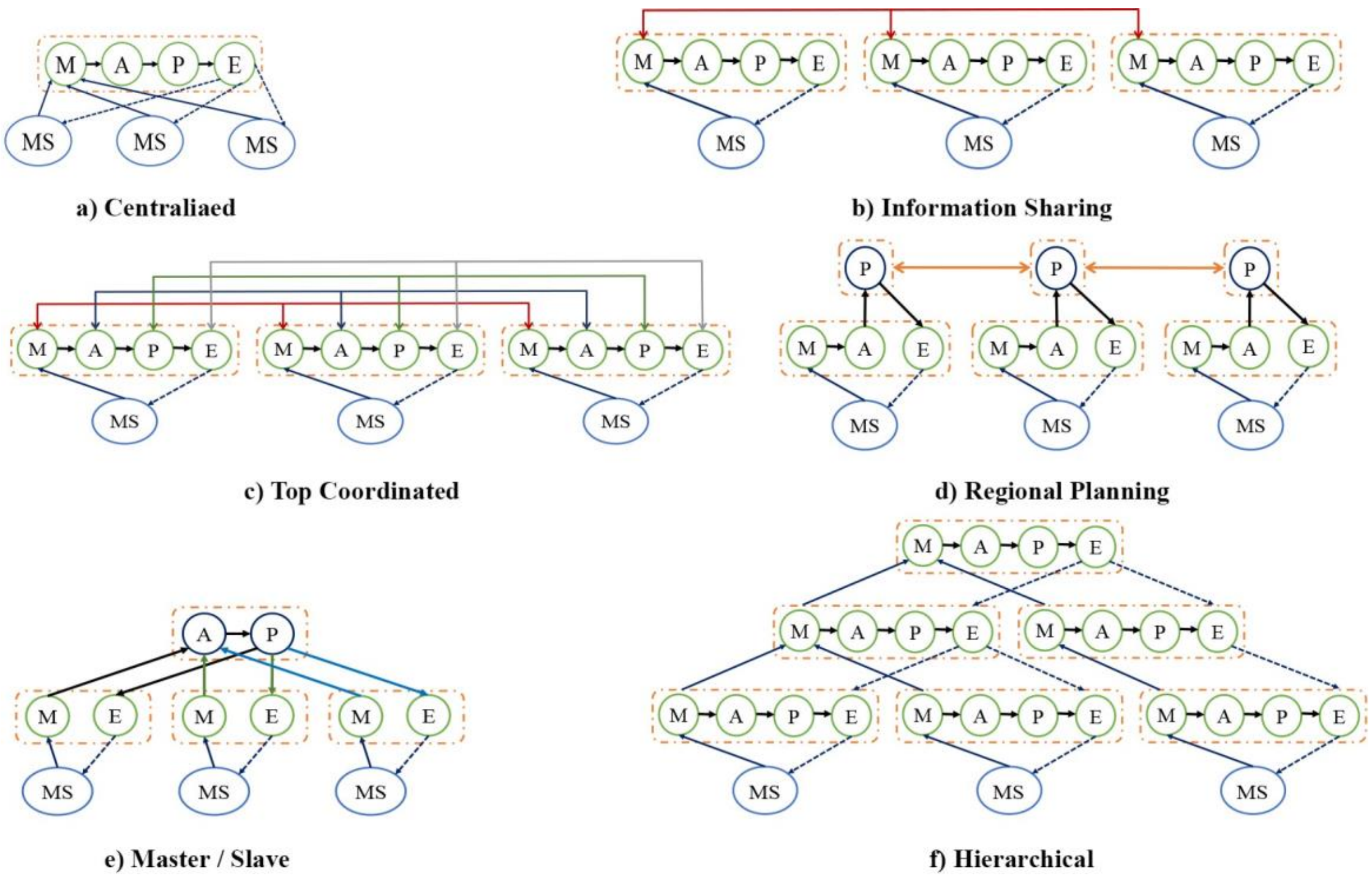

**Figure 3.1** Self-adaptation patterns

## 3.4 Self-adaptive IoT: a Pattern Combination

In order to present the best set of self-adaptive IoT architectures, we should first discover compatibility or inconsistency between various IoT distribution and self-adaptation patterns. Learning aforementioned patterns and comparing their specific characteristics are important towards defining IoT reference architectures for emergency handling.

**Table 3.1** Patterns combination feasibility for IoT

| | Centralized | Collaborative | Connected Intranets | Distributed |
|---|---|---|---|---|
| Centralized | | | | |
| Information Sharing | | | | |
| Top Coordinated | | | | |
| Regional Planning | | | | |
| Master/Slave | | | | |
| Hierarchical | | | | |

Most of adaptation patterns are unmatched with IoT distribution patterns, so that making a combination of them can be infeasible or weak. For instance, a centralized adaptation pattern has a focal MAPE loop that is responsible for MSs adaptation. This pattern can only be combined with centralized IoT distribution pattern since the adaptation should take place in one central processing component.

In information sharing pattern, only monitoring components can communicate with each other to satisfy specific quality attributes. In this pattern, decentralized analysis and planning of each sub-system perform while not coordinating with each other. Hence, it is unmatched with centralized distribution pattern in which no data transfer occurs among monitoring components. Furthermore, information sharing cannot be combined with collaborative and fully distributed patterns, in which the main processing components are in coordination. Therefore, its best combination is with connected intranets distribution pattern as its local intranets are managing their adaptation locally while may share information among each other. However, this combined pattern is not suitable for most of IoT-based systems like our case study in which: *i)* a globally optimal objective is preferred to locally optimal ones; *ii)* a high degree of controllers collaboration is essential; *iii)* energy efficiency is a critical non-functional requirement. It however may be useful for some cases such as self-healing in traffic monitoring systems by means of explicit state information sharing.

Top coordinated is similar to information sharing pattern in which all components are coordinating with their peers on other loops. Due to its highly decentralized structure, it is only matched with fully distributed pattern. However, because of the high degree of coordination between components, this pattern is not the best one for our case study objectives that are greenability and reducing energy consumption.

In regional planning pattern, the local planners coordinate to find the best adaptation solution for a local or global problem. Its characteristics make it only and highly matched with collaborative distribution pattern. It is also suitable for our case study because of its high coordination of planner and low coordination of other local adaptation components to provide fast and energy efficient decisions. Therefore, we chose this combined pattern (collaborative regional planning) to examine some of its quality attributes such as energy consumption and data traffic.

Master/slave adaptation pattern facilitates centralized decision making, and local monitoring and adaptation execution. It simplifies achieving global objectives through central implementation of analysis and planning algorithms. However, in very large distributed systems, it can make communicational overhead and bottle-neck. Obviously, Master/slave can only have a combination with centralized distribution pattern. We chose this combined pattern (centralized master/slave) to explore its level of quality attributes satisfaction. Hierarchical adaptation pattern, provides a layered separation of concerns to manage the complexity of self-adaptation (108). This pattern is best matched with fully distributed pattern that makes an immensely complex architecture. However, this pattern is not suitable for our scenario that is sensitive to global adaptation time and energy consumption. The following section, discusses the most recognized quality attributes for an adaptive IoT-based system.

## 3.5 Quality Attributes

An architectural pattern can have effect on IoT-based systems quality attributes but does not guarantee them. In our previous works, it is argued that the most recognized quality attributes that are supposed to be satisfied by a proper IoT architecture are scalability, security, interoperability, performance, availability, resiliency and evolvability.

*Scalability* is an essential attribute, since IoT should be capable to perform at an acceptable level with a high scale of devices. Commenting on the scalability of IoT as a whole system is difficult, however, it depends on how new resources can be added on demand. Furthermore, *security* gains a high concern in an IoT system, in which different components and entities are connected to each other through a network. *Interoperability* helps IoT heterogeneous components to work together efficiently. It actually depends on how much IoT large-scale heterogeneous devices can communicate directly among each other to gather the required data without having to go through the central component. The level of *performance* depends on how much the processing and storage component is pushed to the edge in a decentralized way.

*Availability* that is the ability of a system to be fully or partly operational as and when required (15) is important for safety critical systems such as disaster management. *Resiliency* and availability are strongly related to each other and against system failure due to components failure. *Evolvability* is tied to Interoperability and concerns that the IoT system should be adaptable to new technologies and applications. This can be realized by system openness to change and extension. In our case study, we assess energy efficiency that is tied to many other quality attributes such as performance and tries to make energy usage as low as possible.

## 3.6 Example of Application

We next describe the application of previously selected self-adaptive IoT architectures on monitoring the recreational forests located in protected mountainous areas in Abruzzo region, Italy. In this section, we design the selected software architectures and run simulations to assess the level of energy consumption and data transfer and to choose the architecture with higher quality. Mountain areas of Abruzzo are especially under the risk of fire and avalanche, for which an accurate disaster prevention and emergency evacuation plan is required. The latter will not happen except with an early disaster detection, congestion estimation, and risk diffusion system using situational awareness sensors (such as cameras, counters, RFID), detectors (such as temperature, smoke, shake) and actuators. The IoT system should have an active component to trigger alarm devices and show endanger people the best path to go through towards the safe place.

The selected area is divided in two different regions (north and south) with their own IoT-based devices. Two safe places are located on the regions border, towards which the crowd should move during an emergency. However, the architectural pattern will clarify how the components can interact to each other in order to satisfy the aforementioned goal. From an architectural viewpoint, the system should provide a map of monitored area on the security agents dashboards. If a disaster is detected, an architectural adaptation will take place to perform the evacuation plan. Since the IoT devices are powered by batteries, here the concern is to choose the best self-adaptive IoT pattern to keep the energy consumption in a minimum level.

***CAPS.*** In order to model the scenarios, we use CAPS modeling environment (73) (95) for disaster management. CAPS is able to create a combined software, hardware, and physical space view, specific for surveillance and emergency evacuation handling. CAPS supports the system with early disaster detection and provides the best architecture to satisfy the desired evacuation time. However, this research has its focus on software architecture modeling language (SAML) for self-adaptive IoT-based systems.

***Energy Consumption.*** The energy consumption denotes the total energy consumed from running the components that are composing each architecture. In order to compute the Energy Consumption (EC in Joule) under different patterns and contextual situations, we will use the utility function presented in the following formula. The values gained from this formula will range from [0,1]. The *Zero* value reflects the worst energy consumption and the *one* value reflects the best energy consumption.

$$U_{EC} = 1 - \left[ \frac{CurrentEnergy - MinEnergy}{MaxEnergy - MinEnergy} \right]$$

Where CurrentEnergy represents the power consumption in Joules for the current pattern and must be always larger than or equal to the minimum energy; MaxEnergy represents the maximum power consumption to be considered or noticed; MinEnergy represents the minimum power consumption to be considered or noticed.

***SAML.*** Within SAML, software components exchange messages through message ports. Each component can declare a set of application data manipulated by actions defined in the behavior of the component. The behavior of each component is represented by a list of events, conditions, and actions. Modes can be defined as well. Figure 3.2 and Figure 3.3 show CAPS software architecture modeling corresponding to aforementioned self-adaptive IoT patterns for forest fire detection and crowd evacuation. It is important to note that these figures are screen-shots of CAPS tool (72) (96).

In the **centralized master/slave** pattern (Figure 3.2), the adaptation logic is being performed by a centralized master component that is responsible for the analysis and planning of adaptations and multiple slave components are responsible for monitoring and execution. This architecture has two adaptation modes:

*1- Normal mode:* in this mode the sensors read CO2 concentration and temperature in each area every 5 seconds. A timer is set in this mode to schedule the reading from the sensors. A message carrying each value is sent from the output message port of the sensor components to the in port of the controller component. The main goal of this application, to be run on a tablet, is to show a 2D-representation of the monitored space providing also contextual data (sensed by RFID systems) on where the crowd is at any time, and how it moves in normal (and emergency) cases. If instead an emergency is detected, the state of the area will be adapted to the critical mode.

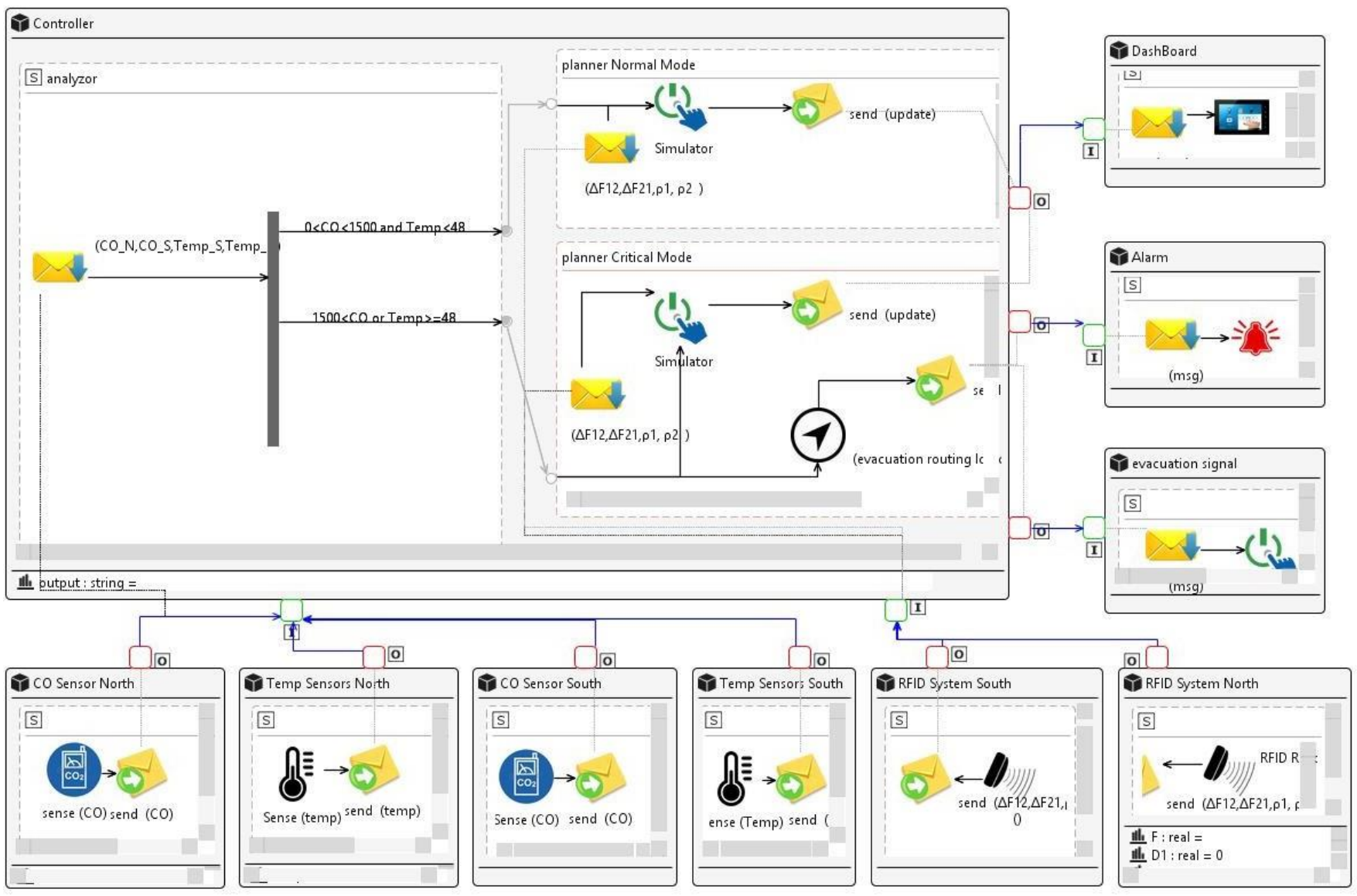

**Figure 3.2** SAML for Central Master/Slave Pattern

2- *Critical mode:* in this mode, an adaptation will take place in monitoring level and sensors value will be read every second. In addition to showing the map on dashboards, a message will be sent to acoustic alarm and evacuation sign actuators of each area to lead people to the safe places. The central controller handles the situation of whole area based on the density and flow of pedestrians from a place to another. we recently published the logic behind the controller functionality as another paper (7).

In **collaborative regional planning** pattern (Figure 3.3) a regional planner decides for each region. The subsystems provide their planner with necessary information and different planners interact with one another to coordinate adaptations that span multiple regions. This architecture has two adaptation modes as well:

1- *Normal mode:* in this mode the sensors read CO2 concentration and temperature in each area every 5 seconds. A timer is set in this mode to schedule the reading from the sensors. A message carrying each value is sent from the output message port of sensor components to the in port of the controller component of corresponding area. The values will be analysed and planned locally. However, the decision making will take place under coordination of other regions planners to support and approve any required execution. In the normal mode, a simulator plans to show to operators the map of the area. The crowdedness of each area and flow of people is sensed by RFID systems. If instead an emergency is detected in a region, the state of that region (or afterwards other regions) will be adapted to the critical mode.

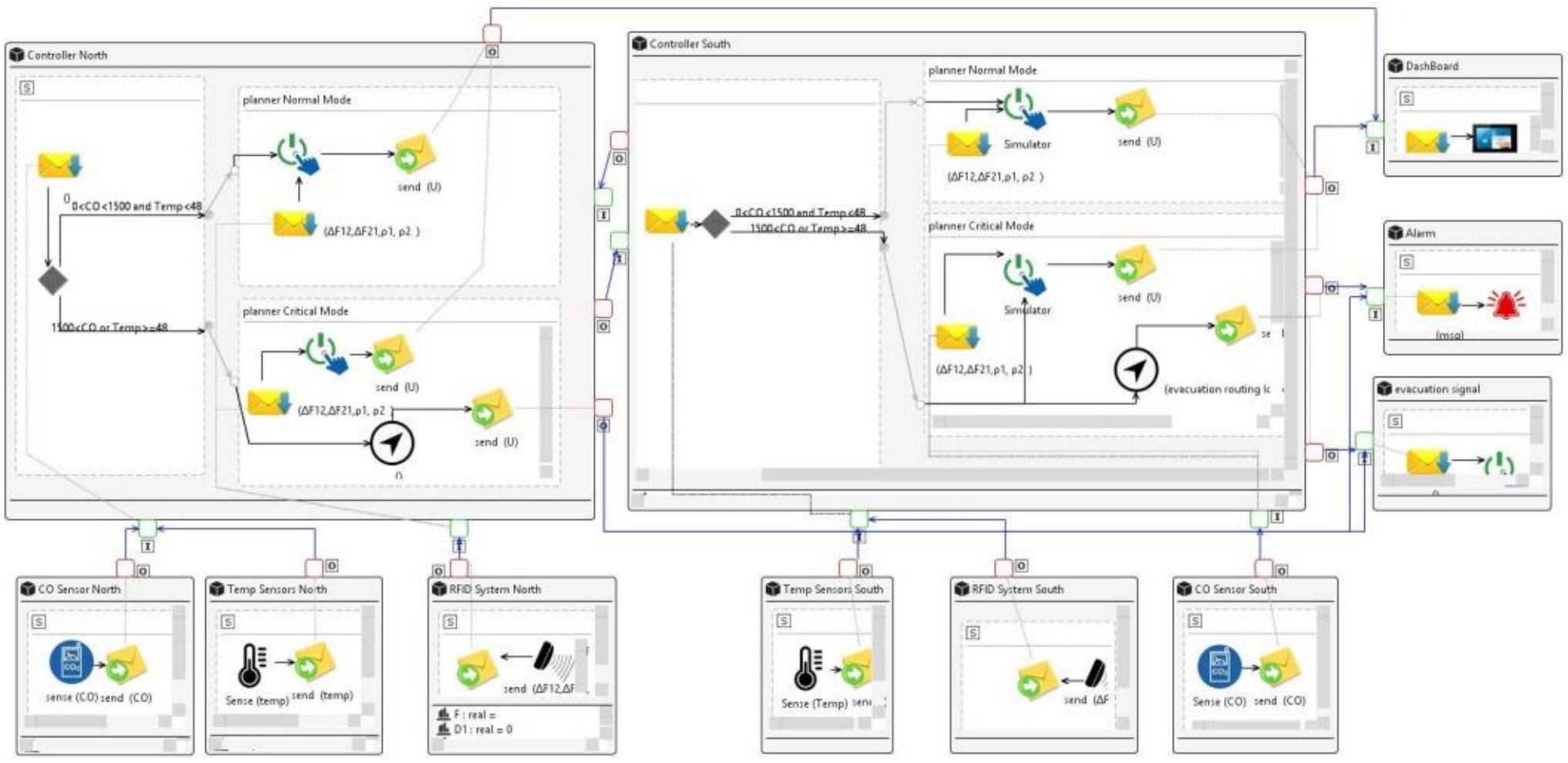

**Figure 3.3** SAML for Collaborative Regional Planning Pattern

2- *Critical mode:* in this mode, an adaptation will take place in the monitoring level and sensors will read the values more frequently (for example every second). In addition to showing the map on dashboards, a message will be sent to acoustic alarm actuators and evacuation sign actuators of the endanger area to lead people to safe areas. Given such a visualization, actuators might be piloted directly from the mobile app in order to direct people through the fastest evacuation path. In this case, the regional controllers in collaboration to each other handle the situation of risky areas based on the density and flow of pedestrians from a place to another (7).

***Simulation.*** In this section we describe the simulation results corresponding to two different selected patterns (central master/slave and collaborative regional planning). We used our CAPS framework for modeling and simulation purposes (72). More precisely, we used CAPS automatic code generation and simulation framework tool to produce two CupCarbon projects (96) and then, we ran the simulator and compared data traffic, energy consumption and battery level of nodes. For all simulation experimentation, we fixed simulation time to 6000 s, and energy max for all nodes to 19159 J. Figure 3.4 shows a screen shot of running centralized master/slave and collaborative regional planning patterns.

In all the patterns applied, the CO2Sensor, Temperature sensor, and RFID systems components are always sending messages and don't receive any. While, dash-board, alarm and evacuation signal are always receiving messages and don't send any. Further, the Controller has traffic in both directions with sending and receiving messages. This explains some zero values that are appeared in Tables 3.5 and 3.6. These tables show the exchanged messages through the IN and OUT ports of the components during running the two patterns in CAPS. They also show the data traffic in Kilo Bytes that occurs at each component in each pattern.

Further, from Tables 3.5 and 3.6, we conclude that the data traffic of Controller-

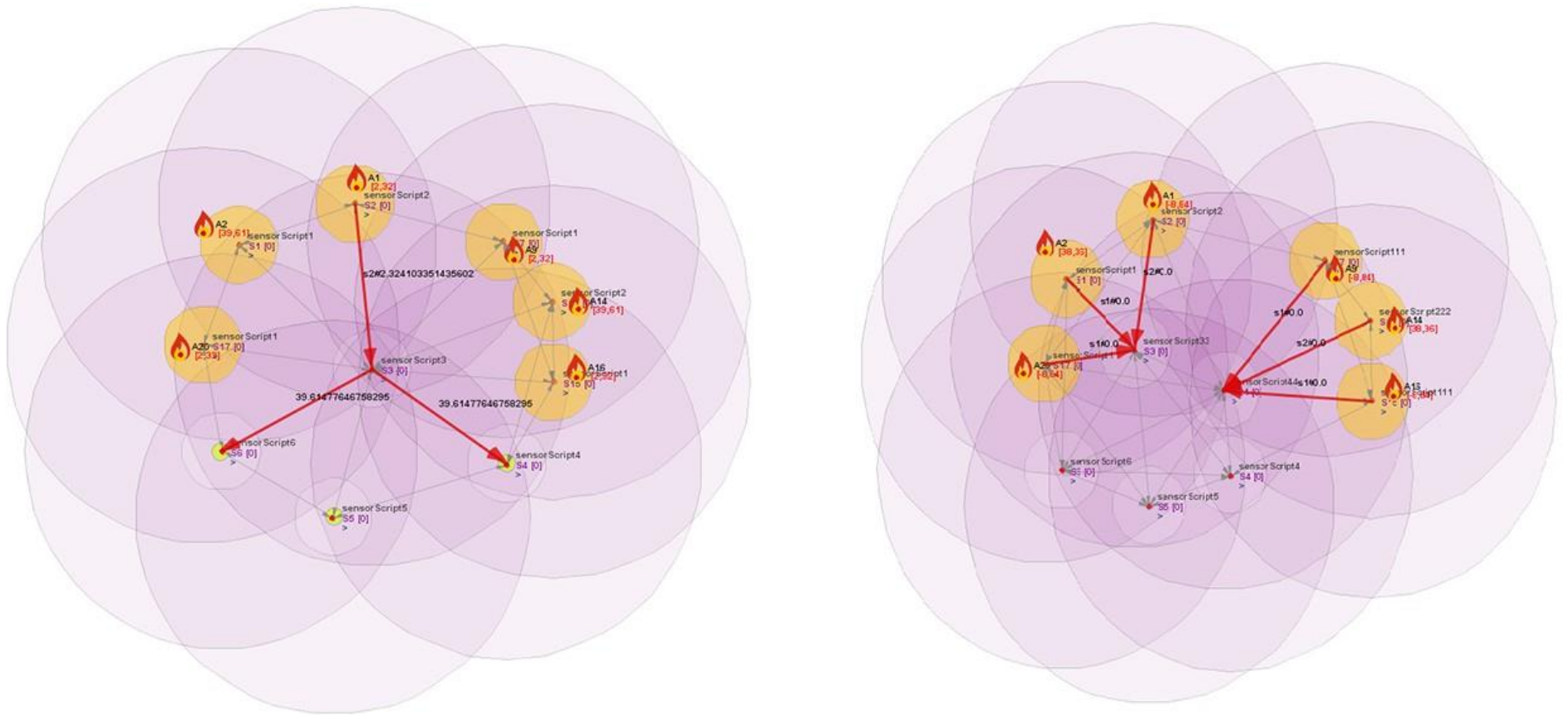

**Figure 3.4** Running project for centralized master/slave (left) and collaborative regional planning (right)

| Component ID | Component Name | # of sent messages | # of received messages | Data traffic in KB |
|---|---|---|---|---|
| S1 | CO2Sensor North | 174 | 0 | 6 |
| S2 | dash board | 0 | 592 | 20 |
| S3 | Controller | 627 | 1307 | 65 |
| S4 | Temperature sensor south | 235 | 0 | 8 |
| S5 | CO2Sensor South | 186 | 0 | 6 |
| S6 | Temperature sensor North | 242 | 0 | 8 |
| S7 | alarm | 0 | 31 | 1 |
| S8 | evacuation signal | 0 | 4 | 1 |
| S9 | RFID systems North | 239 | 0 | 8 |
| S10 | RFID systems South | 231 | 0 | 8 |

**Figure 3.5** Messages exchanged in components during simulation of Centralized Master/slave pattern

| Component ID | Component Name | # of sent messages | # of received messages | Data traffic in KB |
|---|---|---|---|---|
| S1 | Temperature sensor south | 235 | 0 | 8 |
| S2 | Controller South | 976 | 663 | 55 |
| S3 | Controller North | 937 | 640 | 53 |
| S4 | dash board | 0 | 618 | 21 |
| S5 | RFID systems North | 224 | 0 | 8 |
| S6 | evacuation signal | 0 | 3 | 1 |
| S7 | Temperature sensor North | 242 | 0 | 8 |
| S8 | RFID systems South | 239 | 0 | 8 |
| S9 | CO2Sensor South | 189 | 0 | 6 |
| S10 | alarm | 0 | 31 | 1 |
| S18 | CO2Sensor North | 174 | 0 | 6 |

**Figure 3.6** Messages exchanged in components during simulation of Collaborative Regional Planning Pattern

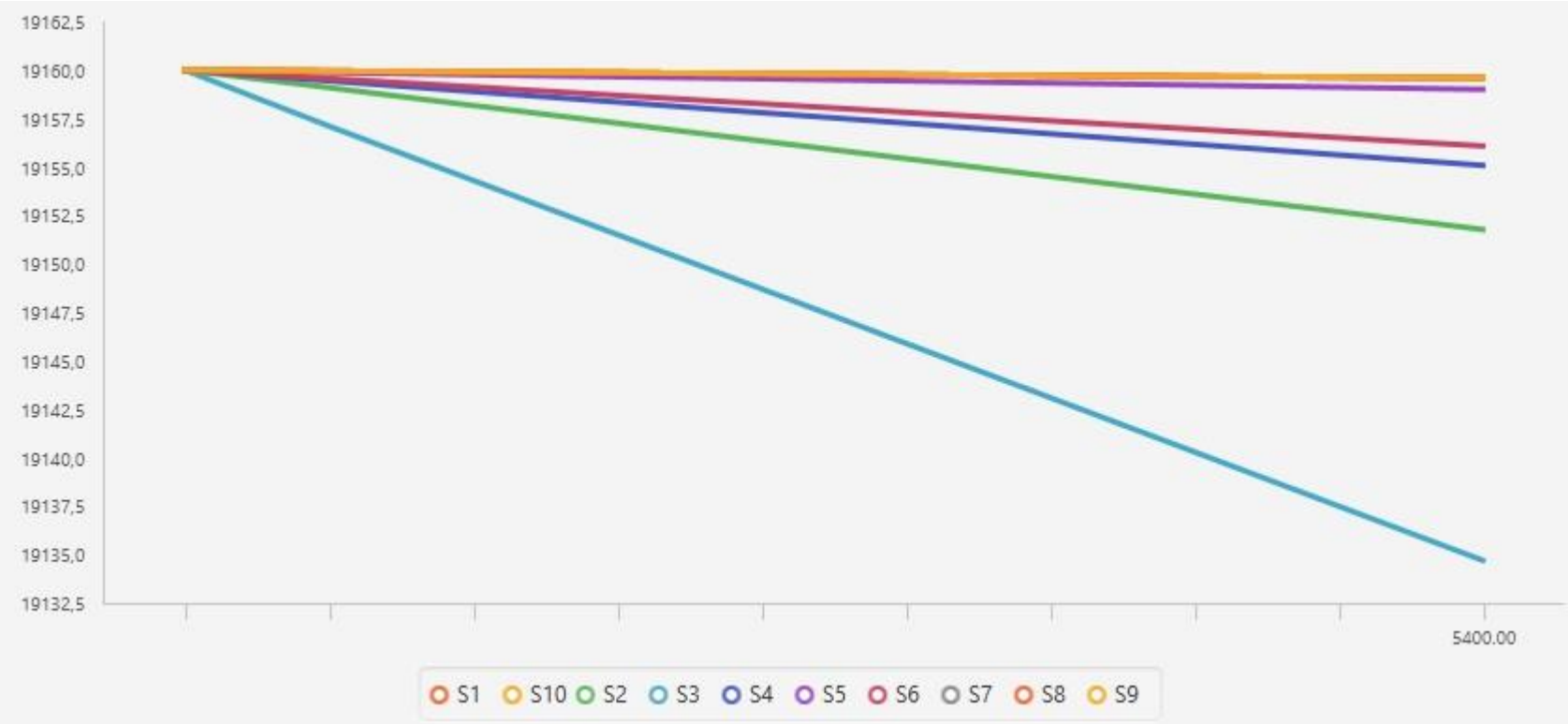

**Figure 3.7** Results for battery level and power consumption in Centralized Master/slave Pattern (Color figure online)

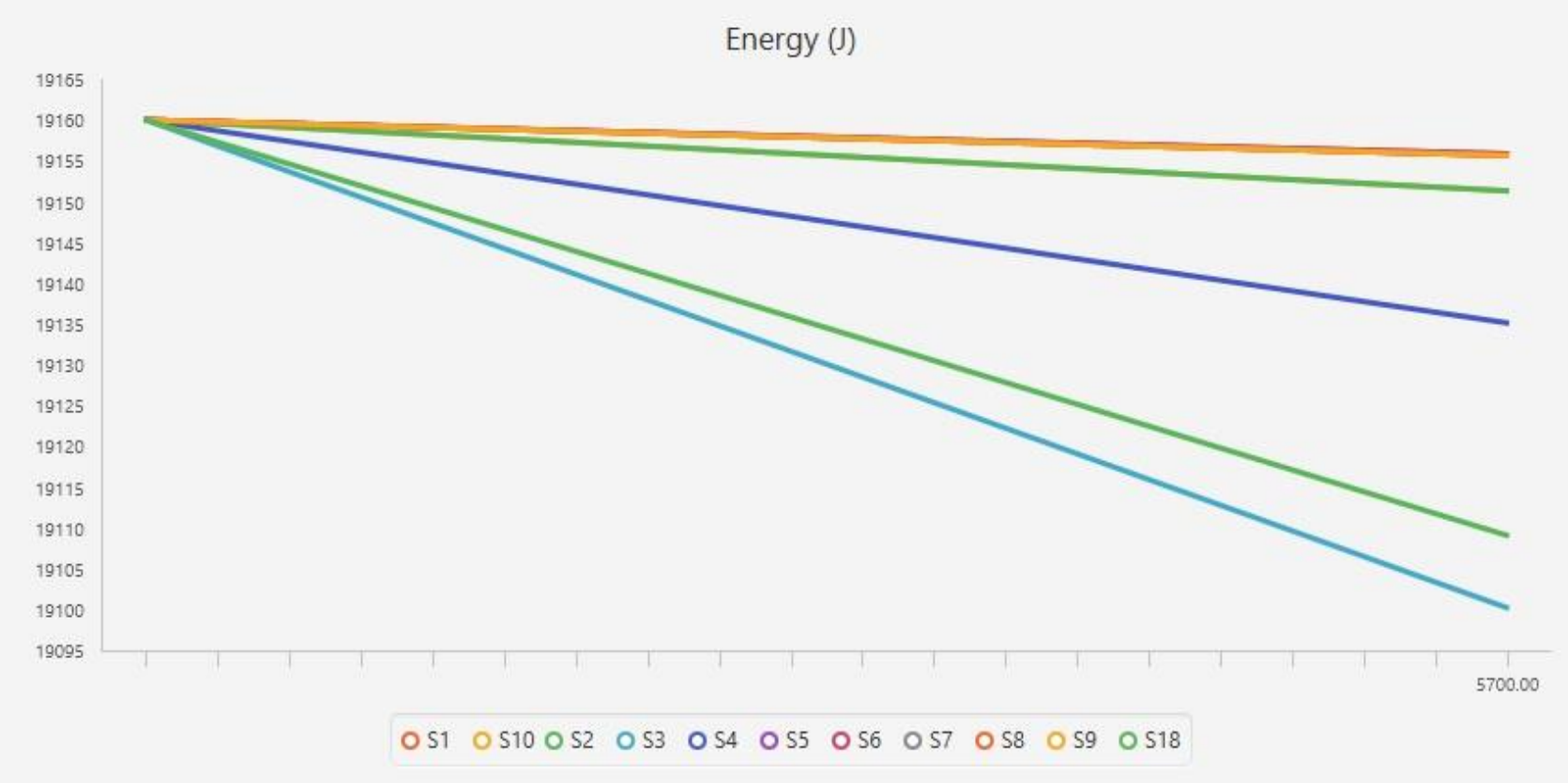

**Figure 3.8** Results for battery level and power consumption in Collaborative Regional Planning Pattern (Color figure online)

North, ControllerSouth, and DashBoard receive the highest traffic (exchanged messages) when the collaborative pattern is running. Therefore, We learn that using the centralized master/slave pattern leads to a lower range of data traffic compared with using collaborative regional planning pattern. This proves how choosing a proper architectural pattern can effect on the efficiency concern.

Regarding to the battery levels and power consumption, Figure 3.7 shows the battery level when running the simulator under the master/salve pattern and Figure 3.8 shows the energy consumption when running the simulator under collaborative pattern. From these figures, considering for example the node S3, which represents the controller in centralized master/slave and ControllerNorth in collaborative regional planning, we notice that S3 received the highest battery level drain when running the system. Thus, considering the same node in collaborative regional planning, we notice how the energy consumption is higher and receives higher battery level drain when running the system, due to increased number of exchanged messages.

## 3.7 Conclusion

The goal of this study is to identify a set of self-adaptive IoT architectures with respect to the philosophy and granularity of IoT distribution and adaptation patterns. We analyzed the various architectural patterns to find the best matched ones for IoT-based crowd monitoring and emergency handling. Two patterns have been chosen to model and simulate, in order to argue the best one for the quality of energy consumption and data transfer. The results show that the energy consumption in collaborative regional planning pattern is higher and receives higher battery level drain due to increased number of exchanged messages.

Within the next chapter, we focus on controller components by presenting and evaluating an optimization algorithm for critical situations.

# Part II

# Integrating the Netflow Algorithm in IoT Architecture

## Introduction to Part II

This part is written based on the following peer reviewed articles:

- **IoT Flows: A Network Flow Model Application to Building Evacuation**, *Published in: AIRO Springer Book: A View of Operations Research Applications in Italy, 2019.*
  *DOI:* `https://doi.org/10.1007/978-3-030-25842-9_9`

- **Applying a network flow model to quick and safe evacuation of people from a building: a real case**, *Published in: Proceedings of the GEOSAFE Workshop on Robust Solutions for Fire Fighting (RSFF), 2018.*
  *DOI:* `http://ceur-ws.org/Vol-2146/paper02.pdf`

- **An IoT Software Architecture for an Evacuable Building Architecture**, *Published in: 52nd Hawaii International Conference on System Sciences, 2019.*
  *DOI:* `https://hdl.handle.net/10125/59508`

- **Real-time Emergency Response through Performant IoT Architectures**, *Published in: International Conference on Information Systems for Crisis Response and Management, 2019.*
  *DOI:* `https://hal.archives-ouvertes.fr/hal-02091586`

**Abstract.** This part describes the design of an Internet of Things (IoT) system for building evacuation. As previously mentioned, there are two main design decisions for such systems: *i)* specifying the platform on which the IoT intelligent components should be located; and *ii)* establishing the level of collaboration among the components. This part specifically presents a real-time / design time emergency evacuation handling system based on internet of things (IoT) technologies discussed in previous part. In real-time, the component operates as the core of an Internet of Things (IoT) infrastructure aimed at crowd monitoring and optimum evacuation paths planning. In this case, a software architecture facilitates achieving the minimum time necessary to evacuate people from a building. In design-time, the component helps discovering the optimal building dimensions for a safe emergency evacuation, even before (re-) construction of a building.

The space and time dimension are discretized according to metrics and models in literature, as well as original methods. The component formulates and solves a linearized, time-indexed flow problem on a network that represents feasible movements of people at a suitable frequency. Accurate parameter setting makes the computational time to solve the model compliant with real-time use. Furthermore, for safety-critical systems such as evacuation, real-time performance and evacuation time are critical. The approach presented in this part aims to minimize computational and evacuation delays and uses Queuing Network (QN) models.

Applications of the proposed IoT system and its core algorithm to handle safe evacuation test in two venues (Palazzo Camponeschi and Alan Turing building in L'Aquila, Italy) are described, and diverse uses of the methodology are presented. Experiments were performed that tested the effect of segmenting the physical space into different sized virtual cubes. Experiments were also conducted concerning the distribution of the software architecture. By evaluation, it is shown that the model has the capability to optimize the safety standards by small changes in the building dimensions and can guarantee an optimal emergency evacuation performance.

**Keywords.** *Network optimization*, *Linear programming*, *Emergency Evacuation*, *IoT*, *Performance*, *Queuing networks*.

**Overview.** In order to design and use a safety oriented building that facilitates emergency evacuation in case of a disaster, a number of issues should be addressed: what are the building restrictions that may make trouble for people and prevent them to save their lives (bottlenecks and obstacles)? Instead, what building characteristics may improve their safety and quickness towards a safe place? Up to what capacity a building can facilitate a smooth flow of people? In post-construction phase, how crowd evacuation can be facilitated by showing the best paths towards safe areas? Overall, how a proper building architecture can save lives in emergency situations?

Despite an emergency can barely be imagined static, actual evacuation plans are still based on static maps with pre-selected routes through which pedestrians should move in case of disaster. This approach has several drawbacks: *i)* it leads all pedestrians into the same routes, thus concentrating people and creating crowd; *ii)* it ignores abrupt congestion, obstacles or dangerous areas; *iii)* it disregards individual behaviors, and does not distinguish special categories (e.g. elderly, children, disabled); *iv)* it does not elaborate different training indications for security operators in different scenarios; *v)* it does not provide evacuation managers and operators either with a comprehensive understanding of possible scenarios, or with real-time situation-awareness. However, in this context, IoT systems can be exploited to support rules for quicker and safer evacuation: by tracking people in a building, possible congestion can be detected, best safety paths can be periodically calculated, and consequently evacuation time under changing emergency conditions can be minimized.

This result is obtained by an analytical model based on a flow optimization problem that can be efficiently solved. The model has a combinatorial nature, as it decomposes both space (building plan) and time dimension into finite elements: *unit cells* and *time slots*. The building topology is then represented as a graph with nodes corresponding to cells, and arcs to connections between adjacent cells. Collected data are used to create a second acyclic digraph, indexed on time, that models all the feasible transitions between adjacent cells at any given time slot, given the current occupancy status of each cell. Minimizing the total evacuation time corresponds then to solve a mathematical program that, in the final refinement, has the form of a linear optimization problem.

In the release here described, the system we propose aims at providing operators with a continuously monitored evaluation of the shortest time required to evacuate people at any time present in a given building. This feature has been approved by the Fire Brigade of the City of l'Aquila as a possible mean to adjust visitor entrance in a public area (in this chapter, a gallery indoor space) according to safety conditions. The main system features are: *i*) continuous update of solutions, so evacuation guidelines can be adjusted according to visitors positions evolving over time; *ii*) continuous path feasibility check, so that unfeasible routes can automatically be discarded; *iii*) possibility (not discussed in this chapter) of incorporating the analytical model into a mobile app that supports emergency units to evacuate both closed and open spaces.

In this part, not only we detail the IoT system and the optimization model, but also, testing different scenarios based on real data, we describe the methodology followed to set system parameters and find an efficiency/accuracy compromise that makes system design compliant with real-time specifications. All that given, our overall view, backed up by ongoing work, is to integrate the mathematical model and algorithm here proposed with an *IoT infrastructure* and a *mobile app*. It is worth mentioning that the potentiality of the IoT system is widely discussed in Chapter 6.

**Chapter 4**

# IoT Flows: A Network Flow Model Application to Building Evacuation

This chapter presents a real-time emergency evacuation handling system based on internet of things (IoT) technologies discussed in previous chapters. The IoT infrastructure has a core computational component that is in charge of minimizing the time necessary to evacuate people from a building. Such component is formulated as a network flow problem. The chapter is organized as follows. Literature is briefly discussed in Section 4.1. Section 4.2 presents the flow model and Section 4.3 refers on how model parameters should be set up to deal with real cases. The application of the model to a real exhibition venue is presented in Section 4.4 and conclusions are finally drawn in Section 4.5.

## 4.1 Literature Review

Evacuation routing problems (ERP) for large scale roads and buildings are complex and subject to various modeling issues. In this domain, a pioneering work is due to Choi et al (26), who modeled a building evacuation problem by dynamic flow maximization where arc capacities may depend on flows in incident arcs. Although dating back to the Eighties and limited to theoretical analysis, the paper provides in our mind a good starting point and deserves consideration in the light of the progress done in Linear Programming solution tools. Chen et al. (25) propose a flow control algorithm to compute evacuation paths according to building plan and total number of evacuees. The model aims at minimizing total evacuation time while assigning an optimal number of evacuees to each evacuation path. However, as network size increases, the associated problem can no longer be solved in real-time. Recent research bases evacuation planning on a transshipment problem. For instance, Schloter et al. (94) study two classical flow models in order to deal with crowd evacuation planning: one algorithm aims at finding the best transshipment as a convex combination of simple lex-max flows over time; a second one computes earliest arrival transshipments that maximize the flow towards the sink for every point in time. Other papers propose a hybrid optimization-simulation approach. To quote an example, Abdelghany et al. (1) integrates a genetic algorithm with a microscopic pedestrian simulation assignment model. The genetic algorithm looks for an optimal evacuation plan, while simulation guides the search by evaluating the quality of the plans generated.

In fact, one crucial issue addressed by recent literature is the ability of finding good solutions in so short time as required by a practical computational core of a real-time IoT system. Today's solvers can foster this achievement, as they can get easily rid of very large problems in fractions of seconds: on one hand, dealing with more variables helps obtain enough resolution to model the necessary details (in terms of both discretization and non-linearity); on the other hand, quick re-computation allows to cope with data that dynamically change over time.

## 4.2 Flow Model For Emergency Handling

The following network construction basically follows (26) and (70). The topology of the building to be evacuated is described by a graph $G = (V, A)$ that in (26) is called *static network*. Nodes of $G$ correspond to the unit cells $i$ obtained by embedding the building into a suitable grid that will be discussed in Section 6. In general, cells may have different shapes or sizes: for the purpose of our work what is important is that every cell can approximately be traversed, in any direction, in a single time slot. Cell 0 conventionally represents the outside of the building, or in general a safe place. Safe places can of course be disconnected areas, but as their capacity is assumed large enough to guarantee safety, we will represent them all by a single cell (therefore what we assume about cells traversing time does not apply to cell 0). Arcs of $G$ correspond to passages between adjacent cells: the passage has full capacity if cells share a boundary not interrupted by walls, and a reduced capacity otherwise. With no loss of generality, arcs are supposed directed. Let us denote:

$T = \{0, 1, \ldots, \tau\}$, set of unit time slots;

$y_i^t$ = state of cell $i \in V$ at time $t \in T$, that is, the number of persons that occupy $i$ at $t$: this number is a known model parameter for $t = 0$ (in particular, $y_0^0 = 0$) and a decision variable for $t > 0$;

$n_i$ = capacity of cell $i$: it measures the maximum nominal amount of people that $i$ can host at any time (in particular, $n_0 \geq \sum_i y_i^0$); this amount depends on cell shape and size; if cells can be assumed uniform one can set $n_i = n$ for all $i \in V, i = 0$.

$x_{ij}^t$ = how many persons move from cell $i$ to an adjacent cell $j$ in $(t, t+1]$: this gives the average speed at which the flow proceeds from $i$ to $j$;

$c_{ij} = c_{ji}$ = capacity of the passage between cell $i$ and cell $j$: this is the maximum amount of people that, independently on how many persons are in cell $j$, can traverse the passage in the time unit (independence on cell occupancy means neglecting system congestion: we will consider this issue later).

The flow model uses an acyclic digraph $D$ with node set $V \times T$ and arc set

$$E = \{(i,t) \rightarrow (j,t+1) : ij \in A, t \in T\}$$

Referred to as $\tau$-*time* or *dynamic network* in (26), $D$ models all the feasible transitions (moves between adjacent cells) that can occur in the building in the time horizon $T$. Transitions are associated with the $x$-variables defined above, whereas $y$-variables define the occupancy of each room (and of the building) from time to time. The $x$- and $y$-variables are declared integer and subject to the following constraints:

$$y_j^t - y_j^{t-1} - \sum_{i: ij \in A} x_{ij}^{t-1} + \sum_{i: ji \in A} x_{ji}^{t-1} = 0 \qquad j \in V, t \in T, t > 0 \qquad (4.1)$$

$$0 \leq x_{ij}^t + x_{ji}^t \leq c_{ij} \qquad t \in T, ij \in A \qquad (4.2)$$

$$0 \leq y_i^t \leq n_i \qquad t \in T, i \in V \qquad (4.3)$$

Equation (4.1) is just a flow conservation law: it expresses the occupancy of cell $j$ at time $t$ as the number $y_j^{t-1}$ of persons present at time $t-1$, augmented of those that during interval $(t-1, t]$ move to $j$ from another cell $i = j$, minus those that in the same interval leave cell $j$ for another room $i = j$. Box constraints (4.2), (4.3) reflect the limited hosting capability of the elements of $G$.

**Maximizing outflow in a given time.** To model the relation between time and people outflow, one can try to maximize the number of persons evacuated from the building within $\tau$:

$$\max \quad y_0^\tau \qquad (4.4)$$

This is the *Max Flow Problem* (MFP) considered in (26). To find the minimum total evacuation time, one can solve an MFP for different $\tau$, looking for the least value that yields a zero-valued optimal solution. To reduce computation time, this optimal $\tau$ can be computed by logarithmic search. The method can thus provide the decision maker with the Pareto-frontier of the conflicting objectives $\min\{\tau\}$, $\max\{y_0^\tau\}$. The linear structure of the model allows its solution with a large number of variables. Adding variables can help improve model granularity by reducing space and time units (e.g., counting people every 4 seconds instead of every minute). More importantly, it can also help approximate the non-linearities of arc capacities. In fact, $c_{ij}$ constant in (4.2) fails to model congestion, that is a situation in which the speed at which the system empties is a decreasing function of room occupancy. A more accurate model of congestion requires arc capacity to be a concave decreasing function of room occupancy. Linearizing this function is quite standard in applications: for simplicity, let us refer to the three-pieces linearization of Figure 4.1. One can rewrite

$$y_i^{t-1} = u_i^{t-1} + v_i^{t-1} + w_i^{t-1} \qquad x_{ij}^t = \varphi_{ij}^t + \chi_{ij}^t + \psi_{ij}^t \tag{4.5}$$

with $u_i^{t-1}$, $v_i^{t-1}$, $w_i^{t-1}$ non-negative and subject to upper bounds

$$u_i^{t-1} \leq n_i \qquad v_i^{t-1} \leq n_i - n_i \qquad w_i^{t-1} \leq n_i - n_i \tag{4.6}$$

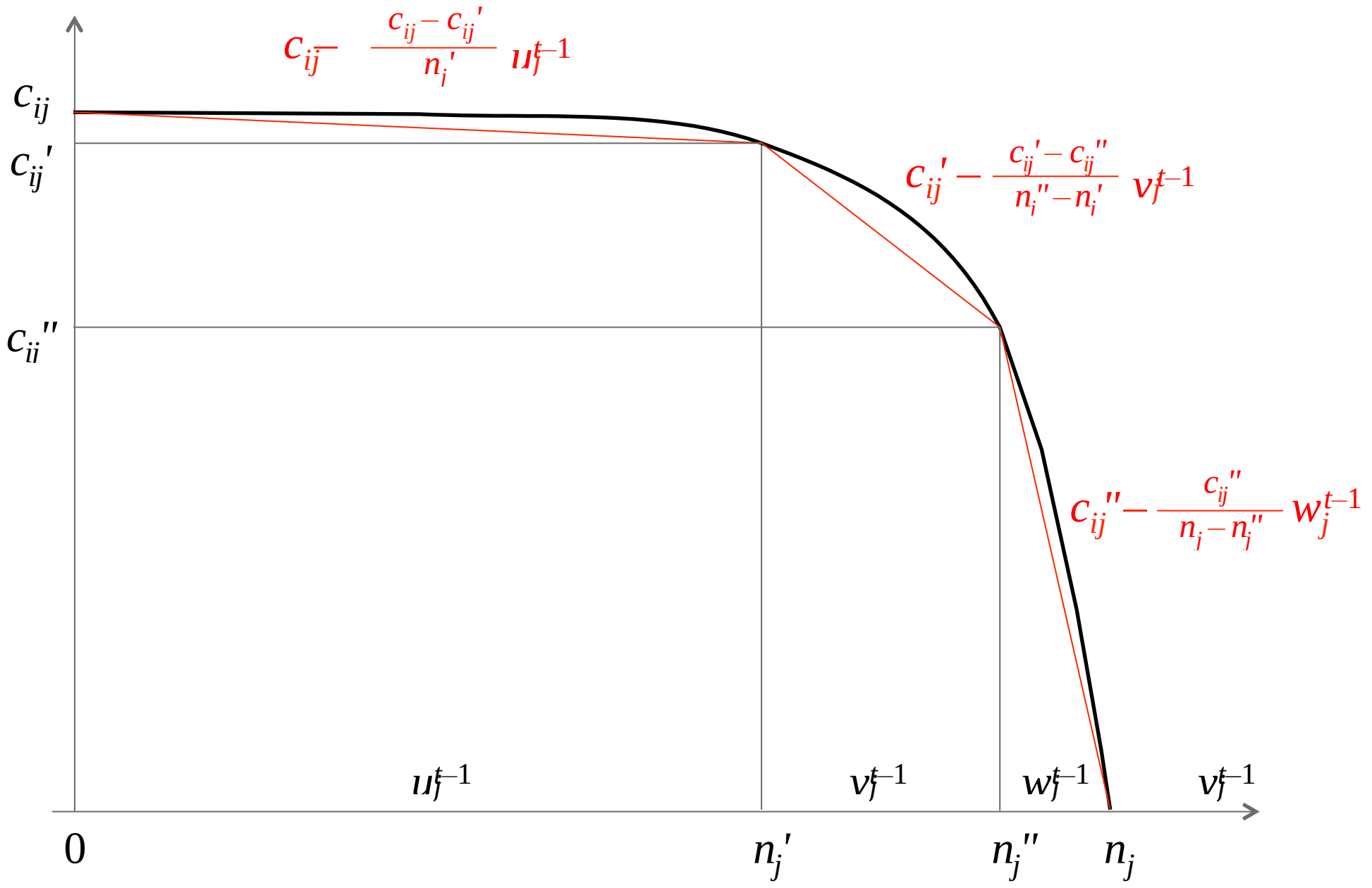

**Figure 4.1** Congestion curve and a linearization.

Arc capacity constraints (4.2) are then replaced by

$$\begin{aligned} 0 \leq \varphi_{ij}^t &\leq c_{ij} - \frac{c_{ij} - c_{ij}}{n_j} u_j^{t-1} \\ 0 \leq \chi_{ij}^t &\leq c_{ij} - \frac{c_{ij} - c_{ij}}{n_j - n_j} v_j^{t-1} \\ 0 \leq \psi_{ij}^t &\leq c_{ij} - \frac{c_{ij}}{n_j - n_j} w_j^{t-1} \end{aligned} \tag{4.7}$$

Consistency of the $\varphi$ , $\chi$ and $\psi$ variables with the $x$ flow variables requires $\chi = 0$ ($\psi = 0$)

if $\varphi$ (if $\chi$) does not saturate its capacity. To see this, we rephrase (4.7):

$$
\begin{array}{llll}
0 \le \varphi^t_{ij} & \varphi^t_{ij} + a_{ij} u^{t-1}_j & \le c_{ij} \\
0 \le \chi^t_{ij} & \chi^t_{ij} + a_{ij} v^{t-1}_j & \le c_{ij} \\
0 \le \psi^t_{ij} & \psi^t_{ij} + a_{ij} w^{t-1}_j & \le c_{ij}
\end{array}
\tag{4.8}
$$

where

$$a_{ij} = \frac{c_{ij} - c_{ij}}{n_j} \qquad a_{ij} = \frac{c_{ij} - c_{ij}}{n_j - n_j} \qquad a_{ij} = \frac{c_{ij}}{n_j - n_j}$$

and $a_{ij} < a_{ij} < a_{ij}$. We then observe a simple property hold by optimal basic solutions, that can easily be generalized to any piecewise linear approximation of the congestion curve. Suppose that, in a feasible solution, $\bar{u}^t_i < n_i$ and $\bar{v}^t_i > 0$. Let

$$\delta = \min\{n_i - \bar{u}^t_i, \bar{v}^t_i\}$$

Then a solution with $u^t_i = \bar{u}^t_i + \delta$, $v^t_i = \bar{v}^t_i - \delta$ and the other components unchanged is also feasible and no worse than the given one.

*Proof.* In fact, by definition of $\delta$, $u^t_i \le n_i$ and $v^t_i \ge 0$. Moreover, by the first of (4.5) the occupancy of $i$ at $t$ remains unchanged. As far as the implication on $\varphi, \chi$ is concerned, the sum of the relevant arc capacities is increased by $\delta(a_{ij} - a_{ij}) > 0$. Thus it is possible to compensate a decrease of $\bar{\varphi}^t_{ij}$ with an identical increase of $\bar{\chi}^t_{ij}$ resulting by the second of (4.5) in an equivalent flow $y_j x^t$.

## 4.3 Setting Model Parameters

To get a reliable model, parameters must be set to numbers that reflect reality. Those numbers depend on several considerations, the most relevant being: model granularity, walking velocity in various conditions (on a flat, on staircases etc.), door (and staircase) entrance capacities, room (and staircase) capacities.

### 4.3.1 Granularity

The issue of model granularity touches both spatial and temporal units, and affects the shape, size and neighborhood function of the unit cells in which the building is decomposed, as well as the slots that form the evacuation time horizon.

**Cell shape.** As described in the previous sections we embed the building plan into a grid, whose cells are assumed isometric: that is, can be crossed in any direction in the same amount of time. We cannot consider each room as a cell, since the rooms vary in size and our model needs unified cells. The reason is that the time people pass from the center of a cell to the center of a neighbor cell should be always equal. That amount will define the time slot duration, and cells will be regarded as virtual unit-rooms that communicate one another via physical or virtual (i.e., open space) doors. Grid geometry can vary. Ideal isometric cells are circles, but circles are not embeddable into a grid with adjacent cell sides. Hexagon cells are a good compromise between isometry and plan embedding. However, in our case study we found room sizes and shapes well compatible with a *quasi-square* grid where each room is split into an integer number of cells. Although almost square, those cells are not isometric and therefore transitions in different directions require different time. This implies some caution in constructing the dynamic network $D$, as it will be explained later.

**Cell size.** Cell size has an evident effect on the resulting spatial patterns, and consequently on both computational efficiency and model accuracy: the largest the cell, the lesser vertices in $G$ and the lower the refresh frequency at which people positions are updated. Given the speed at which people moves and data are acquired, relatively low refresh frequencies are not an issue; instead, an exact partition of each room in identical cells may result in a huge network. Thus, one can in general approximate the diverse room shapes by $a \times b$ rectangles as large as possible, while minimizing the error introduced. Various ways can be adopted to measure approximation error: the most natural is the difference between real and approximated room area, in which case, for room $k$ of size $p_k \times q_k$, the error is given by

$$e_k(a, b) = q_k[p_k \text{mod}(a)] + p_k[q_k \text{mod}(b)] - [p_k \text{mod}(a)][(q_k \text{mod}(b)]$$

As we need isometric cells and look for a uniform approximation, we set $a = b$ and find $a$ minimizing $\max_k\{e_k(a, a)\}$, meanwhile limiting the total number of approximating cells to some predefined $m$: hence we choose among the values of $a$ that fulfill $\sum_k p_k/a \;\; q_k/a \leq m$. A brief description of the method implementation is presented in Section 6.

### 4.3.2 Walking velocity and door/cell capacity

**Walking velocity**. The cornerstone on which the length of each unit time slot in $T$ | and consequently its reciprocal, the monitoring frequency | is established, is the *free flow walking velocity*, i.e. the speed at which humans prefer to walk in non-congested and non-hampered conditions. Clearly, its value varies for different categories of people (child, adult, elderly, disable etc.) and slope (flat, upstairs, downstairs). This parameter is important to perceive the distance that an individual can possibly walk during a specific period of time, and its evaluation contributes to define the cells in which an area is to be divided in order to best approximating travel time. Table 4.1 reports different estimates of pedestrian free flow velocity found in literature.

For our optimization purposes, a single global velocity is needed in homologous ar-eas: in our case, we distinguish flat zones and stairways. For the former, we consider the average free flow walking speed for a flat surface equal to 1.2 m/s (Ye's estimate). For the latter, we observe that in our study the flow is rarely upward and all people located in first floor are supposed to go downstairs to safe places located at the building ground floor: for this situation we stay at Kratchman's estimate that is 0.61 m/s. Therefore, the speed in flat zones is almost twice that of on downward stairways. Different speeds in different building areas will be treated by varying network granularity, as it will be clear in the following; the approach can easily be generalized to diverse situations.

**Table 4.1** Pedestrian free flow velocity.

| **Flat** (m/s) under 65 | over 65 | **Reference** | **Stairs** (m/s) up | down | **Reference** |
|---|---|---|---|---|---|
| 1.36 | | Fruin 1971 (41) | 0.56 | 0.65 | Fruin 1971 (41) |
| 1.36 | | Weidmann 1993 (107) | 0.61 | 0.69 | Weidmann 1993 (107) |
| 1.25 | 0.97 | Knoblauch et al. 1996 (58) | $0.8 \pm 0.19$ | | Kratchman 2007 (59) |
| 1.042 - 1.508 | 0.8 9 - 1.083 | TranSafety Inc. 1997 (102) | 0.59 | 0.76 | Jiang et al. 2009 (50) |
| 1.20 | | Ye et al. 2008 (111) | $0.81 \pm 0.13$ | | Fang et al. 2012 (38) |
| | | | 0.83 | 0.86 | Patra et al. 2017 (80) |

**Door capacity**. The capacity of a door depends on such various aspects as user composition, door type (always open, open when used, turnstile), crowdedness and, last but not least, door width. A study by Daamen et al. (31) focuses on the relationship between door capacity, user composition and stress level, arguing an average 2.8 persons per second for a 1-meter width door (p/m/s). From literature review, door capacities turn out in a range between 1.03 P/m/s and 3.23 P/m/s. None of the aforementioned studies consider bidirectional flow through doors, and so do we in our optimization model: in fact, although bidirectional flows are in principle possible, they will never occur in an optimal solution even without considering flow rate reductions due to collisions. Taking advantage of the aforementioned review and considering the bidirectional flow through doors, We carry out our study using 1.2 p/m/s for every door, meaning that a maximum number of 4.8 persons can pass through a 1-meter width door per time slot (4 seconds). Staircase capacity is treated differently. Fruin (41) measures a maximum flow capacity of 5.5881 persons/s for a 5.6 width staircase, i.e., about 1 p/m/s. Weidmann (107) obtains 0.85 p/m/s for the same parameter. We stay at Fruin's estimate and use 1 p/m/s for each staircase in our example.

**Cell capacity**. The *pedestrian density* is the number of persons per square meter monitored at any time. This information is crucial for crowd safety and evacuation performance, as movements are dramatically reduced in highly dense areas. As density increases, pedestrian movements become constrained and flow rate consequently decreases. According to UK fire safety regulations, the maximum allowed density corresponds to 0.3 square meters per standing person, a value that increases to 0.5 for public houses, to 0.8 for exhibition space, to 1.0 for dining places, to 2.0 for sport areas and to 6 for office areas. In our case study | gallery indoor space | the maximum capacity of each cell is calculated by assuming 0.8 square meters per visitor, that is 1.25 persons per square meter.

## 4.4 Application

We next describe an application of the model of Section 4.2 to the safe evacuation of Palazzo Camponeschi, a building in L'Aquila (Italy) now and then used for exhibitions. Safety conditions of the building, which consists of 31 rooms including corridors (Figure 4.2), are supervised by L'Aquila Fire Brigade and Civil Protection department. Rooms sizes vary in a large range, and consequently the average time of a person to cross them from door to door. The variegate building structure, as well as data on people attendance collected during events, made this study-case ideal for illustrating a general methodology for system sizing and development.

**Cellular approximation of rooms.** As explained in §4.3.1, we split each room in unit cells, each behaving as a (virtual) square room that can be traversed in a unit time slot. In practice, we embedded the building plan into a square grid as shown in Figure 4.2. Red-marked cells are dummy nodes added to adjust pedestrian velocity on stairs (half that on flat surface, §4.3.2) to the unified monitoring time. To decide cell size, we look at both the error introduced by room approximation and the number of nodes in the resulting graph $G$. The latter is in an inverse proportion of cell size (bottom diagram in Figure 4.3); the former varies irregularly with cell size (upper diagram of Figure 4.3). We considered square cells up to $3.5 \times 3.5$ meters (the short edge of the smallest room) and allowed no more than 500 nodes of $G$; then we selected the size that minimizes the largest error for all rooms. As shown in Figure 4.3, $2 \times 2$ cells give the best approximation and imply 462 graph nodes. With $3.5 \times 3.5$ cells, the error rises a little bit over 40 (that is, twice that of $2 \times 2$) but $G$ results in 116 nodes only. Summarizing, $3.5 \times 3.5$ leads to larger error but less CPU time; conversely, $2 \times 2$ causes larger CPU time but smaller error. On the one hand, having a large error reduces the system accuracy since the the exact position of people in the building is not detected. On the other hand, having a high CPU time is against the real-time requirements of the system. We tested scenarios with both cell sizes in order to find the best efficiency/accuracy compromise.

**IoT Infrastructure.** To monitor and control the physical space, an IoT infrastructure consisting of sensing, computation, actuation and other elements is established. For the purpose of our experiment, we took advantage of $CO_2$ and temperature sensors to detect potential disaster, and of RFID tags/readers to gather the data necessary to feed the algorithm, as well as dynamically sensing/tracking crowd geo-location/movement in the physical environment (77; 76). In our case, RFID tags are embedded into tickets that visitors should keep during their visit.

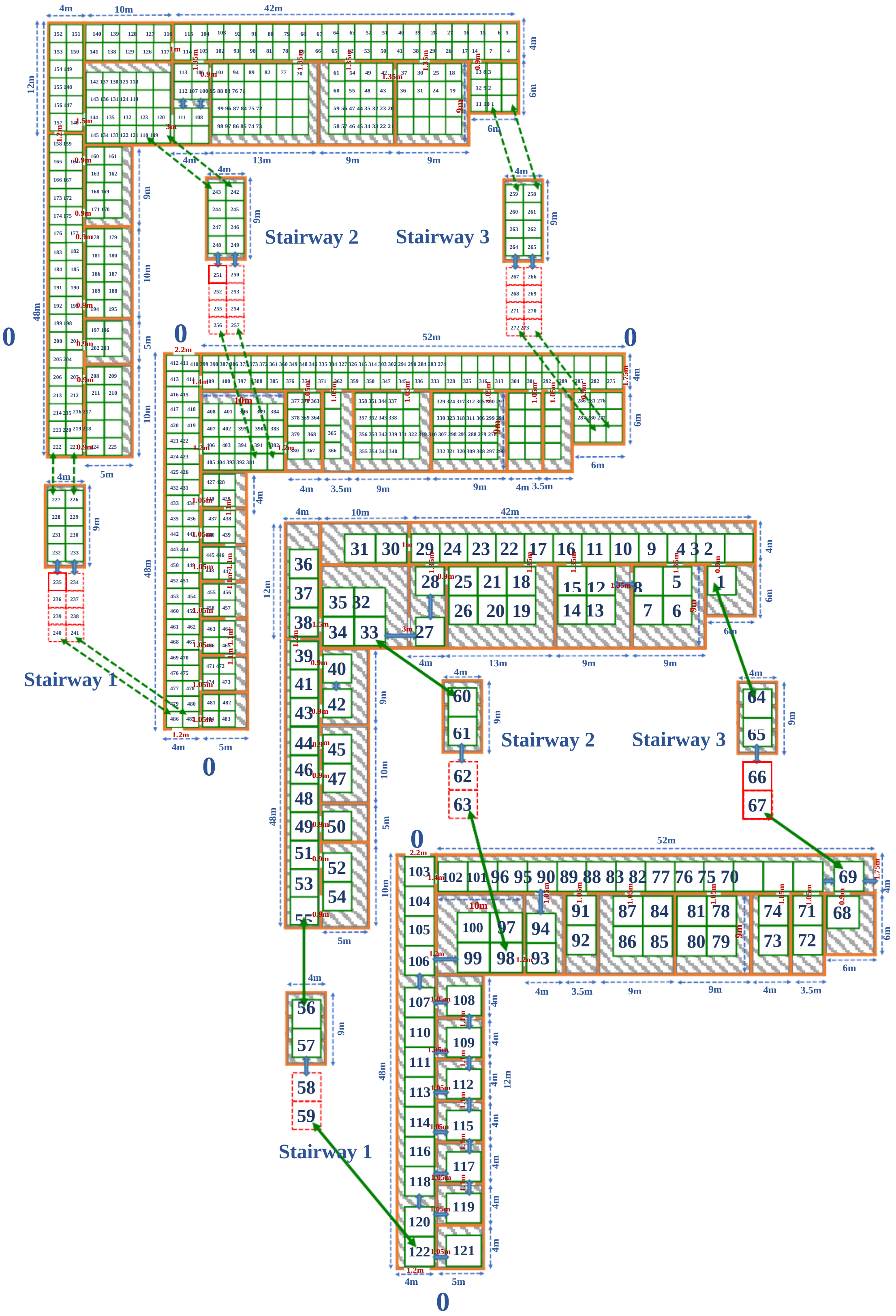

**Figure 4.2** Plan embedding of Palazzo Camponeschi into square grids with low (down-right) and high (up-left) resolution. For each resolution, ground floor down-right, first floor up-left. The area that is not covered by cells (error) is shown in gray.

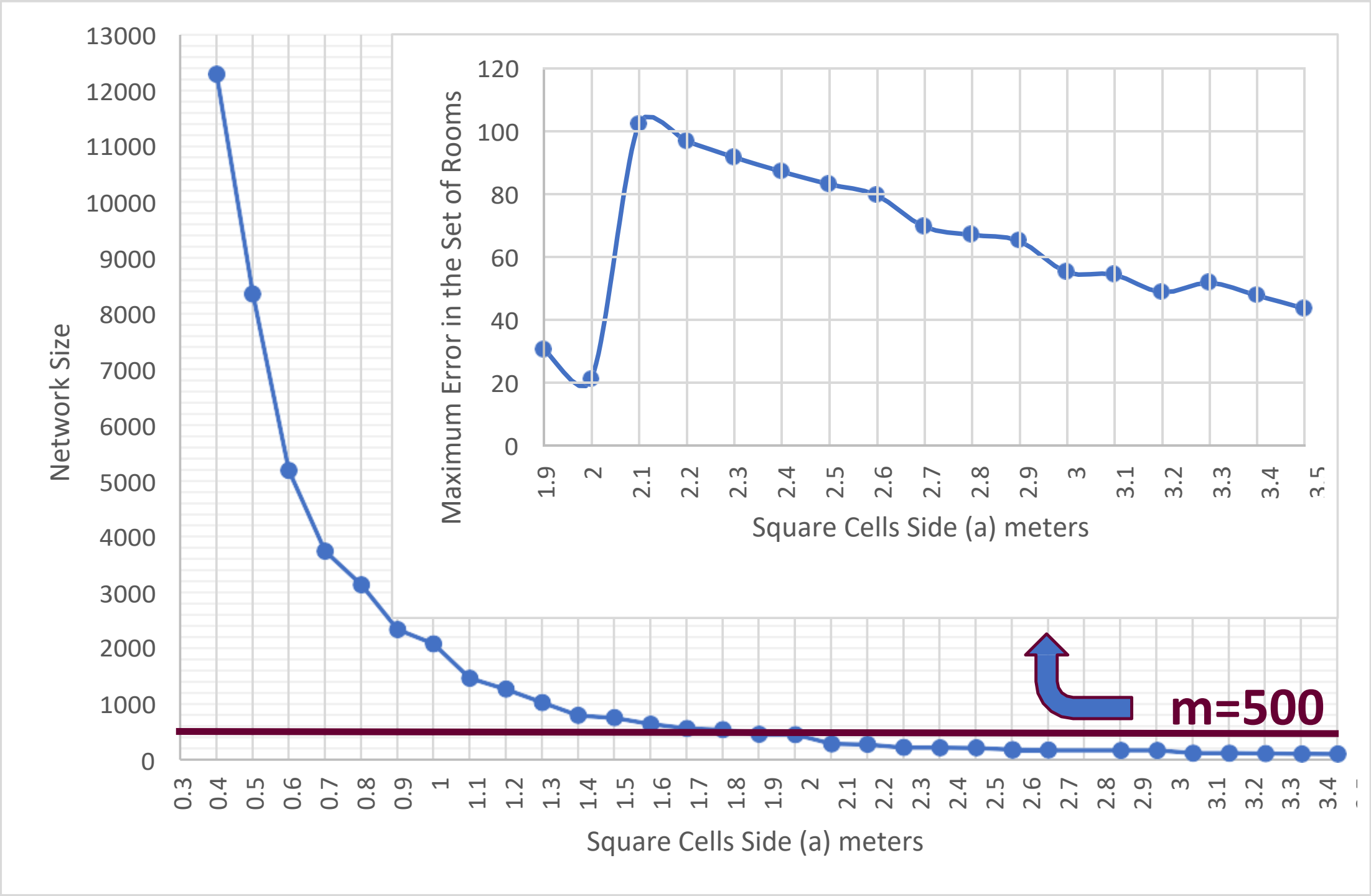

**Figure 4.3** Optimal cell size: maximum error (up right) and network size as a function of cell size.

In normal situation, the sensors read $CO_2$ concentration, temperature and crowd geolocation frequently. This monitoring frequency, or its reciprocal that is the single cell crossing time, is derived from cells dimension and free flow speed: by the average free flow speeds, we obtained time slots of 2.92 seconds for $3.5 \times 3.5$ cells, and of 1.67 seconds for $2 \times 2$ cells. The main goal of this application, to be run on a tablet, is to show a 2D-representation of the monitored space enriched with RFID-sensed contextual data that report where visitors are at any time located, and how they move in normal cases. In this mode, the optimal flow algorithm is periodically run to estimate the minimum evacuation time required under current conditions, a value that can be used to regulate visitor access so as to comply with safety conditions.

If a disaster detected, temperature/$CO_2$ detectors are read more frequently. In addition to map on dashboards, actuators of evacuation signs in each area show the best evacuation routes based on the discussed network model.

**The network model**. In the case here studied, plan embedding results in a graph with 116 (+6 dummy) nodes using $3.5 \times 3.5$ cells, or 462 (+24 dummy) nodes using $2 \times 2$ cells. Figure 4.4 corresponds to the lower resolution of Figure 4.2 down-right, and includes node 0 as safe place. With $2 \times 2$ cells the network is very large and is not displayed. Adjacent cells are linked by arcs which allow flow inside the building. All arcs are assumed bidirectional except those towards the safe place. Time slots are as described above: 2.92 seconds for $3.5 \times 3.5$ cells and 1.67 seconds for $2 \times 2$ cells. Door capacities vary according to size and features. In our case, no more than 1.2 persons can pass through a 1-meter arc (door or cell connection) per second.

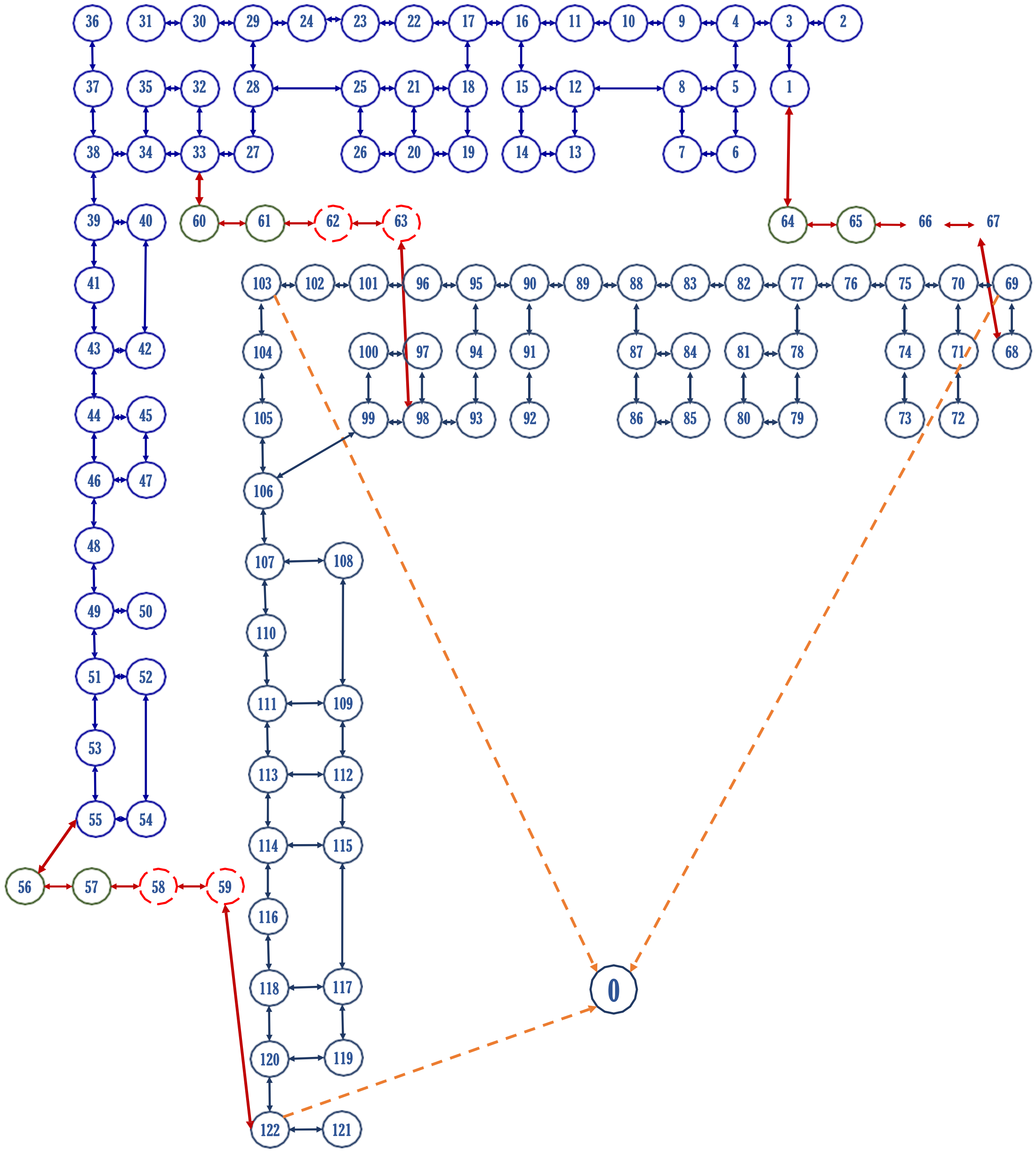

**Figure 4.4** Network associated with the plan of Figure 4.2 down-right.

**Simulations**. We next report the outcome of simulations using the optimization model under various conditions in terms of building occupancy, and compare it with static emergency maps based, as in usual circumstances, on shortest path route prescription. Simulations were first run for both cell sizes to evaluate the best choice. In all tests, we computed the minimum time required of $N$ persons, randomly distributed in the building rooms, to reach a safe place. The code for simulation was written in OPL language and problems were solved by CPLEX version 12.8.0. We ran all the experiments on a Core i7 2.7GHz computer with 16Gb of RAM memory under Windows 10 pro 64-bits.

*Scenario 1.* In the first simulation we suppose an initial occupancy of $N = 528$. This datum comes from an experiment performed in L'Aquila during the *Researchers' Night* event on 29 September 2017, when the simultaneous presence of 528 people in Palazzo Camponeschi was recorded as peak value. We solved problem (4.1)-(4.8) for $\tau = 1, 2, \ldots$ until a solution of value $N$ is found. Table 4.2 reports the number of evacuees at each $\tau$ and the computation time of each solution step. Computation is done for both *low-* and *high-resolution* networks (respectively, $3.5 \times 3.5$ and $2 \times 2$ cells). With the low-resolution network, we get the evacuation and CPU times of Table 4.2 left: in terms of evacuation, everyone has reached a safe place in 1′25″; on the other hand, computation requires 2.89 seconds (presolve included) in the worst case, and is therefore totally compliant with real-time applications. Evacuation and CPU times for the high-resolution network are reported in Table 4.2 right. CPU time is now much larger (93.05 seconds) and the model appears inappropriate for real-time use, unless additional computational resource is deployed. We also see that everyone has reached a safe place in 1′27″: hence we can conclude that sufficient accuracy is obtained using the low resolution network.

**Table 4.2** Evacuation and computation time - Scenario 1 (528): *a)* $3.5 \times 3.5$ cells (time slots of 2.92 seconds); *b)* $2 \times 2$ cells (time slots of 1.67 seconds).

| $\tau(a)$ | evacuees $(a)$ | CPU Time $(a)$ | $\tau(b)$ | evacuees $(b)$ | CPU Time $(b)$ | $\tau(b)$ | evacuees $(b)$ | CPU Time $(b)$ |
|---|---|---|---|---|---|---|---|---|
| 1 | 18 | 0.46 sec | 1 | 10 | 5.01 sec | 30 | 309 | 33.27 sec |
| 2 | 37 | 0.50 sec | 2 | 21 | 5.12 sec | 31 | 319 | 34.02 sec |
| 3 | 56 | 0.51 sec | 3 | 31 | 5.11 sec | 32 | 330 | 36.17 sec |
| 4 | 74 | 0.47 sec | 4 | 41 | 5.14 sec | 33 | 340 | 38.40 sec |
| 5 | 92 | 0.51 sec | 5 | 51 | 5.51 sec | 34 | 350 | 40.61 sec |
| 6 | 111 | 0.61 sec | 6 | 62 | 5.78 sec | 35 | 360 | 41.94 sec |
| 7 | 130 | 0.67 sec | 7 | 72 | 5.92 sec | 36 | 371 | 44.37 sec |
| 8 | 148 | 0.89 sec | 8 | 82 | 6.83 sec | 37 | 381 | 43.63 sec |
| 9 | 167 | 0.91 sec | 9 | 93 | 6.66 sec | 38 | 391 | 47.49 sec |
| 10 | 185 | 0.71 sec | 10 | 103 | 7.30 sec | 39 | 402 | 51.52 sec |
| 11 | 204 | 0.93 sec | 11 | 113 | 7.49 sec | 40 | 412 | 54.31 sec |
| 12 | 222 | 1.35 sec | 12 | 124 | 7.85 sec | 41 | 422 | 56.41 sec |
| 13 | 241 | 1.24 sec | 13 | 134 | 8.84 sec | 42 | 433 | 60.02 sec |
| 14 | 260 | 1.39 sec | 14 | 144 | 9.73 sec | 43 | 443 | 60.77 sec |
| 15 | 278 | 1.48 sec | 15 | 154 | 10.28 sec | 44 | 453 | 68.12 sec |
| 16 | 297 | 1.25 sec | 16 | 165 | 11.54 sec | 45 | 463 | 63.99 sec |
| 17 | 315 | 1.31 sec | 17 | 175 | 12.15 sec | 46 | 474 | 65.30 sec |
| 18 | 334 | 1.43 sec | 18 | 185 | 12.90 sec | 47 | 484 | 75.87 sec |
| 19 | 352 | 1.53 sec | 19 | 196 | 13.96 sec | 48 | 494 | 78.24 sec |
| 20 | 371 | 1.63 sec | 20 | 206 | 15.47 sec | 49 | 505 | 81.01 sec |
| 21 | 389 | 1.63 sec | 21 | 216 | 16.95 sec | 50 | 515 | 79.13 sec |
| 22 | 408 | 2.16 sec | 22 | 227 | 18.78 sec | 51 | 524 | 80.81 sec |
| 23 | 426 | 2.34 sec | 23 | 237 | 19.26 sec | 52 | 528 | 93.05 sec |
| 24 | 445 | 2.38 sec | 24 | 247 | 19.75 sec | | | |
| 25 | 463 | 2.53 sec | 25 | 257 | 21.73 sec | | | |
| 26 | 482 | 2.65 sec | 26 | 268 | 23.51 sec | | | |
| 27 | 501 | 2.76 sec | 27 | 278 | 27.08 sec | | | |
| 28 | 519 | 2.83 sec | 28 | 288 | 23.51 sec | | | |
| 29 | 528 | 2.89 sec | 29 | 299 | 30.35 sec | | | |

*Ideal vs. shortest paths.* This first simulation depicts an ideal situation in which flows autonomously choose the best among all the available routes in the building. Of course, managing such an ideal evacuation is not easy and perhaps unpractical. As a general practice, in fact, evacuation is conducted through pre-determined routes. Considering this fact, then, we suppose that the prescribed evacuation routes are the shortest paths from any cell to the safe place. To evaluate this situation we find the subgraph of $G$ formed by the shortest paths from any cell to 0 (as from static evacuation map), construct its time-indexed network and solve problem (4.1)-(4.4) for increasing $\tau$. Evacuating 528 individuals takes of course more time: 1 minute and 42″. Thus, compared to the Netflow model we propose, shortest routes increase, in this case, the optimal evacuation time by 17%.

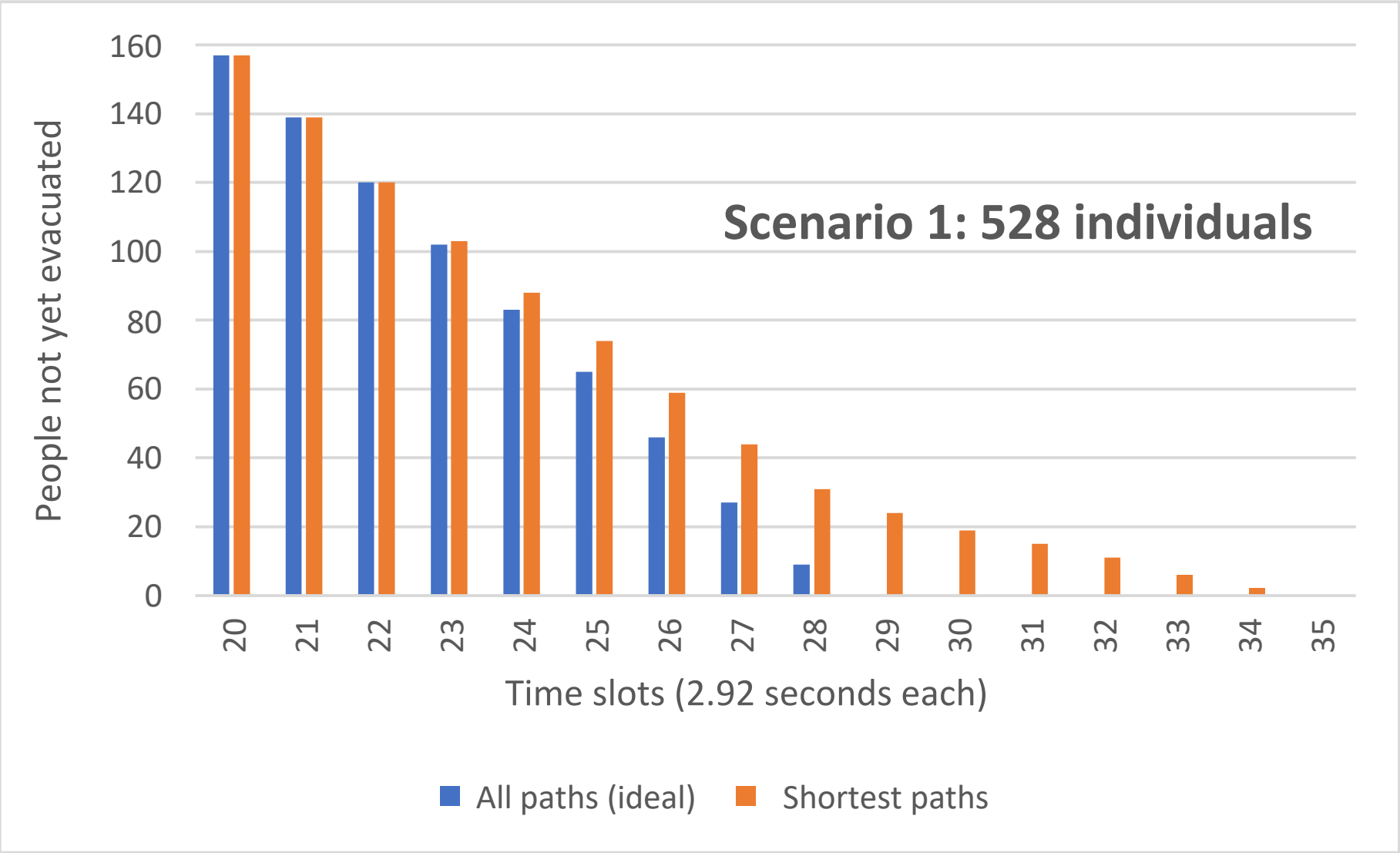

**Figure 4.5** Ideal vs. shortest paths evacuation with $3.5 \times 3.5$ square cells: Scenario 1.

*Scenario 2.* In a second scenario, we repeated the simulation doubling the number of people in the building (1056 individuals). In this case everyone can reach a safe place after 57 time slots, i.e., 2 minutes and 47″. In a second simulation, then, we again suppose that the prescribed evacuation routes are the shortest paths from any cell to the safe place. In this situation, evacuating 1056 individuals takes 3 minutes and 31″. Shortest route evacuation increases in this case the optimum by over 26%.

*Scenario 3.* In a third scenario, we repeated the simulation increasing the number of people to 1584 individuals, three times that of Scenario 1. In this case everyone can reach a safe place after 86 time slots, i.e., 4 minutes and 11″. In a second simulation, then, we again suppose that the prescribed evacuation routes are the shortest paths from any cell to the safe place. In this situation, evacuating 1584 individuals takes 5 minutes and 4″, over 19% more than optimal flows.

In each of the three simulations, we observe plain flows for some time (1′04″ in the first scenario, 2′14″ in the second and 3′19″ in the third). After that time, shortest routes start experiencing congestion, and evacuation is slowed down. The phenomenon is illustrated in the charts of the figures: as one can expect, the tail of people still in the building increases with initial occupancy.

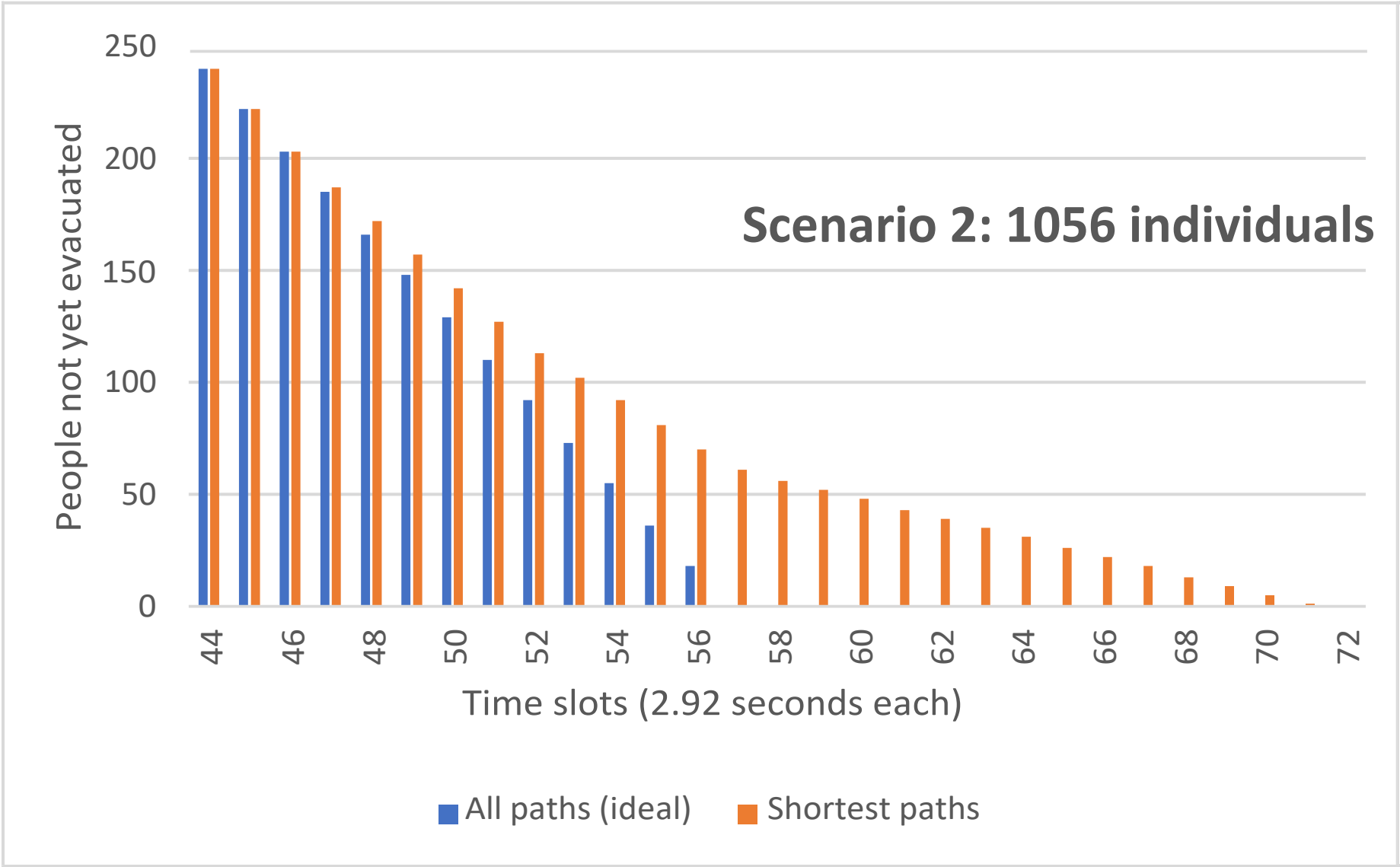

**Figure 4.6** Ideal vs. shortest paths evacuation with $3.5 \times 3.5$ square cells: Scenarios 2.

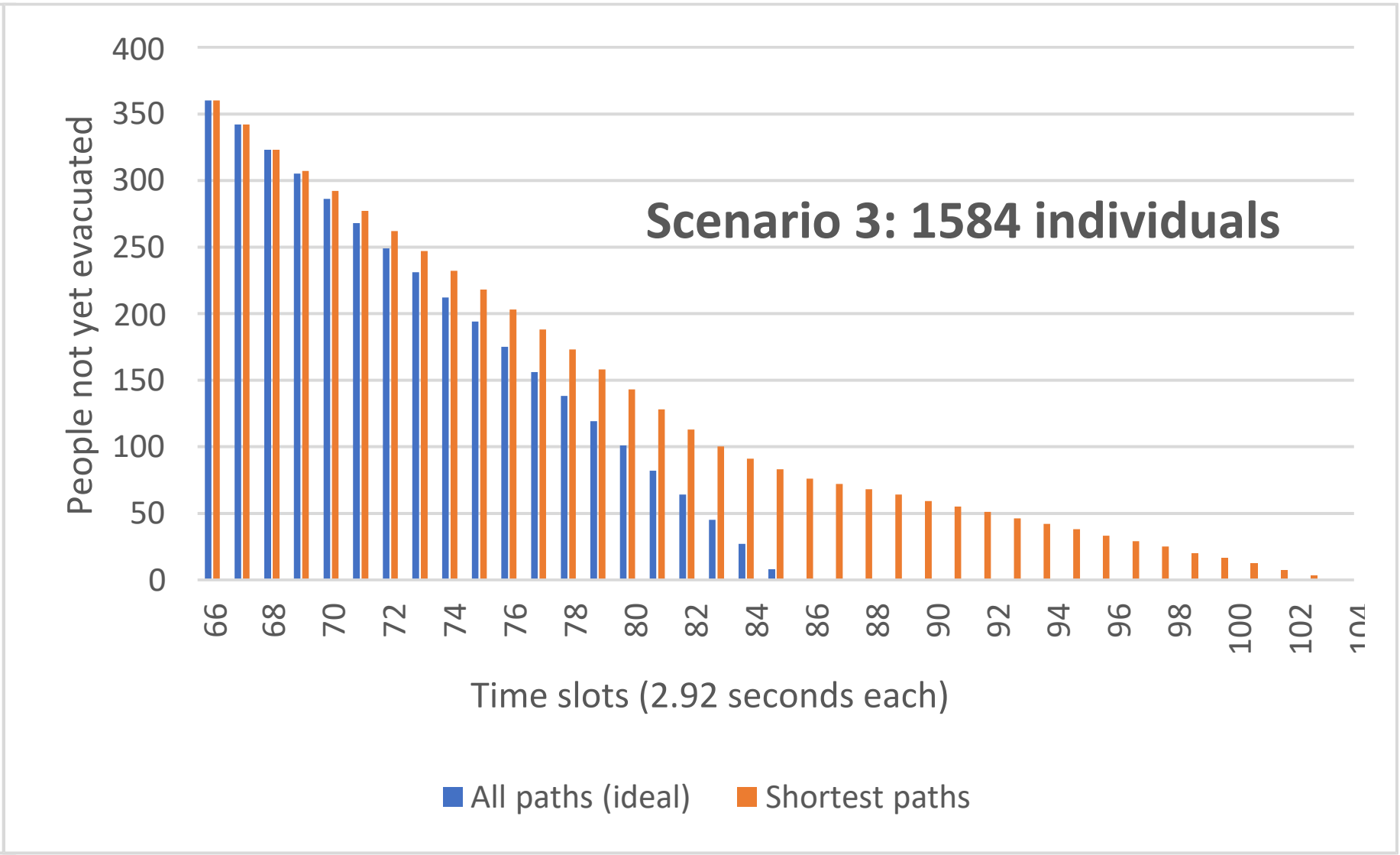

**Figure 4.7** Ideal vs. shortest paths evacuation with $3.5 \times 3.5$ square cells: Scenarios 3.

We conclude with a short note on system usage. Leading people through the ideal evacuation paths needs proper actuation components. Optimal routes can be communicated to people using evacuation signs, lights, smart carpets, human agents, etc. Recent models, simulations and experiments gave evidence to a precious role played by individuals that, in case of emergency, emerge as crowd-leaders, see e.g. (3). In the case study presented here, personnel positioned at almost each intersection was in charge of leading people to safe areas in case of emergency. A possible usage of our system would then envisage the periodical communication of best evacuation paths to personnel via a mobile device, which they are permanently equipped with.

## 4.5 Conclusions and Future Work

We use a run-time network flow model to support the rapid evacuation of people from a building in case of emergency. The model takes as input both static information (such as building dimension and structure, room capacities, door capacities) and run-time information (such as number of people in the building, number of people in specific rooms and corridors) acquired by IoT devices. The building topology is described by a graph. Run-time data are used to create a time-indexed acyclic digraph that models all the feasible transitions between adjacent rooms at any time. Minimizing evacuation time corresponds to solve max flow problems with non-linear capacities that model congestion. Preliminary evaluations have been conducted using data from a real case, and an *ad-hoc* IoT infrastructure has been designed.

The work presented opens ways to investigate a new model with cells occupied by one person at most. In this model, cell occupancy is regulated by individual reaction times, so that different types of individual behaviors can be described and supported. Possible constraints include grouping people (e.g., a couple or a family) into close paths: for instance, one can model specific individuals to be at any time within a prescribed distance. Further algorithmic research is however needed to assess the chances that such a model has to provide decision support in real-time.

More future work can include: *i)* enriching the model with further design/run-time information to make plans more accurate; *ii)* empirically evaluating the model in extensive scenarios, by comparing estimated vs. real data (as, e.g., those generated during evacuation tests); *iii)* completing the IoT infrastructure and app with the computational component here developed.

Next chapter evaluates the functionality of the IoT infrastructure and specifically the proposed algorithm in handling both real-time and design-time emergency management applications.

## Chapter 5

# An IoT Software Architecture for an Evacuable Building Architecture

This chapter continues the work presented in the previous chapter by evaluating the computational component proposed to improve and evaluate emergency handling plans. In real-time, the component operates as the core of an Internet of Things (IoT) infrastructure aimed at crowd monitoring and optimum evacuation paths planning. In this case, a software architecture facilitates achieving the minimum time necessary to evacuate people from a building. In design-time, the component helps discovering the optimal building dimensions for a safe emergency evacuation, even before (re-) construction of a building. The chapter is organized as follows. Literature is briefly discussed in Section 5.1. Section 5.2 defines the concept of evacuability and different emergency handling challenges in real-time and design-time. The self-adaptive architectures for IoT infrastructures are presented in Section 5.3. Section 5.4 presents the flow model whilst Section 6 discusses the static and dynamic risks that may happen during a disaster. Section 5.6 refers on how model parameters should be set up to deal with real cases. The application of the model to a real exhibition venue is presented in Section 5.7 and conclusions are finally drawn in Section 5.8.

## 5.1 Literature Review

Despite that a large body of knowledge has been proposed for surveillance software architectures, the research gap towards emergency evacuation architectural design is undeniable. However, the few related works deal with a small subset of disaster management architecture. Cabrera et al (42) propose an architecture to simulate a large-scale version of a virtual crowd in a building evacuation. Their architecture is composed by two elements: the action server and the client processes. However, their architecture remains high level and limited to control how the existing agents can share information about the 3D virtual scene. Lujak et al (63) propose a distributed architecture for situational aware evacuation guidance in smart buildings. They use WiFi, RFiD and Beacon for identification and sensing purposes. The users smart-phones act as reader of the beacon signals to localize and track the users. However, their architecture remains in an abstract level and a proper evaluation is not provided. Raj et al (87) propose a decentralized client-server architecture for their crowd evacuation system and the architecture includes some processing at the server end. However, only the sensing layer is designed in a decentralized manner (such as using mobile phones as sensors) and computation of danger indexes and evacuation plan remains centralized.

In order to support coordinated emergency management in smart cities based on the localization of first responders during crisis events, Palmieri et al (79) present a hybrid cloud architecture to manage computing and storage resources needed by command and control activities in emergency scenarios. Their first responder localization service relies on a novel positioning approach which combines the strength of signals received from landmarks placed by first responders on the crisis site with information obtained from motion sensors. Despite the level of distribution and its impacts on system non-functional requirements are not clear, their research can be considered as a proper complementary work to our architectural model, by adapting the geolocation of first responders to track people during an evacuation.

## 5.2 Building Evacuability: envisioned solution

Emergency evacuation handling for large scale roads and buildings is complex. Nowadays, evacuation plans appear as static maps, designed by civil protection operators, that provide some pre-selected routes through which pedestrians should move in case of emergency. The static models may work in low congested spacious areas. However, the situation may barely be imagined static in case of a disaster.

The static emergency map of the physical space for which our model is run (see Section 5.7) is shown in Figure 5.1. These kind of maps expose several limitations such as: *i)* ignoring abrupt congestion, obstacles or dangerous routes and areas; *ii)* leading all pedestrians to the same route and making that area highly crowded; *iii)* ignoring the individual movement behavior of people and special categories (e.g. elderly, children, disabled); *iv)* lack of providing proper trainings for security operators in various scenarios; *v)* lack of providing a comprehensive understanding for evacuation manager and operators by a real-time situational awareness. The advent of IoT architectures may support a quicker and safer evacuation. By simply tracking people in an indoor area, possible congestions can be detected and best evacuation paths can be periodically recalculated, or minimum evacuation time under ever-changing emergency conditions can be evaluated.

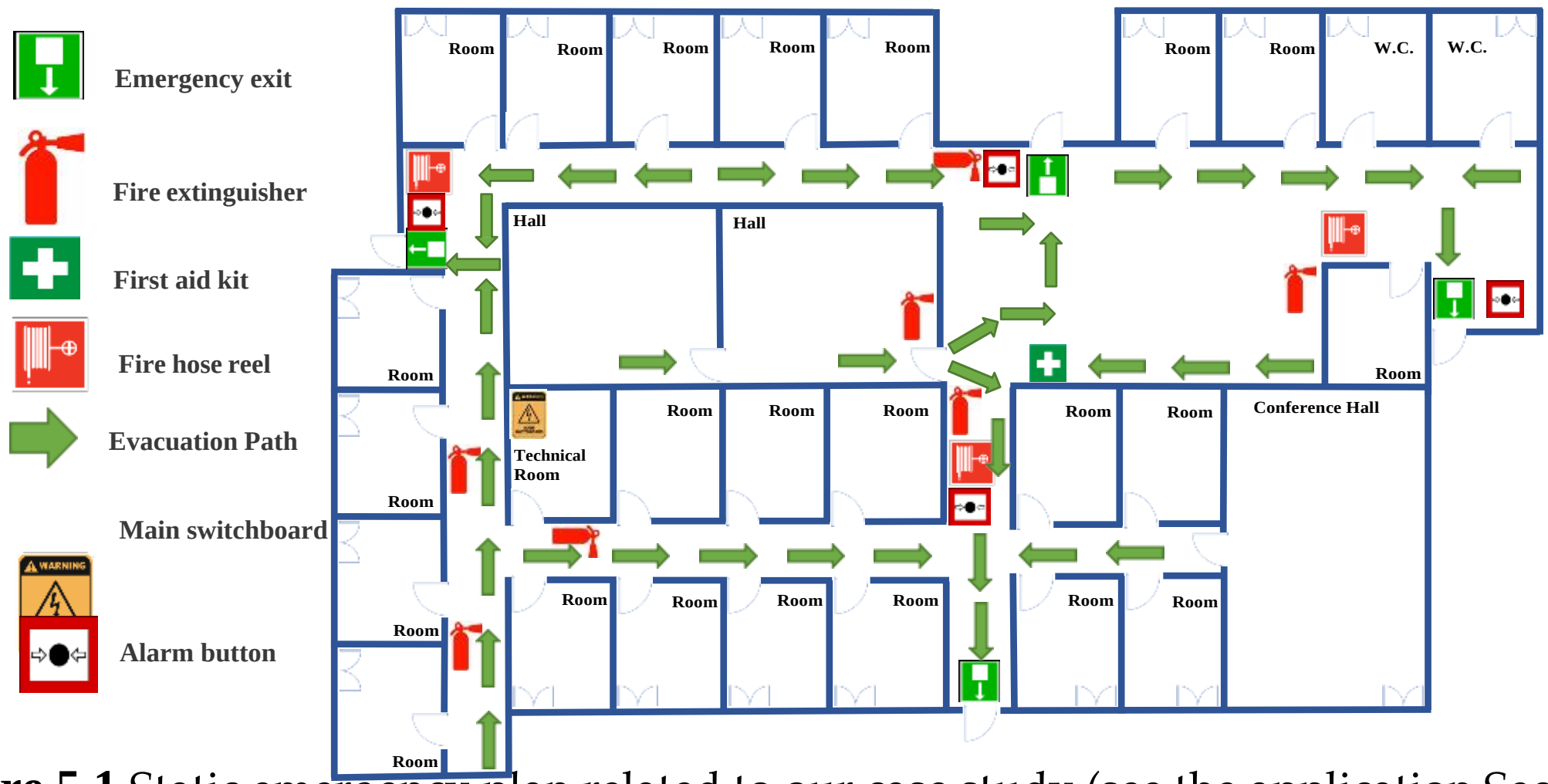

**Figure 5.1** Static emergency plan related to our case study (see the application Section).

We foresee design-time and run-time solutions. At *design-time*, a building architecture can be subject to safety evaluations even before its (re-) construction. We advocate the use of simulations as a feasible solution to assess the evacuability of buildings and feasibility of evacuation plans. However, a strong mathematical model should support the simulation tool. At design-time, an IoT-based evacuation system provides: *i)* Safety considerations for building architecture in early (re-) construction phase; *ii)* Finding out the building dimensions that lead to an optimum evacuation performance; *iii)* Bottleneck discovery that is tied with the building characteristics; *iv)* Comparing various routing optimization models to pick the best match one as a base of real-time evacuation system; *v)* Problem solutions for different time horizon provide a Pareto frontier that relates available time to the best possible people outflow in the given conditions; *vi)* Visualizing dynamic evacuation executions to demonstrate a variety of scenarios to security operators and train them.

At real-time, the IoT architecture we propose supports the gathering of data that will be used for dynamic monitoring and evacuation planning. At *real-time*, an IoT-based evacuation system provides: *i)* Optimal solutions that can be continuously updated, so evacuation guidelines can be adjusted according to visitors position that evolve over time; *ii)* Paths that become suddenly unfeasible can automatically be discarded by the system; *iii)* The model can be incorporated into a mobile app supporting emergency units to evacuate closed or open spaces.

## 5.3 Self-adaptive IoT Architectures for Emergency Handling

An IoT-based emergency evacuation architecture is defined as a safety critical IoT infrastructure to be used to collect and analyze data to perform proper actuation. In order to engineer such a high quality IoT application, a proper architecture should be designed with the ability to adapt itself to environment transformation, and in a proper level of elements distribution.

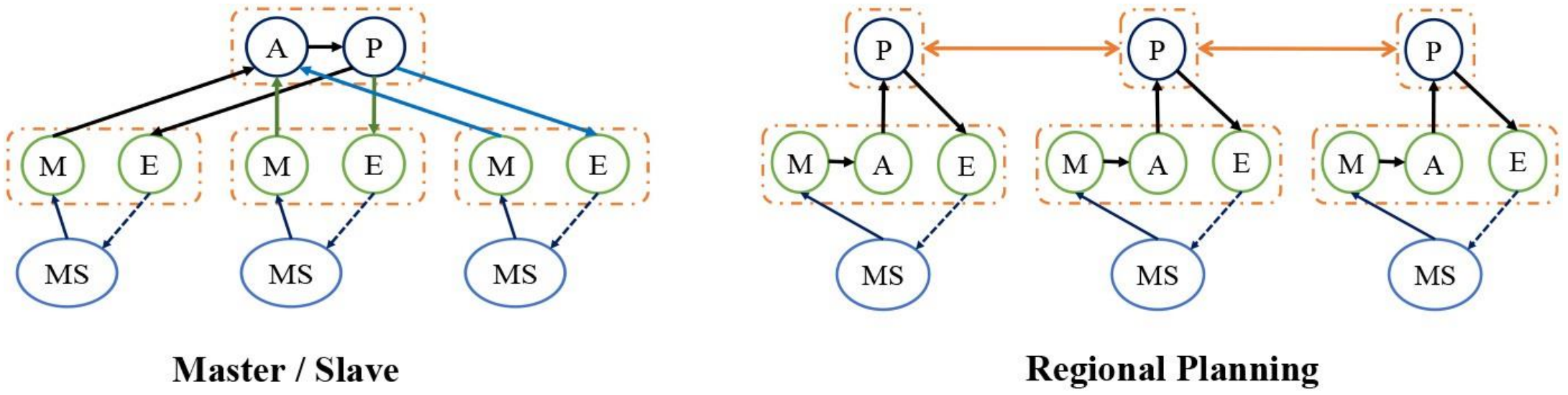

**Figure 5.2** Selected self-adaptation architectural patterns.

In our previous work (77), we classified IoT distribution patterns as: centralized, collaborative, connected intranets, and distributed based on a layered architectural style that consists of *Perception*, *Processing and Storage*, and *Application* layers.

Furthermore, Muccini et al (76) analyze a set of IoT distribution and self-adaptation patterns to identify their suitable architectural combinations. Here self-adaptation is based on a control loop such as MAPE (*Monitoring*, *Analysis*, *Planning*, *Execution*), that is a model objected on imposing automatic control on dynamic behavior of a system and has been used in various fields such as software engineering. However, we realized that most of adaptation patterns are unmatched with IoT distribution patterns, so that making a combination of them can be infeasible or weak. Furthermore, among feasible combined patterns, only two of them satisfy non functional requirements for IoT based emergency evacuation systems, that are fault-tolerance, performance, interoperability, scalability and energy efficiency. Therefore, in this work, we make a concrete use of those two suitable architectural patterns: *collaborative regional planning* and *centralized master/slave*.

Figure 5.2 shows the aforementioned self-adaptation control patterns. In the figure, managed subsystems (MS) comprise the application logic that provides the systems domain functionality. The managing subsystems instead manage the managed subsystems and comprise the adaptation logic. In the *collaborative regional planning* pattern, the local planners coordinate to find the best adaptation solution for a local or global problem. This pattern is suitable for our case study because of its high coordination of planners and low coordination of other local adaptation components (M, A, E) to provide fast and energy efficient decisions. The *centralized master/slave* pattern facilitates centralized decision making, and local monitoring and adaptation execution. This pattern is chosen as well since it simplifies achieving global objectives through central implementation of analysis and planning algorithms.

For both aforementioned patterns, the computational component adopted, will thus become the central element that, while inputting situational awareness information, will provide evacuation recommendations. This central computational component has a mathematical logic in behind that is proposed as an algorithm in the following section.

## 5.4 A Flow Model to Minimize Total Evacuation Time

In order to avoid repetition, we removed the algorithm explanations that are already discussed in section 4.2.

## 5.5 Risk Consideration

Despite that final objective of emergency handling plans are to minimize the evacuation time, safety risks have a critically important role. In the model discussed in Section 5 (see section 4.2), cells adjacency is modeled via a graph, to be able to model real-time changes in which the risky or infeasible paths are automatically discarded by deletion in the graph. However, due to low predictability essence of a disaster, considering risk aspects in design-time is complex.

Some disasters (such as earthquake) have a momentary impact on buildings, in which the risk appears as ruined areas or unavailable paths, so that it can be modeled via static changes in the graph in a specific time step. For the other category of disasters (such as fire), the risk can be propagated over time. In this case, different cells may be influenced by their neighbor during evacuation time steps. In our algorithm, dynamic propagation means that graph $G$ changes (with a form of arc removals) over time, that is, set $A$ is progressively reduced and has the form $A^t$. This has an effect on digraph $D$ and set $E$, that becomes in turn $E^t$. Consequently, all the constraints that depend on $A$ or $E$ will then be rewritten for $A^t$ and $E^t$. It is worth mentioning that, decision variables are reduced as well: $x_{ij}^t$ is in fact defined for all $ij$ in $A^t$ (which are generally less than those in $A$).

In both aforementioned categories of risk: *i)* the risky cells should be evacuated as quick as possible; *ii)* no one should be entered inside them; *iii)* disaster suppression equipment should be brought to the risky cells. In our example of application, we simulate a situation of static risk, however, due to the page and time limits of, we address the dynamic risk concept in our future work.

## 5.6 Setting Model Parameters

To get a reliable model, parameters must be set to numbers that reflect reality. Those numbers depend on several considerations, the most relevant being: model granularity, walking velocity in various conditions, door entrance capacities, cell capacities. We set most of parameters the same as previous chapter (see Section 4.4). The only difference is that, in this chapter we set door capacity in a range between optimistic and pessimistic values to observe its impact on evacuation time.

### 5.6.1 Door capacity

The capacity of a door depends on such various aspects as user composition, door type (always open, open when used, turnstile), crowdedness and, last but not least, door width. A study by Daamen et al. (31) focuses on the relationship between door capacity, user composition and stress level, arguing an average 2.8 persons per second for a 1-meter width door (p/m/s). They argue a door capacity range between 1.03 P/m/s and 3.23 P/m/s, resulted from a literature review. Taking advantage of the aforementioned review, we carry out our case study simulations considering the pessimistic (1.03 p/m/s) and optimistic (3.23 p/m/s) values in order to assess this parameter impact on the evacuation time. Therefore, a maximum number of 5 persons in pessimistic and a maximum number of 16 persons in optimistic situation can pass through a 1-meter width door per time slot (5 seconds), whilst the capacity is proportional to door width.

## 5.7 Example of Application

Using the measures discussed in the previous section and previous chapter, we next describe an application of the model to evacuability assessment of Alan Turing, a building at l'Aquila University (Italy) normally used for exhibitions. By setting optimistic and pessimistic parameters (see Section 5.6), in this section, we run various simulations to assess the application of our model on: *i)* discovering the optimal evacuation time that results from crowd routing via ideal evacuation paths and compare it with the evacuation time that derives from static shortest path; *ii)* evaluating the evacuation time in a static risk situation (see Section 5.5); *iii)* providing guidelines in order to adapt the building architecture with a better safety condition.

The building consists of 29 rooms and 4 main corridors. Rooms sizes vary in a large range, and so the average time of a person to cross them from door to door is required. As explained in §5.4, we split each room in unit cells, each behaving as a (virtual) quasi-square room that can be traversed in a unit time slot. In practice, we embedded the building plan into a quasi-square grid as shown in Figure 5.3.

It is worth mentioning that, in this chapter we do not take advantage of unified square grids concept (see the previous chapter) and simply try to divide the physical space in quasi-square grids with the same approximated capacity. The next chapter investigates on the same case study and tests various unified square cells on the system performance.

The embedding results in a graph with 112 nodes (Figure 5.4) corresponding to the cells of Figure 5.3 and including node 0 as safe place. Adjacent cells are linked by 264 arcs which allow people to flow inside the building. All arcs are assumed bidirectional except the four towards the safe place. A time slot corresponds to the time required for crossing one cell: using average free flow speeds from the previous chapter and considering cell size, we obtained time slots of 5 seconds each, and therefore the monitoring frequency. All doors have 1-meter width, so with a similar pessimistic / optimistic capacity. As a rule of thumb, no more than 5 persons in pessimistic and no more than 16 persons in optimistic situation can pass through a 1-meter width door (or free space) per monitoring frequency. In all simulation scenarios, we computed the minimum time required to $N$ persons, randomly distributed in the building rooms, for reaching a safe place. The code for simulation was written on OPL language and solved on CPLEX version 12.8.0. We ran all the experiments on a Core i7 2.7GHz computer with 16Gb of RAM memory under Windows 10 pro 64-bits.

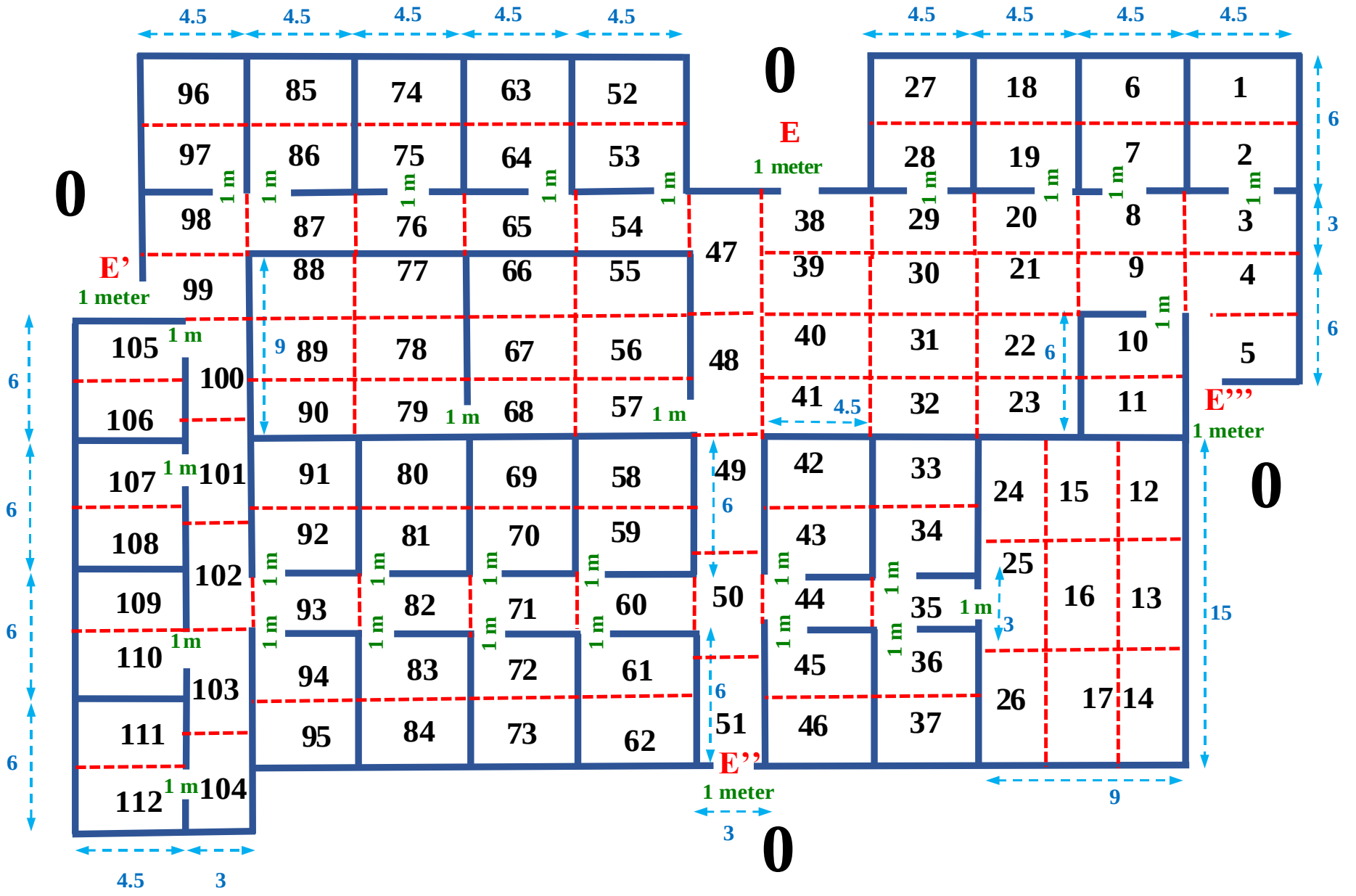

**Figure 5.3** Embedding of Alan Turing building architecture into a square grid.

**Ideal evacuation paths V.S. static shortest paths** In the first simulation, we suppose an initial occupancy of $N = 1008$ (based on real data), which relatively represents the area as highly crowded. We solved problem (4.1)-(4.8) for $\tau = 1, 2, \ldots$ until a solution of value $N$ is found.

**Pessimistic.** Table 5.1 reports the number of evacuees at each $\tau$ and the computation time (CPU) of each resolution step 2for pessimistic path capacity scenario. In terms of evacuation time, everyone has reached the safe place in 4 minute and 15”. As shown in the table, computations require 6.47 seconds (presolve included) in the worst case and are therefore totally compliant with real-time applications. This first simulation depicts an ideal situation in which flows autonomously choose the best among all the available routes in the building. Of course, managing such an ideal evacuation is not

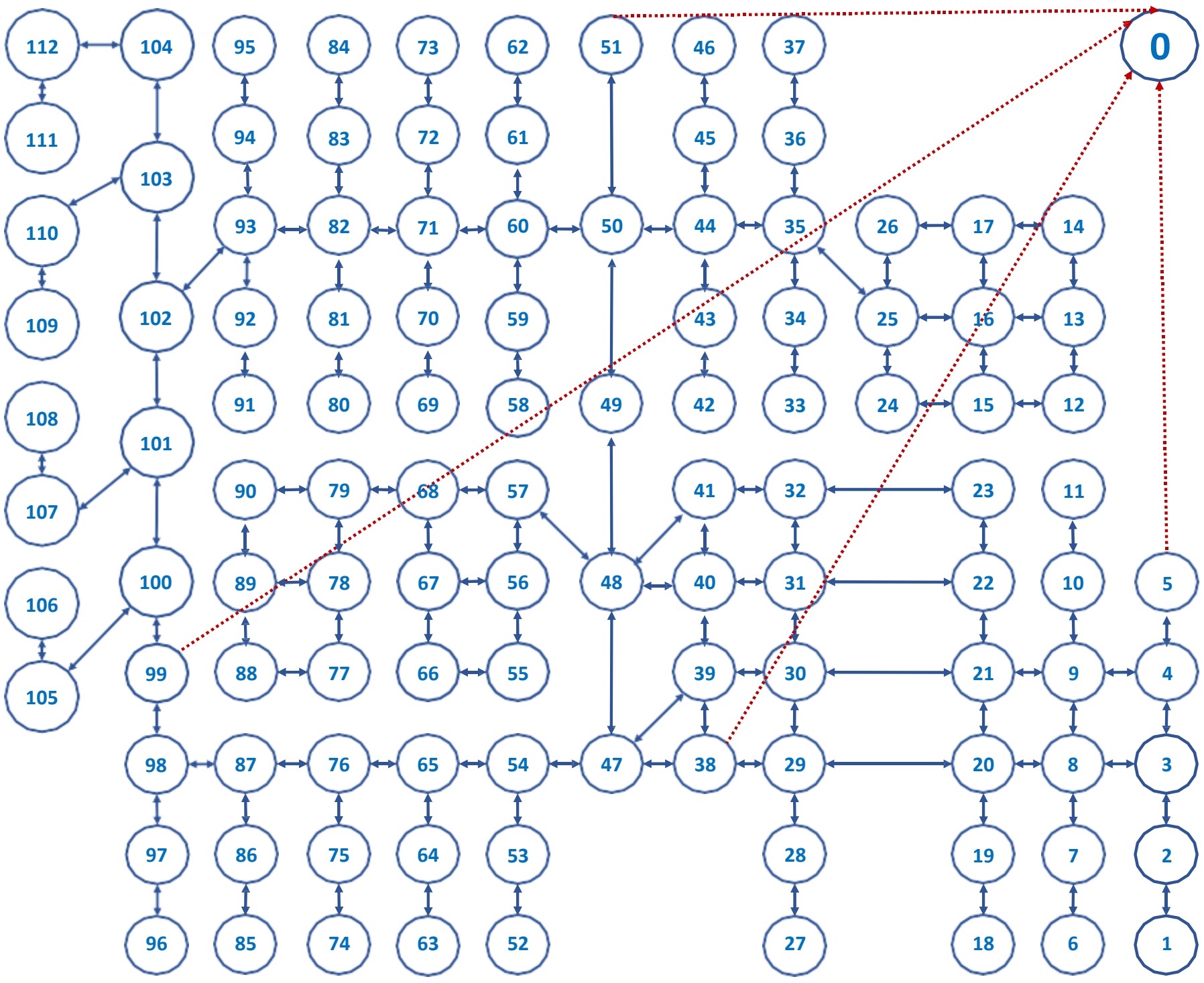

**Figure 5.4** Network associated with the plan of Figure 5.3.

**Table 5.1** Ideal evacuation paths - pessimistic path capacity scenario

| $\tau$ | evacuees | CPU Time | $\tau$ | evacuees | CPU Time |
|---|---|---|---|---|---|
| 1 | 20 | 0.28 sec | 27 | 540 | 2.61 sec |
| 2 | 40 | 0.31 sec | 28 | 560 | 2.77 sec |
| 3 | 60 | 0.44 sec | 29 | 580 | 2.84 sec |
| 4 | 80 | 0.53 sec | 30 | 600 | 2.96 sec |
| 5 | 100 | 0.47 sec | 31 | 620 | 3.10 sec |
| 6 | 120 | 0.58 sec | 32 | 640 | 3.53 sec |
| 7 | 140 | 0.60 sec | 33 | 660 | 3.32 sec |
| 8 | 160 | 0.61 sec | 34 | 680 | 3.54 sec |
| 9 | 180 | 0.71 sec | 35 | 700 | 3.91 sec |
| 10 | 200 | 0.76 sec | 36 | 720 | 3.42 sec |
| 11 | 220 | 0.83 sec | 37 | 740 | 4.14 sec |
| 12 | 240 | 0.88 sec | 38 | 760 | 4.16 sec |
| 13 | 260 | 1.01 sec | 39 | 780 | 4.17 sec |
| 14 | 280 | 1.09 sec | 40 | 800 | 4.19 sec |
| 15 | 300 | 1.12 sec | 41 | 820 | 4.30 sec |
| 16 | 320 | 1.44 sec | 42 | 840 | 5.13 sec |
| 17 | 340 | 1.28 sec | 43 | 860 | 5.07 sec |
| 18 | 360 | 1.33 sec | 44 | 880 | 5.12 sec |
| 19 | 380 | 1.57 sec | 45 | 900 | 5.27 sec |
| 20 | 400 | 1.61 sec | 46 | 920 | 5.36 sec |
| 21 | 420 | 1.73 sec | 47 | 940 | 5.49 sec |
| 22 | 440 | 1.88 sec | 48 | 960 | 6.01 sec |
| 23 | 460 | 2.02 sec | 49 | 980 | 6.35 sec |
| 24 | 480 | 2.08 sec | 50 | 1000 | 6.25 sec |
| 25 | 500 | 2.19 sec | 51 | 1008 | 6.47 sec |
| 26 | 520 | 2.35 sec | | | |

easy and perhaps unpractical. As a general practice, in fact, evacuation is conducted through pre-determined routes. Therefore, in a second instance, we consider the static emergency plan in which the prescribed evacuation routes are the shortest paths from any cell to the safe place. In this situation and for pessimistic scenario, evacuating 1008

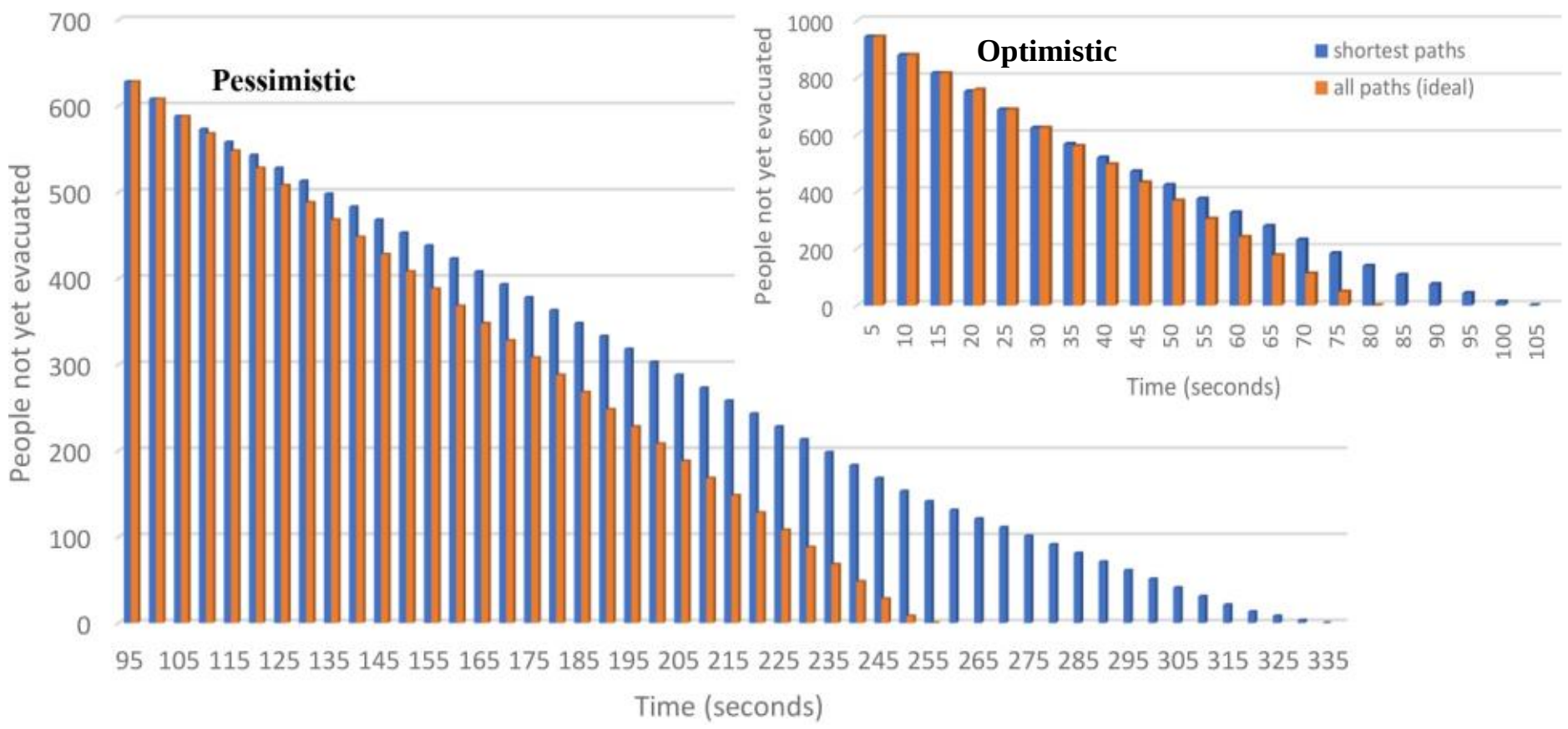

**Figure 5.5** Ideal evacuation and evacuation along shortest paths.

individuals takes of course more time: 5 minutes and 35 seconds.

**Optimistic.** In a second scenario, we repeated the simulation using optimistic parameters with a higher paths capacity. In this case everyone can reach the safe place after 16 time slots, i.e., 1 minutes and 20″. Also in this case, computation time is short, being always under 3 seconds including presolve. In a second instance, we again suppose that the prescribed evacuation routes are the shortest paths from any cell to the safe place. In this situation, evacuating 1008 individuals takes 1 minute and 45″.

By comparing two simulations that are run in each scenario, we observe that people flows plainly for some time (1 min and 45″ in pessimistic and 30″ in optimistic scenario). After that time, shortest routes start experiencing congestion, and evacuation is slowed down. The phenomenon is illustrated in the charts of Figure 5.5: as one can expect, the tail of people still in the building increases with initial occupancy.

**Risk consideration.** In another scenario, we assume that two emergency exits (of the four) are blocked due to a static emergency risk (such as earthquake). In this case, the evacuation time increases to 8 minutes and 25 seconds in pessimistic and to 2 minutes and 40 seconds in optimistic path capacity (Figure 5.6).

**Optimum dimensions (emergency exits)** In another scenario, considering the same occupancy ($N = 1008$), we performed continuous simulations by increasing/decreasing the emergency exits width, in order to observe its improvement or deterioration impact on the evacuation time.

Looking at the results shown in Figure 5.7, for both optimistic and pessimistic path capacities, the evacuation obviously takes longer by decreasing the emergency exits width. The interesting point is that, the evacuation time horizon highly slopes downward by making the exits wider, *but up to a certain dimension: 2.3 m for pessimistic and 2.2 m for optimistic condition*. For exits wider than these measurements, the evacuation time remains constant. The reason is that, having exits that are wide enough, the evacuation time will not be a function of congestion on exits, rather it will depend on the traveling time and congestion in corridors and internal doors. By performing continuous simulation for other areas, one can make the building architecture optimally evacuable and safe.

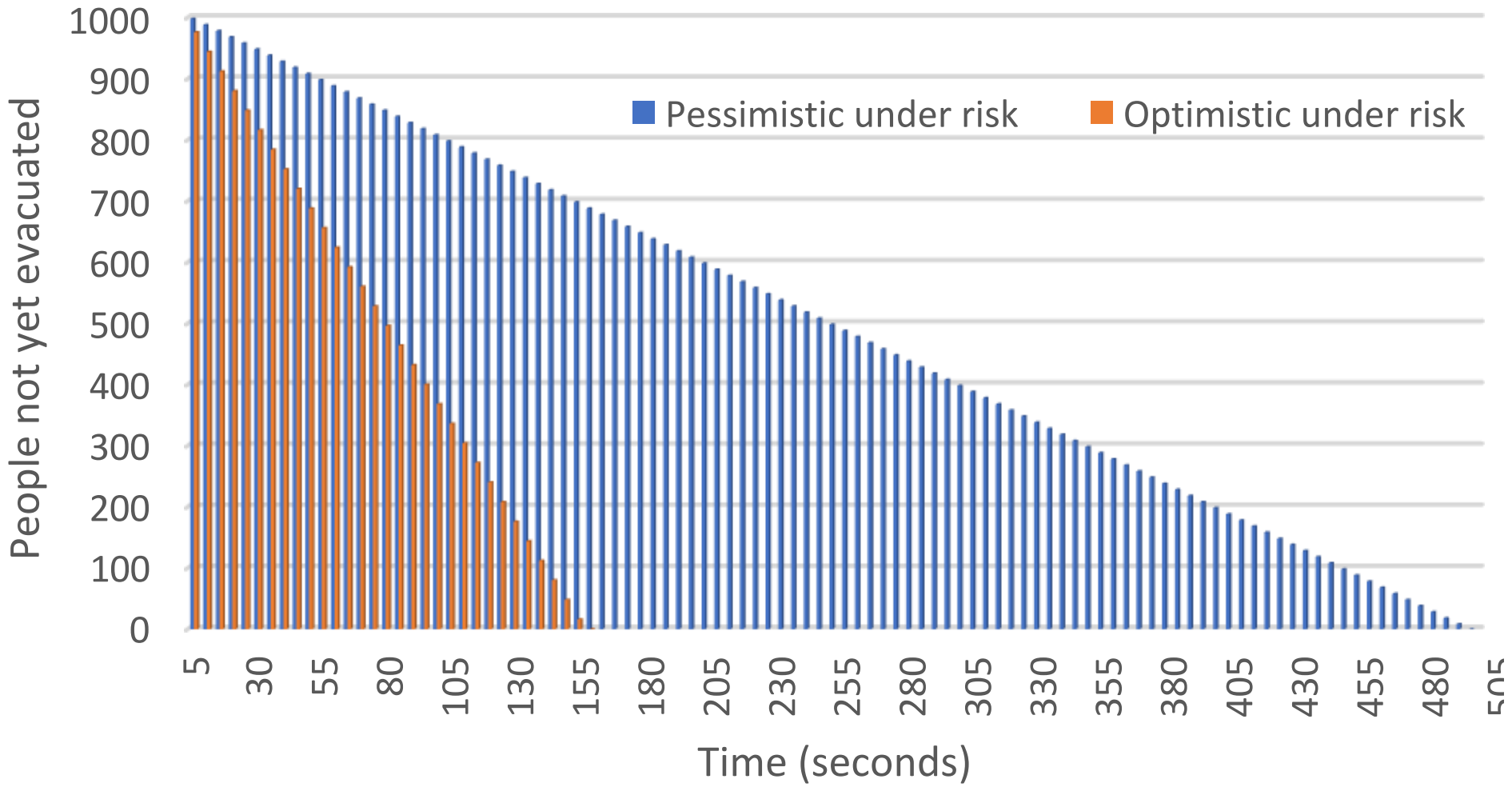

**Figure 5.6** Evacuation time under a static risk for pessimistic and optimistic paths capacity

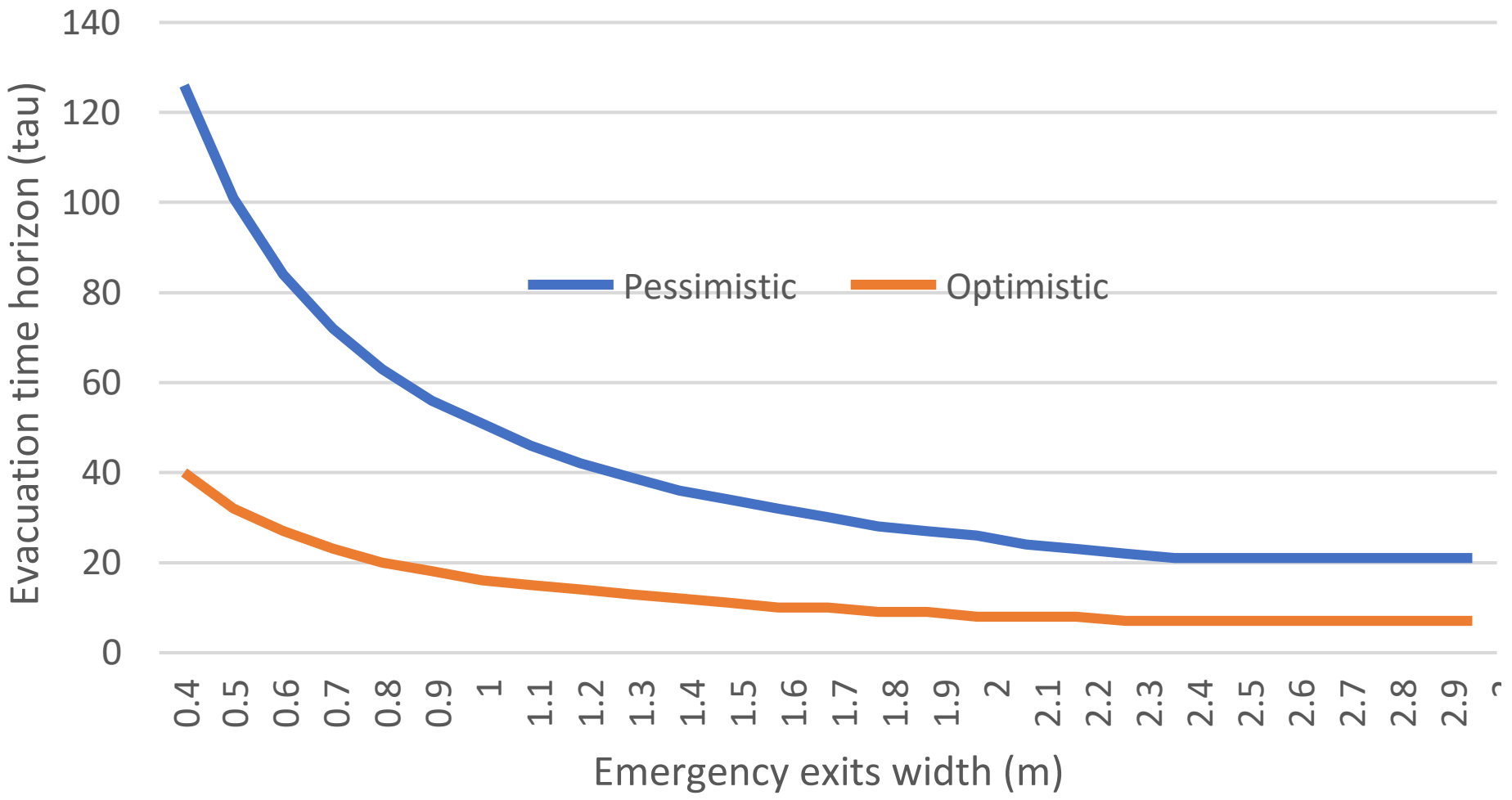

**Figure 5.7** Evacuation time variations w.r.t. emergency exits width

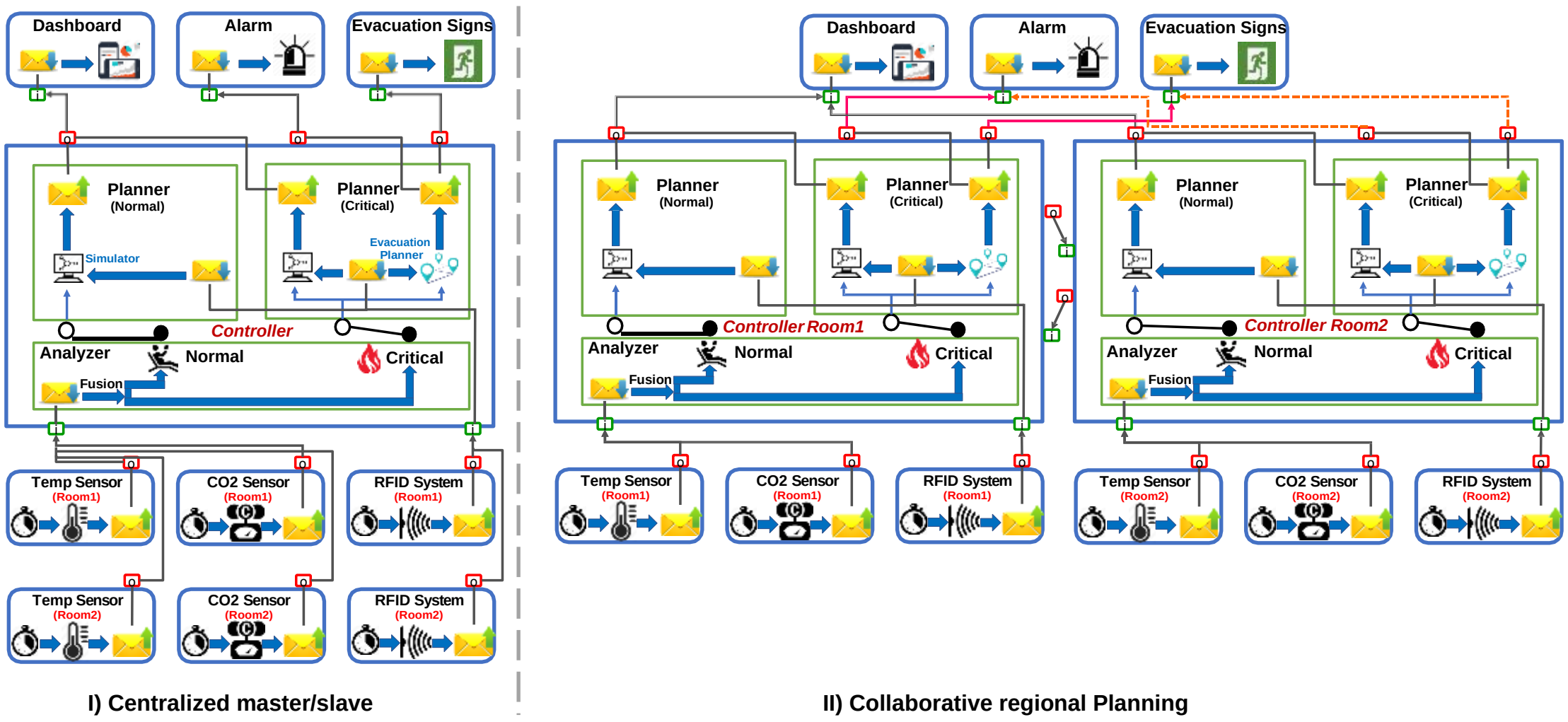

**Figure 5.8** Architectural patterns.

### Software Architecture of the Example of Application

From an architectural viewpoint, the system should provide a map of monitored area on the security agents dashboards. If a disaster is detected, an architectural adaptation will take place to perform the evacuation plan. Figure 5.8 shows self-adaptive software architectures corresponding to the patterns proposed in Section 5.3. In **centralized master/slave** pattern (Figure 5.8, left), the adaptation logic performs by a centralized master component that is responsible for the analysis and planning of adaptations and multiple slave components are responsible for monitoring and execution in the entire building. In **collaborative regional planning** pattern (Figure 5.8, right) a regional planner decides for each region. The subsystems provide their planner with necessary information and different planners interact with one another to coordinate adaptations that may span multiple regions.

Both architectures have two adaptation modes: Normal mode and Critical mode. *Normal mode:* in this mode the sensors read CO2 concentration and temperature in each area every 5 seconds. A timer is set in this mode to schedule the reading from the sensors. A message carrying each value is sent from the output message port of the sensor components to the in port of the central controller in master/slave pattern. In regional planning pattern, the values will be analyzed and planned locally in each area; however, the decision making will take place under coordination of other areas planners to support and approve any required execution. The main goal of this application, to be run on a tablet, is to show a 2D-representation of the monitored space providing also contextual data (sensed by RFID systems) on where the crowd is at any time, and how it moves in normal (and emergency) cases. If instead an emergency is detected, the state of the area will be adapted to critical mode.

*Critical mode:* in this mode, an adaptation will take place in monitoring level and sensors value will be read more frequently. In addition to show the map on dashboards, a message will be sent to acoustic alarm and evacuation sign actuators of each area to lead people to the safe places. In centralized master/slave pattern, the central controller handles the situation of whole area based on the network flow model. In collaborative regional planning, instead, the regional controllers in collaboration to each other handle the situation of risky areas based on the proposed algorithm. However, an architecture can qualitatively be better than another. For instance, in our previous work (76) we argued that the energy consumption in collaborative regional planning pattern is higher than centralized master/slave and it receives higher battery level drain due to increased number of exchanged messages.

**IoT infrastructure.** The IoT infrastructure, whose architecture is sketched in Figure 9, consists of various elements, such as sensing, computation and actuation. For the purpose of this example, we used RFID tags and readers, Beacons, CCTV cameras, and people counters to track crowd movements in the real environment. Each abovementioned device operates according to its own particular principles and constraints, i.e. RFID technologies and Beacons require to equip pedestrians with special RFID or BT tags, while the use of CCTV cameras and people counters is not constrained to any additional device. Furthermore, mobile phones are becoming increasingly powerful equipped by a set of embedded sensors to be used to detect a disaster or track people in indoor/outdoor areas. In this line, Some tools such as *Mission Track* let inform and control the localization of people in a risky situation. The system actually facilitates the notification of an incident detected by any observer through a mobile app. In computation phase, some simple analytics (such as crowdedness detection) will typically be done on the edge, then more detailed analysis (such as routing planning) can be performed on the cloud. In actuation step, our *evacuation guidance system*, interacts to people via an optical arrow that shows the direction to follow on the wall. This solution has a competitive performance w.r.t intelligent carpet, whilst with a lower price and a slight risk of exposure to disaster. Among other advantages, the system can be directly connected to the controller to perform our desired logic, while other evacuation guidance systems on the market predisposed their own system not providing API for manipulation. The actuation can be performed using users' mobile phones as well. We currently design a mobile application for cultural heritage in which, in case of emergency, the position of people are detected from their device Bluetooth and the evacuation is shown to individuals on their smart phone.

Nevertheless, due to legal barriers towards carrying out a real evacuation with our novel tool, we used our IoT infrastructure to monitor the movement of people in the considered physical space to feed the simulation with some gathered data, such as maximum simultaneous presence of crowd. In this line, the initial occupancy that is considered for our simulation scenarios comes from an experiment performed at University of L'Aquila during an exhibition on 15 January 2018, when the simultaneous presence of 1008 people in Alan Turing building was recorded as peak value.

## 5.8 Conclusion

This work uses the network flow model proposed in previous chapter for supporting the rapid evacuation of people from a building in case of emergency, as well as providing safety measurements for complex buildings architectural design. The chapter evaluates various optimistic and pessimistic scenarios useful for both design-time and real-time applications.

The following chapter evaluates the performance of the same IoT software system that is associated with the model setting and design decisions.

# Chapter 6

# Real-time Emergency Response through Performant IoT Architectures

For safety-critical systems, such as evacuation, real-time performance and evacuation time are critical. The approach aims to minimize computational and evacuation delays and uses Queuing Network (QN) models. The approach was tested, by computer simulation, on a real exhibition venue in Alan Turing Building, Italy, that has 34 sets of IoT sensors and actuators. Experiments were performed that tested the effect of segmenting the physical space into different sized virtual cubes. Experiments were also conducted concerning the distribution of the software architecture. Section 6.1 gives an overview on QN and software architectures. Section 6.2 describes the design of a reference IoT architecture and its corresponding QN for emergency handling. The application of our approach is presented in Section 6.3.

## 6.1 Overview

There is a large body of previous work in the three topics that shape our research: IoT software architecture, queuing networks, and optimization algorithms. However, their application to the emergency management has been rarely explored.

### 6.1.1 Related Work

QNs have been widely and successfully applied to the hw/sw performance assessment domain (30; 84) and several implementations have been developed by providing editors and analysis environments with QN models. Many existing approaches use QNs as first-class entities for performance analysis (9; 103; 4; 10). Despite the wide adoption in the performance domain, QNs have started to be exploited for non-functional assessment in the context of IoT systems only in recent years.

El Kafhali et al (34) proposed an analytic model for a fog/cloud-based Medical IoT system showing how to reduce the cost of computing resources while guaranteeing performance constraints. They used the QN concept to predict the system response time and estimate the minimum required number of PandS resources to meet the service level agreement. However, they do not provide any kind of high or low level architectural model. Huang et al (49) propose a theoretical approach of performance evaluation for IoT services, which provides a mathematical prediction on performance metrics during design before system implementation. The authors formulate an atomic service by a queuing system in order to model IoT systems by a queuing network and obtain performance metrics. Whilst using QN, this paper does not address any modeling based on software architecture to be assessed by performance indices. Whilst few related works have been found on IoT systems modeling with QNs, we did not find any previous work on modeling emergency evacuation systems by QNs.

### 6.1.2 IoT software architectures

IoT architectures are generally composed of three main layers Muc-ecsa namely *Perception, Processing and Storage (PandS)*, and *Application*:

- The Perception layer represents the IoT physical sensors that collect information. For emergency management, this layer hosts a large number of different types of sensors, e.g. temperature, smoke and movement detectors.
- The Application layer determines the class of services provided by the IoT system. For emergency management, this layer hosts a large number of different types of actuators, e.g. dashboards, evacuation signs and alarms.
- The PandS layer is the central entity of an IoT system that stores and analyses data gathered by the perception components to be accessed by other entities for their applications. Based on the PandS design philosophy, this layer can be divided into various sub-layers to set up the *IoT patterns* as follows:

**Centralized**. In a centralized pattern, data coming from the perception layer are processed by a central component that makes decisions on actuation. This central component can either be a local controller or, for massive PandS requirements, the cloud. Based on this pattern, if a device wishes to use an IoT service, it must connect to the central PandS component. A centralized architecture simplifies things through a central implementation of analysis and planning algorithms.

**Collaborative**. In this pattern, data are processed and stored separately (locally and/or remotely) but with the potential collaboration of other local/remote PandS components of the IoT system. In this pattern, a network of local intelligent components can communicate in order to form and empower IoT services. The advantage is that, should a local PandS component fail, a service would still be provided.

Given the above, we designed a set of QNs for IoT architectures that are described in following sections.

### 6.1.3 Queuing networks

We use Queuing Network (QN) to model our software architectures (which has an algorithmic core) and adapt it to reduce the delay to a minimum. Thus, in order to estimate performance indicators and avoid the performance degradation issues associated with IoT architectural patterns, we rely on QN models. QNs have emerged as powerful instruments to model and estimate the performance of hardware and software systems. They ground on theoretical foundations based on an algebraic approach to computer system modelling proposed by Lazowska in 1984 (60), where the computer system is represented as a network of delay and/or queuing *stations* (i.e. topology of the QN). Different *classes of jobs* may flow through the QN, each representing different types of user requests (i.e. dynamics within the QN). While flowing through the QN, each task requires a certain amount of service, namely *service demand* (mentioned as *CPU time* in this chapter) to each visited station, depending on the job class the task belongs to. It is worth mentioning that, service demands represent input parameters that must be specified during QN design. Beside service demands, *workload intensities* must be specified, that is the rate at which tasks of each job class enter the QN. For example, a request (of a certain job class) every 2.5 seconds.

Once a QN has been designed, it can be solved analytically or by simulation, carrying out performance indices of interest such as system/stations response time and throughput for both the overall system and single classes of jobs.

### 6.1.4 Algorithm

In order to avoid repetition, we removed the algorithm explanations that are already discussed in section 4.2.

## 6.2 Designing Performant Architectures for Emergency Handling

### 6.2.1 Software Architecture for Emergency

Figure 6.1 shows an example of an IoT-based environment for emergency response: CCTV cameras detect peoples position and movement that is used to feed the algorithm running in a PandS component. The algorithm decides on the actuation set based on the situation. In normal situations, the system shows, on a tablet, a 2D-representation of the monitored space and shows where crowds are located and how they move at any time. In this mode, the optimal flow algorithm is periodically run to estimate the minimum evacuation time required under current conditions. This value can be used to regulate visitor access to a venue in order to comply with safety conditions. If an emergency happens, in addition to the tablet map, alarm actuators are activated and evacuation signs in each area show the best evacuation routes based on the network model described above.

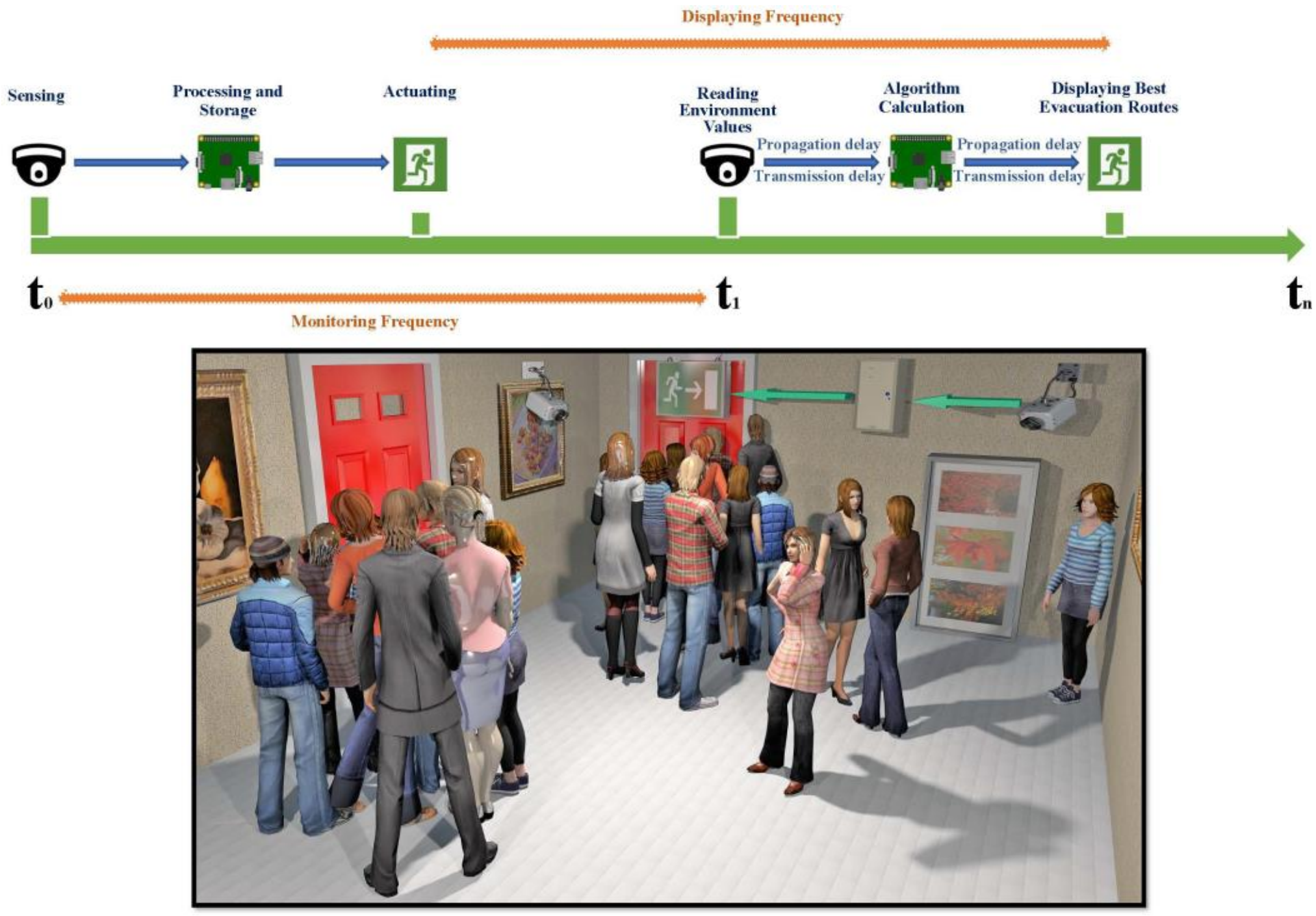

**Figure 6.1** IoT Infrastructure for Emergency Handling.

Figure 6.2 shows the corresponding software architecture. As depicted, additional sets of sensors can be embedded for emergency detection to further enable controllers to decide about normal or critical mode and activate a special set of actuators. As shown in the upper part of Figure 6.1, in addition to the computational delay of the PandS component, the sensors take some time to detect peoples position, transmit these data, and display the best evacuation routes. Reducing these delays to a minimum improves the system's functionality: since people can follow the given instructions more quickly and more individuals will be in a better evacuation position at the next monitoring time-spot. It is worth mentioning that reducing the aforementioned delays is a function of software architectural patterns, to be improved by properly relating the IoT components to one another. The following section presents a Queuing Network (QN) method that is designed on top of the software architecture, so as to facilitate assessing the performance of our IoT-based emergency handling system.

### 6.2.2 Queuing Networks for Emergency

Fig. 6.3 shows a QN representing the performance model that we will exploit later in the chapter for our case study. The QN conforms to the architectural patterns of Fig.

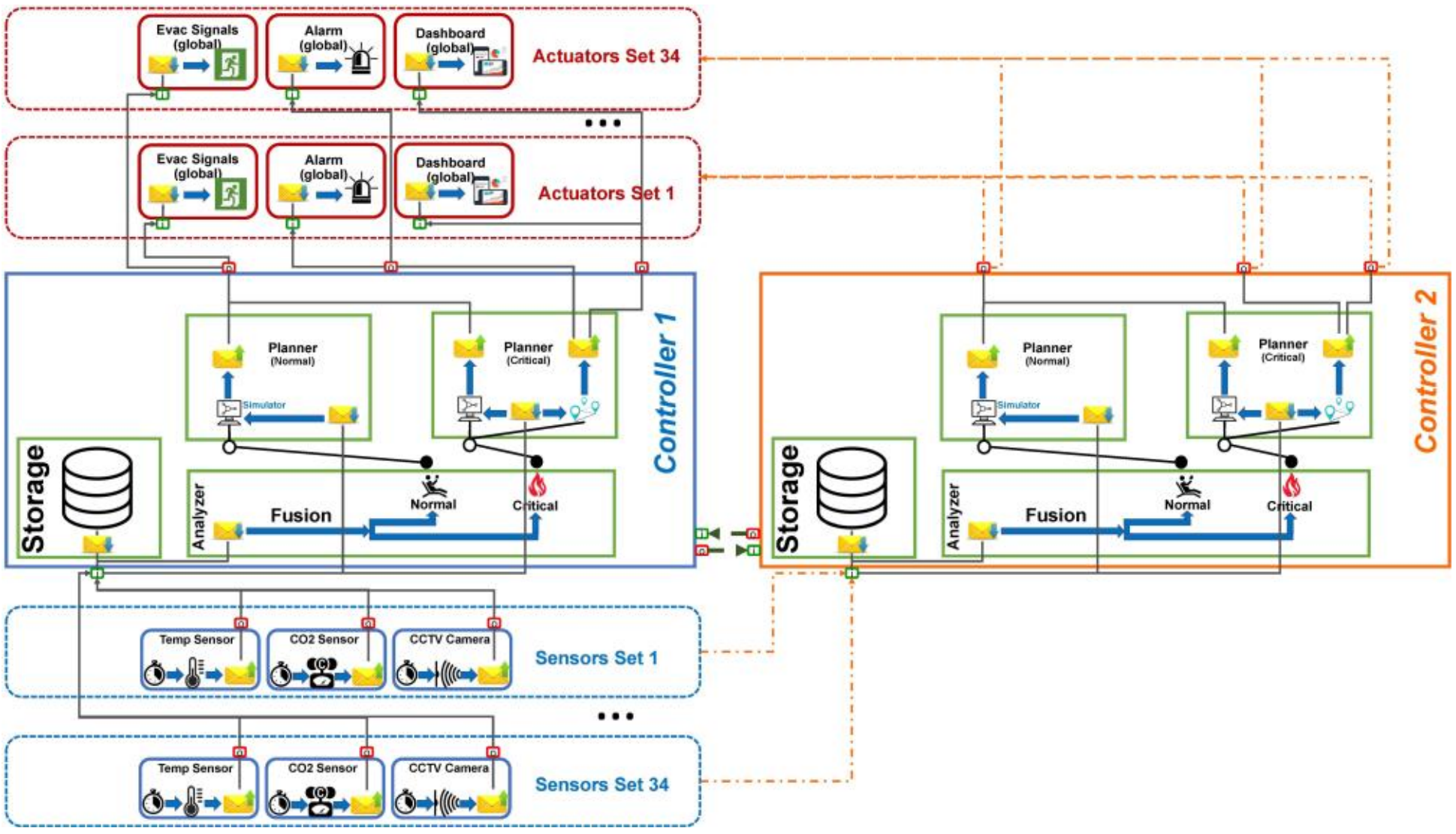

**Figure 6.2** Architectural Patterns. Only Controller 1 Active: Centralized - Both Controllers Active: Collaborative.

6.2, in fact, from a topological perspective:

- *CCTVs* corresponds to a specific kind of sensors, i.e. cameras.
- *Dashboard*, *Alarm* and *EvacuationSigns*, correspond to the three kinds of actuators.
- *CentralController* and *PeerController* correspond to the two PandS components.
- *PL2IL* and *IL2AL* represent networks between sensors and controllers and between controllers and actuators, respectively.
- The two QN constructs on the left and right of the figure, namely *Sampling* and *Done* represent the entry and exit points of the QN, respectively. In particular, *Sampling* indicates the point where data sampled by sensors is generated and *Done* represents the point where actuation ends.

Concerning dynamics, we devise the following control flow within IoT architectures: *i)* Data are sampled by sensors and forwarded through a network to controller(s); *ii)* A control layer aims at achieving the goal of evacuating people during emergency, through an actuation plan that is forwarded to actuators; *iii)* The actuation plan is implemented by actuators, thus possibly achieving the common goal.

The above control flow can be translated into QN language, by identifying a minimal set of 6 different kinds of tasks within a QN for IoT system performance (namely *SMAPEA* loop), sequentially executed as follows: *1.) Sense*: Raw data retrieval. *2.)Monitor*: Raw data aggregation and refinement for analysis at controller level. *3.)Analyze*: Interpretation of monitored data. *4.)Plan*: Building an actuation strategy. *5.)Execute*:

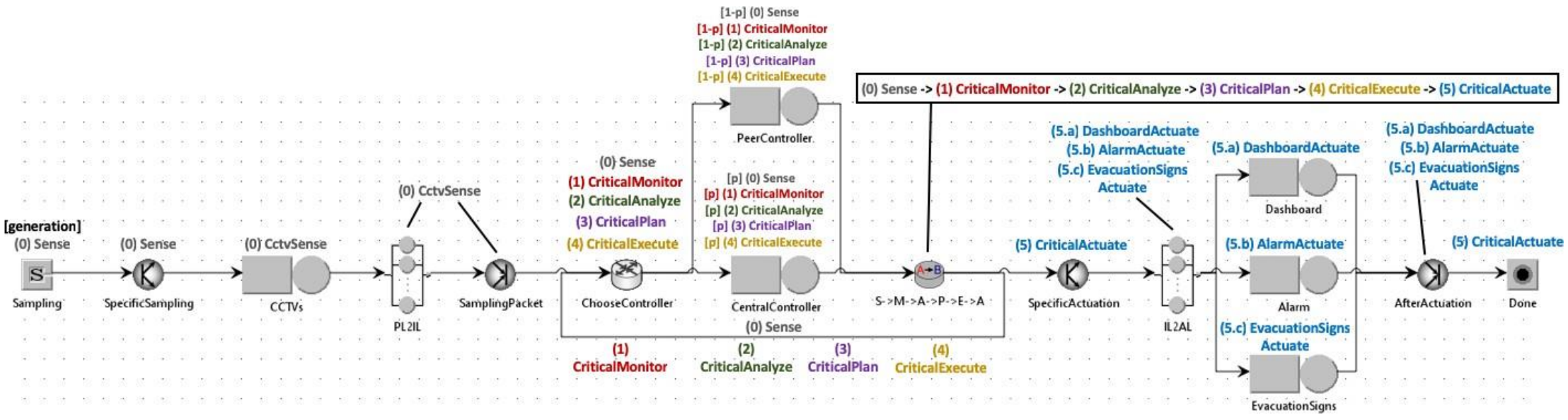

**Figure 6.3** SMAPEA Queuing Network for the Alan Turing building case study.

Pre-processing the actuation strategy towards actuation. *6.)Actuate*: Practically undertaking the actuation by implementing the planned execution.

In order to implement the SMAPEA loop, a *class-switch* that transforms each SMAPEA task to a subsequent task is introduced (see the element of Fig. 6.3). Moreover, routing probabilities for the switch must be properly defined. In particular, right after the switch, any SMAPEA task type is routed back to controllers through the *ChooseController* router (except *Actuate* tasks that are routed to *IL2AL* for actuation). By exploiting a probability-based routing strategy for *ChooseController* both Centralized and Collaborative patterns can be implemented, in fact the former may route any task to *CentralController*, whilst the latter may route tasks to *CentralController* or *PeerController* with 50% probability.

Notice that, in the QN of Fig. 6.3, fork/join nodes have been introduced, namely *SpecificSampling* (fork), *SamplingPacket* (join), *SpecificActuation* (fork), *AfterActuation* (join), aimed at modelling the fact that *Sense* and *Actuate* tasks involve the specific sensor and actuator sets, respectively. For example, each *CriticalActuate* task is split into three new tasks, namely *DashboardActuate*, *AlarmActuate* and *EvacuationSignsActuate*, because in case of emergency *i)* the best evacuation paths must be displayed on dashboard, *ii)* an acoustic alarm must be triggered and *iii)* evacuation signs must be properly turned on.

### 6.2.3 Algorithm Settings for Emergency

**Cell Size Setting and its Impact on Architecture**. We use the same cell size setting method as chapter 4. The cell size has an obvious effect on the resulting spatial patterns, and consequently on both the computational efficiency and model accuracy: the larger the cell, the fewer vertices in $G$ and the lower the refresh frequency at which people's positions are updated. Given the speed at which people move and the data that are acquired, relatively low refresh frequencies are not an issue. Instead, partitioning each room into identical cells may result in a huge network with consequently high CPU time. This issue has a direct impact on software architectural patterns since an operation which requires a huge amount of processing time on a low capacity machine is not suitable for real-time applications. Therefore it should be processed on a more powerful (potentially remote) PandS component.

In general, one can approximate the diverse room shapes by $a \times b$ rectangles as large as possible, while still minimizing any consequent error. Various ways can be adopted to measure approximation error: the most natural is the difference between real and approximated room area, in which case, for room $k$ of size $p_k \times q_k$, the error is given by

$$e_k(a, b) = q_k[p_k \bmod(a)] + p_k[q_k \bmod(b)] - [p_k \bmod(a)][(q_k \bmod(b)]$$

As we need isometric cells and look for a uniform approximation, we set $a = b$ and find $a$ minimizing $\max_k\{e_k(a, a)\}$, meanwhile limiting the total number of approximating cells to some predefined $m$: hence we choose among the values of $a$ that fulfill $\sum_k p_k/a \;\; q_k/a \leq m$. A brief description of the method implementation is presented in the Application section.

**Time and Space Decomposition Setting and its Impact on Architecture**. In order to run the algorithm in distributed PandS components (collaborative pattern), we assessed the space- and time- decomposition feasibility. In the former case, we can give, for instance, two controllers the responsibility of two distinct areas, and let them share border information. Since the global objective cannot generally be satisfied by summing up two local objectives (that is running the algorithm in two local controllers instead of a central one), this method mostly leads to a non-optimal solution obtained by "gluing" together the two distinct areas. In the latter case, we conjectured that the optimal flow obtained at time $t$ can be extended to $t + 1$ and stay optimal. This means that, for instance, one PandS element is in charge of solving the algorithm for $t_{even}$, and for $t_{odd}$ (Figure 6.4) whilst sharing a level of data. Assessing this decomposition and taking into account our main requirement of optimally is described in the Application section. Here we can observe the ability of the algorithm to be run in a distributed way or centralized.

## 6.3 Application

We use the same case study as many other chapters of this thesis. Our proposed model has been applied to the evacuation of the Alan Turing building, in Italy, which is sometimes used for exhibitions. The considered building consists of 29 rooms, 4 main corridors and 34 sets of IoT sensors and actuators (Figure 6.4). In order to investigate our approach we address four research questions. The first two questions are centered around the algorithm and its levels of granularity and distribution:

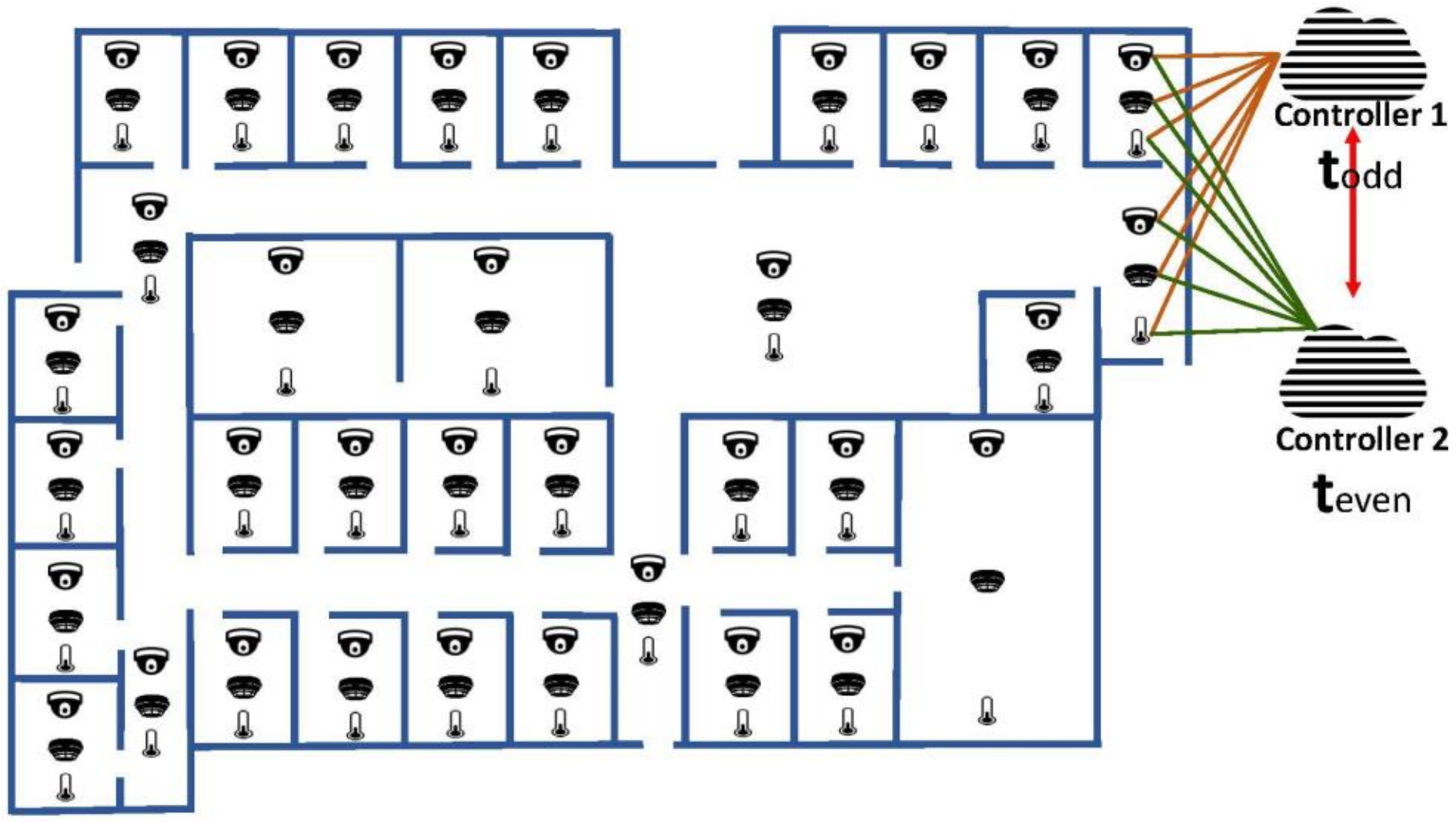

**Figure 6.4** Alan Turing Building IoT Infrastructure.

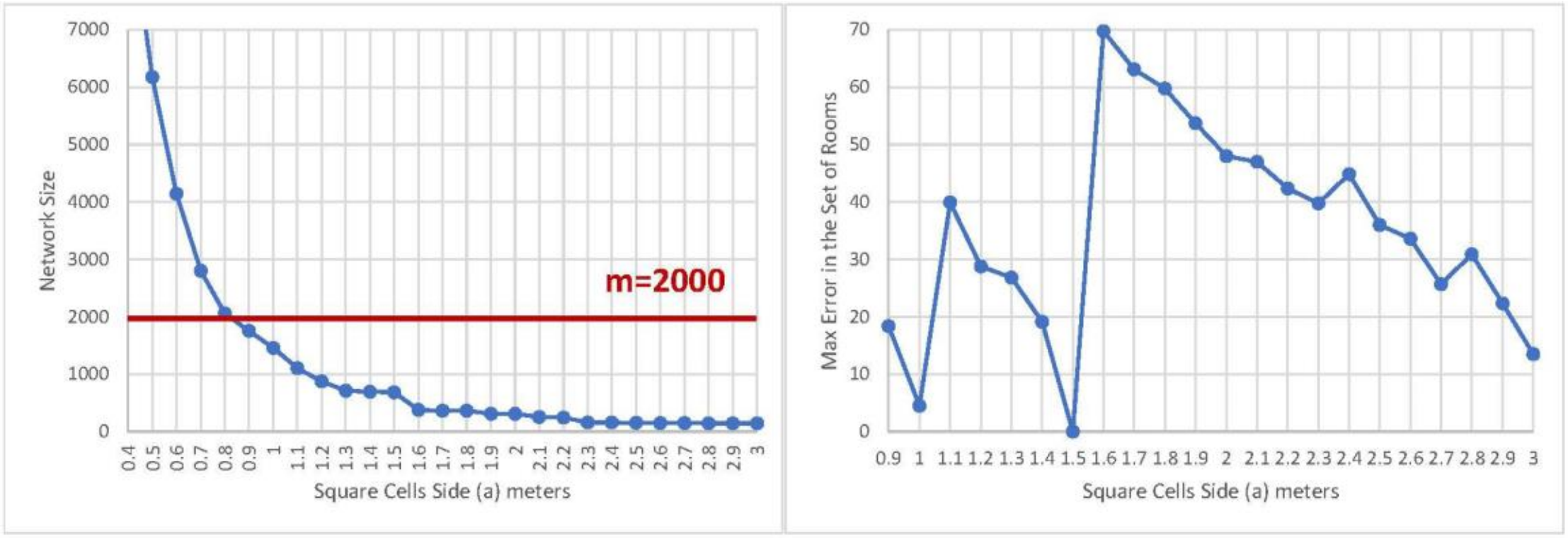

**Figure 6.5** Optimal cell size: maximum error (right) and network size as a function of cell size (left).

- **RQ1**: what are the best cells sizes to divide the building surface, and how does the size affect the evacuation times and computational delays?
- **RQ2**: does running the algorithm in a time-decomposed way increase or decrease the computational and operational delays? What are the designed software architectures corresponding to these results?

The next two research questions focus on software system delays (using Queuing Networks):

- **RQ3**: what level of delay is associated with each QN model?
- **RQ4**: which software architectural design decisions facilitate real-time applications?

### 6.3.1 Answers to RQ1: cellular approximation of physical space

We split each room in unit cells, each behaving as a (virtual) square room that can be traversed in a unit time slot. In practice, we embedded the building plan into a square grid as shown in Figures 6.6 and 6.8. To decide the cell size, we look at both the error introduced by room approximation and the number of nodes in the resulting graph $G$. The latter is in an inverse proportion of cell size (left diagram in Figure 6.5); the former varies irregularly with cell size (right diagram of Figure 6.5). We considered square cells up to $3 \times 3$ meters (the short edge of the smallest room) and allowed no more than 2000 nodes of $G$; then we selected the size that minimizes the largest error for all rooms. The reason of considering such a big maximum network size is to assess the impact of increasing the number of nodes on CPU time and consequently, on the software architectural pattern. As shown in Figure 6.5, $1.5 \times 1.5$ cells (Figure 6.8) give the best approximation (no error) and involve 687 graph nodes (Figure 6.9). With $3 \times 3$ cells (Figure 6.6) the error rises to 13.5 but $G$ contains only 144 nodes (Figure 6.7). Summarizing, $3 \times 3$ leads to larger error but less CPU time; conversely, $1.5 \times 1.5$ causes larger CPU time but no error. We tested scenarios with both cell sizes in order to find the best efficiency/accuracy compromise.

**Simulation**. Simulations were first run for both cell sizes. The simulation code was written in the OPL language and problems were solved by CPLEX version 12.8.0. All experiments were run on a Core i7 2.7GHz computer with 16Gb of RAM memory under Windows 10 pro 64-bits. In all tests, we computed the minimum time required for 264 persons, randomly distributed in the building rooms, to reach a safe place. This datum comes from an experiment performed in the Alan Turing during the *the Researchers Night* event, when the IoT system recorded the simultaneous presence of 264 people in the building as a peak value. We solved problem (1-4) for $\tau = 1, 2, \ldots$ until a solution of value 264 was found.

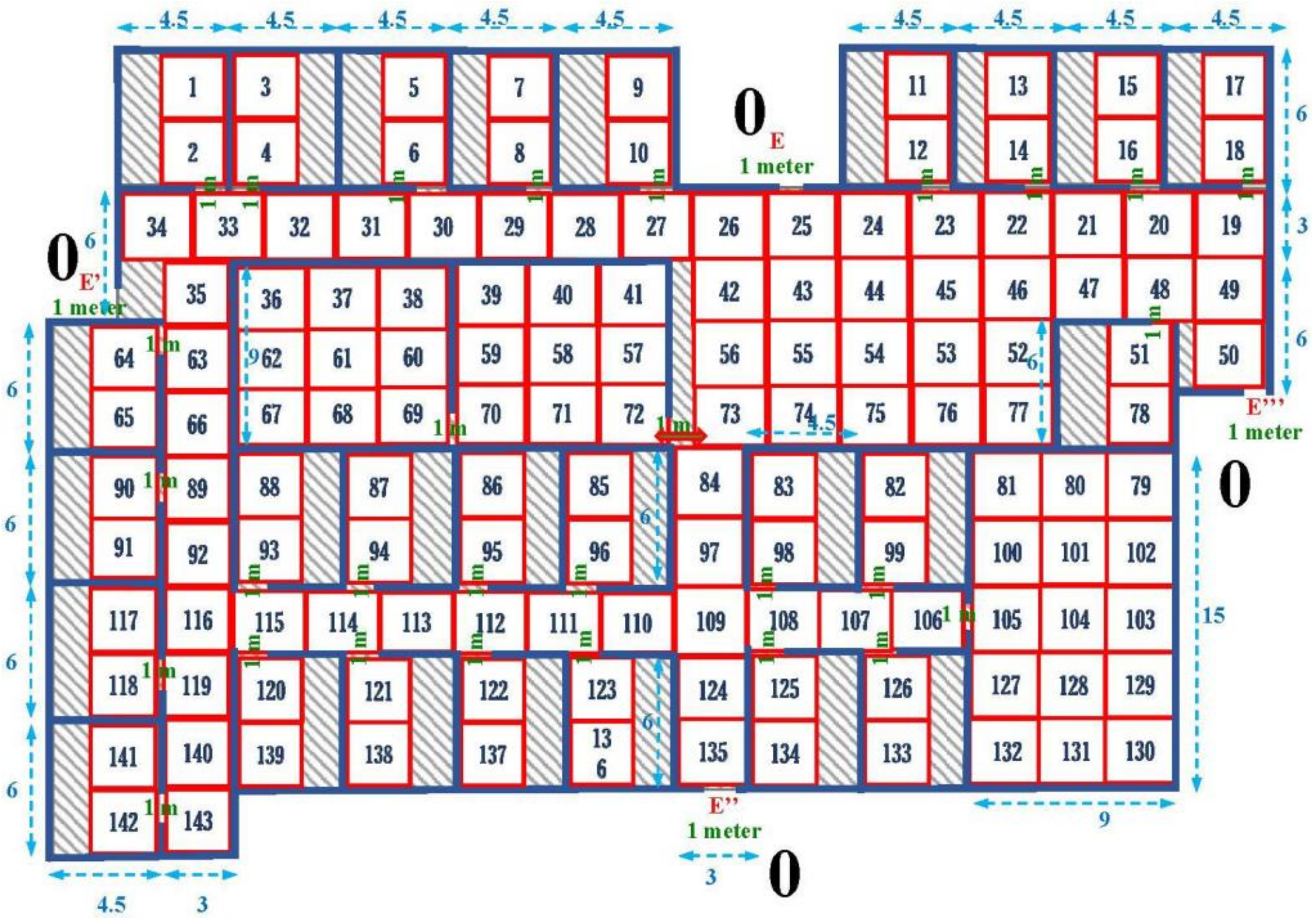

**Figure 6.6** Plan embedding the Alan Turing building into square grids with a low resolution: $3 \times 3$ cells. The area that is not covered by cells (error) is shown in gray.

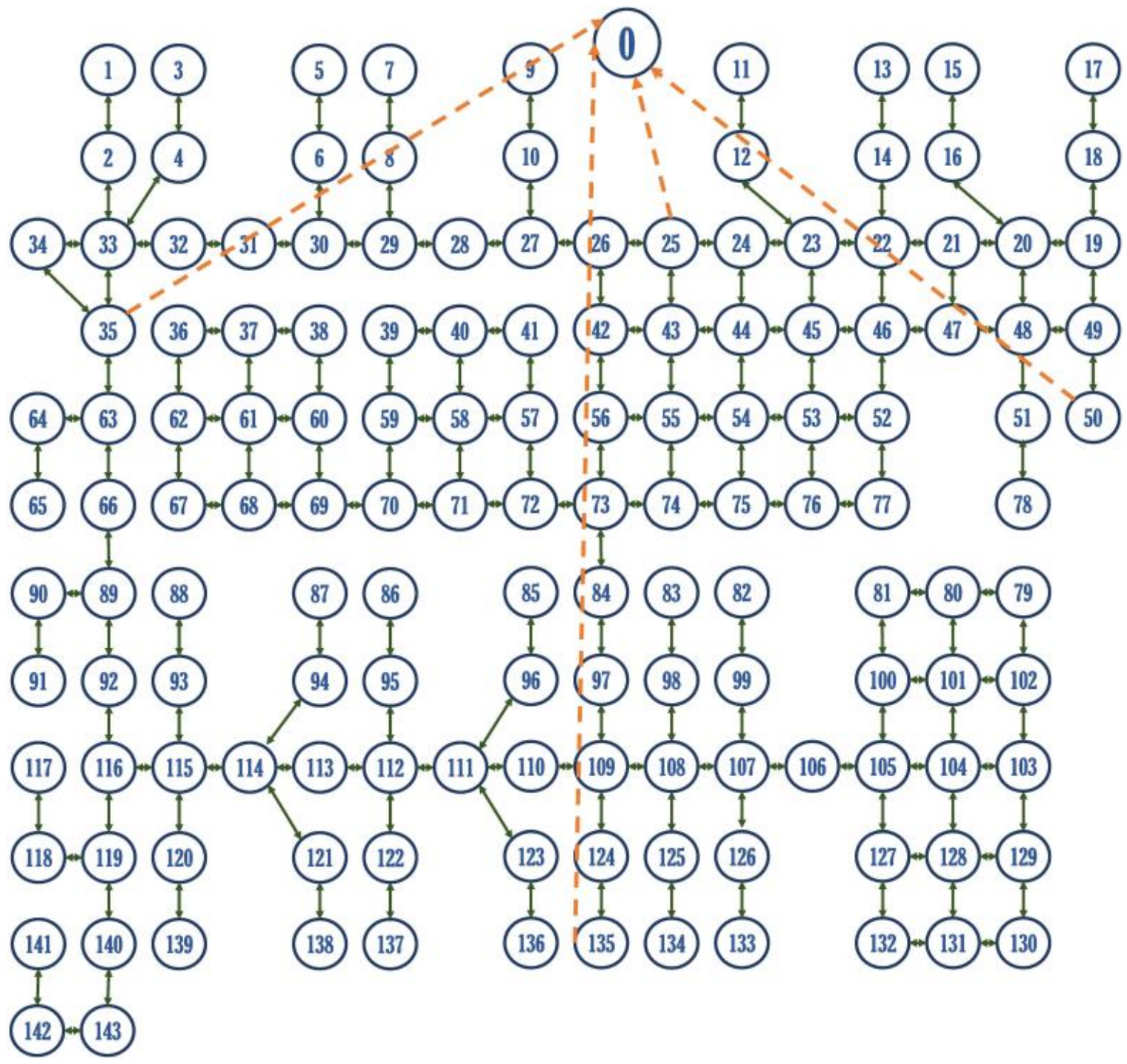

**Figure 6.7** Network associated with the plan of Figure 6.6.

To get a reliable model, some more parameters such as walking velocity under various conditions, door entrance capacities and room capacities must be set to numbers that reflect reality. We set these model parameters based on a literature review (Table 6.1).

**Table 6.1** Evacuation Model Parameters

| Model Parameter | Assigned Value |
|---|---|
| Walking Velocity | 1.2 $m/s$ (Ye et al. 2008) |
| Door Capacity | 1.2 $p/m/s$ (Daamen and Hoogendoorn 2012) |
| Cell Capacity | 1.25 $p/m^2$ (Uk Fire Safety) |

Table 6.2 reports the number of evacuees at each $\tau$ and the computation time of each solution step. Computation is done for both *low*- and *high-resolution* networks (respectively, $3 \times 3$ and $1.5 \times 1.5$ cells). With the low-resolution network, we get the evacuation and CPU times shown in Table 6.2 left: in terms of evacuation, everyone has reached a safe place in 55 seconds; on the other hand, computation requires 2.33 seconds in the worst case, and is therefore totally compliant with real-time applications. Evacuation and CPU times for the high-resolution network are reported in Table 6.2 right. We see that everyone has reached a safe place in 98'75". CPU time is now much larger (382.24 seconds in the worst case) and the model appears inappropriate for real-time use, unless additional computational resources are deployed. Hence we can conclude that sufficient accuracy is obtained using the low resolution network.

### 6.3.2 Answers to RQ2: Time Decomposition

Table 6.3 gives the number of evacuees at each $\tau$ and the computation time of each solution step corresponding to time-decomposed networks (collaborative PandS). With a time-decomposed simulation of $3 \times 3$ cells, we obtain the evacuation and CPU times in Table 6.3 left: in terms of evacuation, everyone has reached a safe place in 75 seconds, however computation requires 0.58 seconds in the worst case, and is therefore compliant with real-time applications. Compared with continuous simulation (central PandS) of the same case that is presented in Table 6.2 left, whilst CPU time is now significantly reduced, the optimal evacuation time is increased by 36 percent. Looking at the time-decomposed simulation results for $1.5 \times 1.5$ cell (Table 6.3 right), CPU time is decreased by at least 123 percent (being 3.11 seconds in the worst case) and the evacuation time remained constant.

Hence, for a high resolution time-decomposed network: whilst the operational delay remains constant, the computational delay improves significantly. However total response time should be observed since the quicker sampling rate tied up with high resolution may cause a negative impact. For a low resolution time-decomposed network:

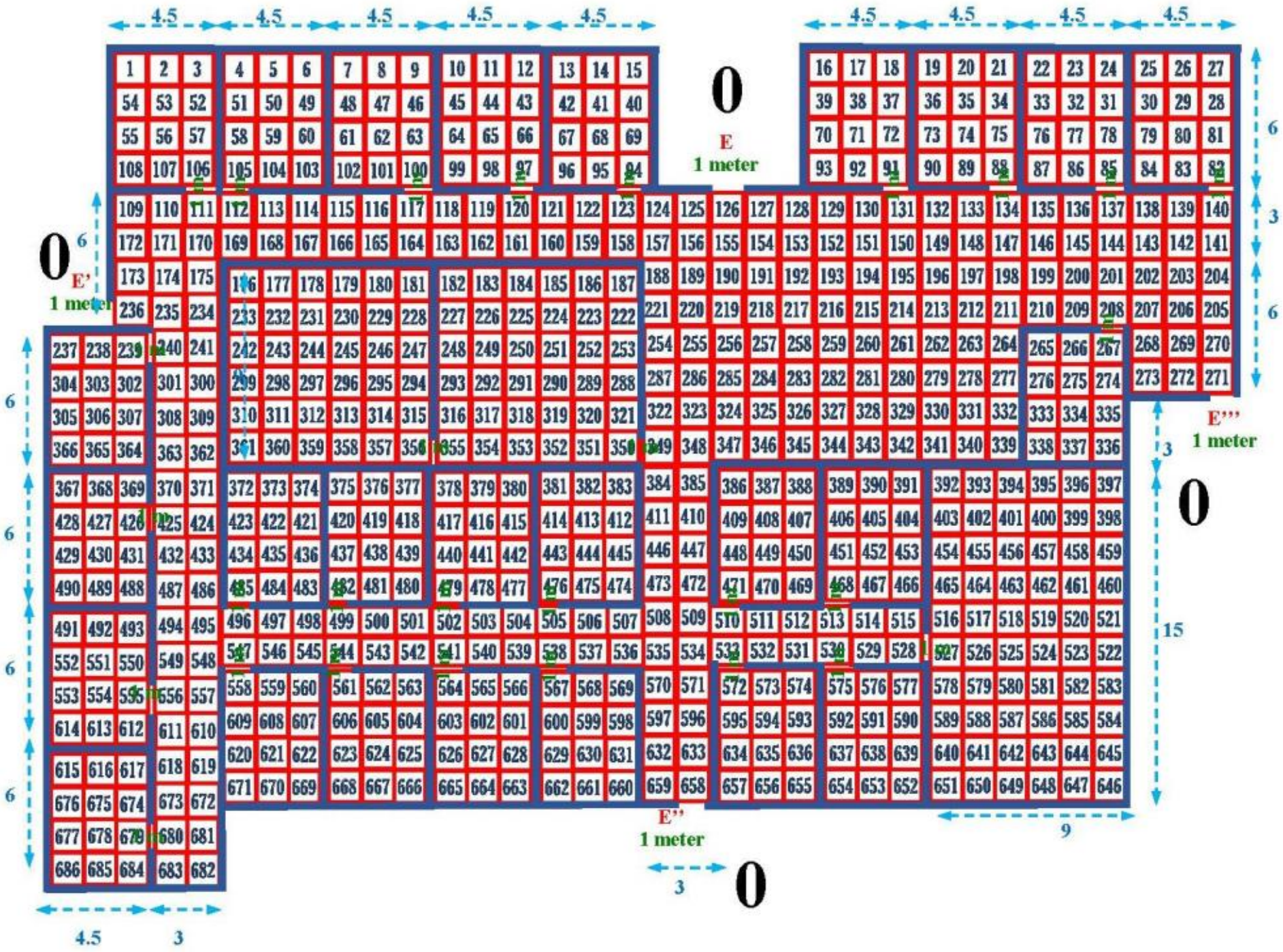

**Figure 6.8** Plan embedding the Alan Turing building into square grids with a high resolution: $1.5 \times 1.5$ cells.

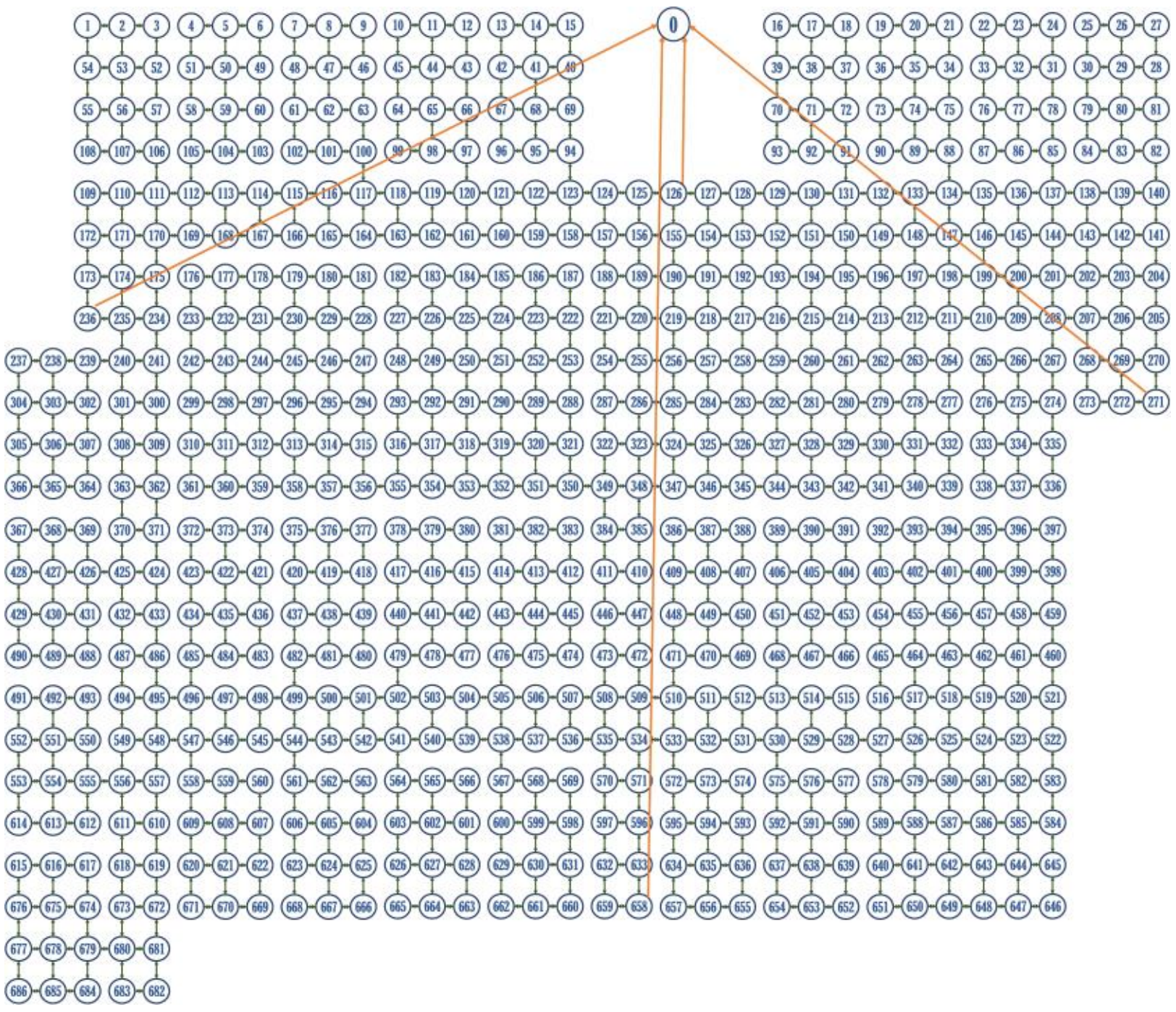

**Figure 6.9** Network associated with the plan of Figure 6.8.

**Table 6.2** Evacuation and computation time: *a*) $3 \times 3$ cells (time slots of 2.5 seconds); *b*) $1.5 \times 1.5$ cells (time slots of 1.25 seconds).

| $\tau(a)$ | evacuees $(a)$ | CPU Time $(a)$ | $\tau(b)$ | evacuees $(b)$ | CPU Time $(b)$ | $\tau(b)$ | evacuees $(b)$ | CPU Time $(b)$ |
|---|---|---|---|---|---|---|---|---|
| 1 | 12 | 0.48 sec | 1 | 6 | 1.85 sec | 41 | 207 | 79.10 sec |
| 2 | 24 | 0.45 sec | 2 | 12 | 1.99 sec | 42 | 209 | 84.37 sec |
| 3 | 36 | 0.50 sec | 3 | 18 | 2.10 sec | 43 | 210 | 87.07 sec |
| 4 | 48 | 0.53 sec | 4 | 24 | 2.35 sec | 44 | 212 | 94.39 sec |
| 5 | 60 | 0.59 sec | 5 | 30 | 2.90 sec | 45 | 213 | 95.04 sec |
| 6 | 72 | 0.62 sec | 6 | 36 | 3.21 sec | 46 | 215 | 97.86 sec |
| 7 | 84 | 0.66 sec | 7 | 42 | 3.72 sec | 47 | 216 | 98.71 sec |
| 8 | 96 | 0.71 sec | 8 | 48 | 4.18 sec | 48 | 218 | 105.83 sec |
| 9 | 108 | 0.79 sec | 9 | 54 | 5.18 sec | 49 | 219 | 120.60 sec |
| 10 | 120 | 0.88 sec | 10 | 60 | 5.53 sec | 50 | 221 | 125.50 sec |
| 11 | 132 | 0.97 sec | 11 | 66 | 6.27 sec | 51 | 222 | 130.07 sec |
| 12 | 144 | 1.02 sec | 12 | 72 | 6.76 sec | 52 | 224 | 128.54 sec |
| 13 | 156 | 1.11 sec | 13 | 78 | 7.92 sec | 53 | 225 | 138.54 sec |
| 14 | 168 | 1.16 sec | 14 | 84 | 8.76 sec | 54 | 227 | 139.20 sec |
| 15 | 180 | 1.42 sec | 15 | 90 | 10.10 sec | 55 | 228 | 154.73 sec |
| 16 | 192 | 1.60 sec | 16 | 96 | 10.54 sec | 56 | 230 | 160.65 sec |
| 17 | 204 | 1.76 sec | 17 | 102 | 12.38 sec | 57 | 231 | 161.58 sec |
| 18 | 216 | 1.82 sec | 18 | 108 | 14.19 sec | 58 | 233 | 160.01 sec |
| 19 | 228 | 1.86 sec | 19 | 114 | 15.17 sec | 59 | 234 | 171.88 sec |
| 20 | 240 | 2.12 sec | 20 | 120 | 17.36 sec | 60 | 236 | 169.26 sec |
| 21 | 252 | 2.29 sec | 21 | 126 | 19.26 sec | 61 | 237 | 178.87 sec |
| 22 | 264 | 2.33 sec | 22 | 132 | 21.55 sec | 62 | 239 | 182.49 sec |
| | | | 23 | 138 | 23.50 sec | 63 | 240 | 201.91 sec |
| | | | 24 | 144 | 24.98 sec | 64 | 242 | 205.39 sec |
| | | | 25 | 150 | 27.67 sec | 65 | 243 | 240.93 sec |
| | | | 26 | 156 | 30.12 sec | 66 | 245 | 213.88 sec |
| | | | 27 | 162 | 34.03 sec | 67 | 246 | 216.48 sec |
| | | | 28 | 168 | 36.00 sec | 68 | 248 | 216.67 sec |
| | | | 29 | 174 | 38.31 sec | 69 | 249 | 230.26 sec |
| | | | 30 | 180 | 40.06 sec | 70 | 251 | 310.11 sec |
| | | | 31 | 186 | 38.93 sec | 71 | 252 | 256.64 sec |
| | | | 32 | 192 | 43.35 sec | 72 | 254 | 329.78 sec |
| | | | 33 | 195 | 47.16 sec | 73 | 255 | 324.49 sec |
| | | | 34 | 197 | 51.70 sec | 74 | 257 | 337.55 sec |
| | | | 35 | 198 | 55.31 sec | 75 | 258 | 348.49 sec |
| | | | 36 | 200 | 63.31 sec | 76 | 260 | 300.19 sec |
| | | | 37 | 201 | 66.90 sec | 77 | 261 | 381.83 sec |
| | | | 38 | 203 | 67.27 sec | 78 | 263 | 382.24 sec |
| | | | 39 | 204 | 71.91 sec | 79 | 264 | 323.34 sec |
| | | | 40 | 206 | 75.90 sec | | | |

**Table 6.3** Evacuation and computation time for time-decomposed scenarios: *a*) $3 \times 3$ cells (time slots of 2.5 seconds); *b*) $1.5 \times 1.5$ cells (time slots of 1.25 seconds).

| $\tau(a)$ | evacuees $(a)$ | CPU Time $(a)$ | $\tau(b)$ | evacuees $(b)$ | CPU Time $(b)$ | $\tau(b)$ | evacuees $(b)$ | CPU Time $(b)$ |
|---|---|---|---|---|---|---|---|---|
| 1 | 12 | 0.48 sec | 1 | 6 | 2.43 sec | 41 | 205 | 1.89 sec |
| 2 | 24 | 0.56 sec | 2 | 12 | 3.11 sec | 42 | 208 | 1.88 sec |
| 3 | 36 | 0.45 sec | 3 | 18 | 2.45 sec | 43 | 210 | 1.89 sec |
| 4 | 48 | 0.54 sec | 4 | 24 | 2.42 sec | 44 | 212 | 1.89 sec |
| 5 | 60 | 0.56 sec | 5 | 30 | 2.31 sec | 45 | 213 | 1.83 sec |
| 6 | 72 | 0.50 sec | 6 | 36 | 2.39 sec | 46 | 215 | 1.85 sec |
| 7 | 84 | 0.51 sec | 7 | 42 | 2.32 sec | 47 | 216 | 1.86 sec |
| 8 | 96 | 0.52 sec | 8 | 48 | 2.53 sec | 48 | 218 | 1.82 sec |
| 9 | 108 | 0.49 sec | 9 | 54 | 2.45 sec | 49 | 219 | 1.89 sec |
| 10 | 120 | 0.51 sec | 10 | 60 | 2.54 sec | 50 | 221 | 1.84 sec |
| 11 | 132 | 0.48 sec | 11 | 66 | 2.39 sec | 51 | 222 | 1.90 sec |
| 12 | 144 | 0.55 sec | 12 | 72 | 2.42 sec | 52 | 224 | 1.84 sec |
| 13 | 156 | 0.50 sec | 13 | 78 | 2.30 sec | 53 | 225 | 1.90 sec |
| 14 | 168 | 0.58 sec | 14 | 84 | 2.48 sec | 54 | 227 | 1.85 sec |
| 15 | 180 | 0.56 sec | 15 | 90 | 2.48 sec | 55 | 228 | 1.92 sec |
| 16 | 192 | 0.51 sec | 16 | 96 | 2.89 sec | 56 | 230 | 1.93 sec |
| 17 | 204 | 0.53 sec | 17 | 102 | 2.24 sec | 57 | 231 | 1.84 sec |
| 18 | 211 | 0.52 sec | 18 | 108 | 1.97 sec | 58 | 233 | 1.85 sec |
| 19 | 217 | 0.41 sec | 19 | 114 | 2.05 sec | 59 | 234 | 1.82 sec |
| 20 | 223 | 0.43 sec | 20 | 120 | 1.90 sec | 60 | 236 | 1.87 sec |
| 21 | 229 | 0.43 sec | 21 | 126 | 1.98 sec | 61 | 237 | 1.83 sec |
| 22 | 235 | 0.52 sec | 22 | 132 | 1.95 sec | 62 | 239 | 1.85 sec |
| 23 | 241 | 0.43 sec | 23 | 138 | 2.04 sec | 63 | 240 | 1.87 sec |
| 24 | 246 | 0.50 sec | 24 | 144 | 1.99 sec | 64 | 242 | 1.91 sec |
| 25 | 249 | 0.39 sec | 25 | 150 | 1.87 sec | 65 | 243 | 1.95 sec |
| 26 | 252 | 0.55 sec | 26 | 156 | 1.86 sec | 66 | 245 | 1.89 sec |
| 27 | 255 | 0.50 sec | 27 | 162 | 187 sec | 67 | 246 | 1.81 sec |
| 28 | 258 | 0.51 sec | 28 | 166 | 1.93 sec | 68 | 248 | 1.89 sec |
| 29 | 261 | 0.53 sec | 29 | 169 | 1.93 sec | 69 | 249 | 1.85 sec |
| 30 | 264 | 0.50 sec | 30 | 172 | 1.85 sec | 70 | 251 | 1.95 sec |
| | | | 31 | 175 | 1.87 sec | 71 | 252 | 1.84 sec |
| | | | 32 | 178 | 1.87 sec | 72 | 254 | 1.85 sec |
| | | | 33 | 181 | 1.89 sec | 73 | 255 | 1.92 sec |
| | | | 34 | 184 | 1.86 sec | 74 | 257 | 1.90 sec |
| | | | 35 | 187 | 1.90 sec | 75 | 258 | 1.81 sec |
| | | | 36 | 190 | 1.96 sec | 76 | 260 | 1.90 sec |
| | | | 37 | 193 | 1.83 sec | 77 | 261 | 1.94 sec |
| | | | 38 | 196 | 1.89 sec | 78 | 263 | 1.81 sec |
| | | | 39 | 199 | 1.82 sec | 79 | 264 | 1.98 sec |
| | | | 40 | 202 | 1.96 sec | | | |

whilst the computational delay decreases, the evacuation delay (that has a priority over all other delays) increases: so that running the algorithm in a collaborative architecture is not recommended, regardless of melioration/ deterioration of total response time.

Taking into account the result of this subsection and keeping evacuation time and CPU time as inputs, the following subsection practically assesses, in terms of total response time (delay), the quality of proposed architectural patterns with respect to four different scenarios, resulting from the different combinations of architectural patterns designed to handle emergency situations:

1. *Centralized with High Resolution (Centralized-HR)*: the critical situation is handled by a continuous simulation of the algorithm, (on a single controller) and physical space is divided into $1.5 \times 1.5$ cells.
2. *Centralized with Low Resolution (Centralized-LR)*: the critical situation is handled by a continuous simulation of the algorithm, (on a single controller) and physical space is divided into $3 \times 3$ cells. Therefore, again all tasks are routed to a central controller.
3. *Collaborative with High Resolution (Collaborative-HR)*: the situation is handled by a time-decomposed simulation of the algorithm, (on collaborative controllers) and physical space is divided into $1.5 \times 1.5$ cells. This means that the two available controllers are intermittently chosen as the destination of routed tasks.
4. *Collaborative with Low Resolution (Collaborative-LR)*: the emergency situation is handled by a time-decomposed simulation of the algorithm, (in a collaborative way) and physical space is divided into $3 \times 3$ cells. Thus, one of the controllers handles the situation for *odd* and the other for *even* time slots.

### 6.3.3 Answers to RQ3: Total Delay Assessment using Queuing Networks Parameterization

In order to realize the software architectures resulting from algorithm simulations, we exploit the QN of Fig. 6.3 that we have previously introduced, with the four different sets of input parameters needed to model the software architectures resulting from algorithm simulations.

*Sense* tasks (by CCTVs) are generated at a certain rate (i.e. every $2.5$ seconds for high resolution and $1.25$ seconds for low resolution networks). These rates are the monitoring frequencies, and is the time taken for an individual to cross a single cell in our crowd monitoring algorithm. Such time intervals represent a key-value, due to their impact on the overall evacuation delay.

Exponential distributions are used to define CPU times. Table 6.4 reports the means of such distributions for our case study, which have been estimated in several ways.

Our focus is on evacuation, i.e. *CriticalPlan-HR* and *CriticalPlan-LR*, which have been estimated by formulating the evacuation handling problem within CPLEX. Other parameters are set as follows:

- Service time distribution means for sensors, networks and actuators, have been obtained by modelling the IoT system for the Alan Turing building with CAPS, our simulation framework (73). CPU times for sensors, actuators and networks, have been calculated in terms of transmission and propagation delays (*td* and *pd*, respectively).
- Service time distribution means for controllers refer to SMAPEA task types. *Sense* tasks have zero CPU time since they do not introduce additional computation for controllers, hence they just need to be transformed into *Monitor* tasks.
- As shown in Table 6.4, *Monitor* and *Execute* tasks are equal and have the same order of magnitude of *Analyze*. To avoid further complexity, we ignore formulating optimization models for such task types, by setting arbitrary values as follows: we assume that aggregating raw data (i.e. *Monitor*) and building a list of atomic actions to execute (i.e. *Execute*) are less demanding than interpreting monitored data (i.e. *Analyze*).

**Table 6.4** SMAPEA Task Types and CPU Times for case Study

| SAMPEA-QN Layer | Service Center | Task Type | | Mean CPU time (exp) | | Notes |
|---|---|---|---|---|---|---|
| | | ID | Name | Centralized | Collaborative | |
| Perception | CCTVs | 1 | CctvSense | 0.01386 | | *By CAPS (td).* |
| Network | PL2IL | 2 | CctvSense | 0.023480633 | | *By CAPS (td+pd).* |
| P&S | Controllers | 3 | Sense | 0 | | *Sense just becomes xMonitor* |
| | | 4 | CriticalMonitor | 0.0045067 | 0.0045067 | *Arbitrarily set.* |
| | | 5 | CriticalAnalyze | 0.00676 | 0.005735 | *By CPLEX.* |
| | | **6** | **CriticalPlan-HR** | 110.1791139 | 2.0164557 | *By CPLEX* |
| | | **7** | **CriticalPlan-LR** | 1.1668182 | 0.5016667 | *By CPLEX* |
| | | 8 | CriticalExecute | 0.0045067 | 0.0045067 | *Arbitrarily set.* |
| Network | IL2AL | 9 | DashboardActuate | 0.013619641 | | *By CAPS (td+pd).* |
| | | 10 | AlarmActuate | 1.013619641 | | |
| | | 11 | EvacuationSignsActuate | 2.013619641 | | |
| Application | Dashboards | 12 | DashboardActuate | 0.000921667 | | *By CAPS (td).* |
| | Alarms | 13 | AlarmActuate | 0.000921667 | | |
| | EvacuationSigns | 14 | EvacuationSignsActuate | 0.000921667 | | |

**Simulation**. We assess the total delay corresponding to each IoT architectural pattern design based on the flow algorithm. The response time (delay) that is analyzed is the mean time spent from starting the sampling to the time that actuation ends.

Table 6.5 reports, for each scenario, the overall system response time (in seconds) in the second column and the architectural design decision in the third column.

**Answers to RQ4: Architectural Design Decision**. Experimental results show that, in LR, the collaborative pattern minimizes system response time. However, this pattern does not satisfy the precondition of optimal evacuation of people from the building. Thus, the centralized pattern may be more appropriate in a critical mode. The drawback is that, managing critical cases with the centralized architecture increases system response time by more than 200% with respect to the collaborative one. The solution could use the HR network. However, HR does not allow the system to fulfill the real-time requirement, due to two factors: *i)* working in HR requires much more CPU time; *ii)* the sampling rate doubles when the system performs in HR. Both factors contribute to a significant worsening of performance that might lead to system saturation (as for the Centralized-HR).

**Table 6.5** Experimental results.

| Pattern | System response time (s) | Architectural design decision |
|---|---|---|
| Centralized with HR | *System saturation* | Violates real-time requirement |
| Centralized with LR | *1.5085* | Practical |
| Collaborative with HR | *26.5597* | Violates real-time requirement |
| Collaborative with LR | *0.5864* | Violates optimally requirement |

As a result of the considerations above, *only one pattern could be adopted, i.e. the Centralized-LR*.

It is worth noticing that, in order to address the sampling rate for HR (i.e. $1.25$), while satisfying the precondition that minimizes operational delay, the CPU time of the controller in the Centralized pattern should have the same order of magnitude of the controllers in the Collaborative-LR. For example, with a CPU time of $0.5016667$ (i.e. the same as the Collaborative-LR), the Centralized-HR would show a system response time of $0.6535$ seconds (i.e. 11.5% more than Collaborative-LR). As a second example, with a CPU time of $2.0164557$ (i.e. the same as Collaborative-HR), the Centralized-HR would experience continuous saturation that might lead to a system response time in the order of hours.

## 6.4 Conclusion

This work uses the Queuing Network concept in order to model the IoT architectural patterns for emergency evacuation, and assess the patterns' corresponding delays. The architecture has a core computational component in the form of network flow, which supports the design decision by providing the model with expected operational and computational delays. Preliminary evaluations using data from a real case, and an *ad-hoc* IoT infrastructure showed the suitability of a centralized software architecture based on a low resolution division of the building surface. The following chapter uses the social agent-based modeling approach to fine-tune the IoT-based emergency infrastructure design based on real human behaviour.

# Part III

# Combining Agent-based Social Simulation and IoT Infrastructure

# Introduction to Part III

This part is written based on the following peer reviewed articles:

- **A Combined Netflow-Driven and Agent-Based Social Modeling Approach for Building Evacuation**, *PRIMA 2019: Principles and Practice of Multi-Agent Systems.*
  *DOI:* `https://doi.org/10.1007/978-3-030-33792-6_30`
- **Agent-based Simulation for IoT Facilitated Building Evacuation**, *Published in: International Conference on Information and Communication Technologies for Disaster Management, 2019.*
  *DOI:* `https://www.researchgate.net/publication/338398264_Agent-based_Simulation_for_IoT_Facilitated_Building_Evacuation`

**Abstract.** In an emergency, finding safe egress pathways in a short period of time is crucial. To do so, we take advantage of a network flow (netflow) algorithm that acts as the core of a real-time recommender system to be used by building occupants and decision-making bodies. However, a purely optimization approach can lack realism since building occupants may not evacuate immediately, stopping to to look for their friends or trying to assess if the alert is for real or just a drill, etc. Furthermore, they may not always follow the recommended optimal paths. Thus, in order to assess the egress in a physical space and to test our evacuation algorithms, we use a simulation-optimization (*S/O*) approach. The approach allows us to test more realistic evacuation scenarios and compare them with an optimal approach. The *S/O* approach uses both a netflow algorithm and an agent-based approach to model and simulate individual human behaviours. People are modeled as agents with specific characteristics, such as social attachment to others, variation in speed of movement, etc. Furthermore, a Belief-Desire-Intention (BDI) agent architecture is used to model the individual differences in people and to more accurately describe the heterogeneity of the building occupants in terms of their current beliefs about the situation and goals. The real geospatial data obtained from three experiments is set as the model input to endorse the validity of the simulations. The results confirm the usefulness of using such *S/O* approach to improve design-time and real-time evacuation systems.

**keywords.** Building evacuation, Agent-based social simulation, Internet of Things,Network optimization.

**Overview.** The increasing topological changes in urban environment have caused the human civilization to be subjected to an increasing risk of emergencies. Hence, designing infrastructures to handle possible emergencies has become an ever increasing need. The safe evacuation of people and personnel from the premises takes precedence when dealing the necessary mitigation and disaster risk management. The evacuation time of people from a scene of an emergency is crucial. In order to reduce the time taken for evacuation, better and more robust evacuation algorithms are developed. Such algorithms are used to model participating agents for their exit patterns and strategies and in order to evaluate their movement behaviour based on performance, efficiency and practicality attributes.

This part extends the work in (17) (blinded for review) that explores the collaboration between Internet of Things architectures and safety critical systems. Specifically we look at the incorporation of a netflow algorithm as developed in previous part that can be used in a computer simulation for designing buildings, and also in real-time building evacuation. The algorithm decomposes both the space (building plan) and the time dimension into finite elements: unit cells and time slots. The space element is monitored by sensors, whose data constantly feed into the algorithm to show the best evacuation routes to the occupants. However, such a system may lack the accuracy since: *i)* a purely optimization approach can lack realism as building occupants may not evacuate immediately; *ii)* occupants may not always follow the recommended optimal paths due to various behavioural and organizational issues; *iii)* the physical space may prevent an effective emergency evacuation.

To deal with the above-mentioned challenges, we introduce a simulation-optimization (*S/O*) approach. The *S/O* is an umbrella term for techniques used to optimize stochastic simulations (5). Our *S/O* approach allows us to test more realistic evacuation scenarios and compare them with an optimal approach. We simulate the optimized netflow algorithm under different realistic behavioral agent-based modeling (ABM) constraints, such as social attachment (66) to others, variation in speed of movement, etc. The part furthermore presents a correlation between evacuation time and the influence of human, social, physical and temporal factors (17).

The work is simulated using the PedSim microscopic pedestrian simulation tool and customized in order to incorporate the aforementioned constraints.

## Chapter 7

# Agent-based Simulation for IoT Facilitated Building Evacuation

The Internet of Things (IoT) has changed our approach to safety systems by connecting sensors and providing real-time data to endangered people in emergency fire situations. In an emergency, evacuating a building in the shortest time possible is essential. Although optimisation methods give a best case scenario, they do not reflect the realism of actual evacuations. We combine a network flow optimisation algorithm with agent based social simulation to provide more realistic evacuation scenarios. Individual occupants are modelled as computational agents in the simulator. Agents are heterogenous and exhibit social attachment to others and variations in speed of movement. A Belief-Desire-Intention (BDI) agent architecture is used to model the cognitive reasoning of individual agents. The work is applied to a real building, equipped with IoT sensors whose data feed into the network flow model. The results show that: the network flow algorithm performs better than the shortest path algorithm and avoids congestion in building bottlenecks; social attachment slows down evacuation; internal adjacently placed doubled doors are better than having a single double width door for increasing evacuation.

## 7.1 Introduction

Rapid evacuation of people from buildings during an emergency is crucial. Traditional static evacuation maps do not adequately take into account the dynamic nature of the physical, organizational and behavioural issues. Pathways become easily congested or blocked, dynamic decision making is difficult, and people do not always follow the advised path.

An Internet of Things (IoT) architecture in a building can help with speedy evac-uation since occupants can be continuously tracked and real-time instructions can be given on the safest evacuation routes. However, before an IoT can be installed a number of design decisions should be taken: *i)* what prevents a safe and timely evacuation? *ii)* how should the physical space be designed or modified to facilitate a safe evacuation? *iii)* what IoT components should be used and how should they be distributed to monitor crowds? *iv)* what algorithm should be used as the core of the IoT system to handle crowd routing? *v)* how should the IoT infrastructure be adapted to deal with environment, human behaviour and occupants' interaction with the system?

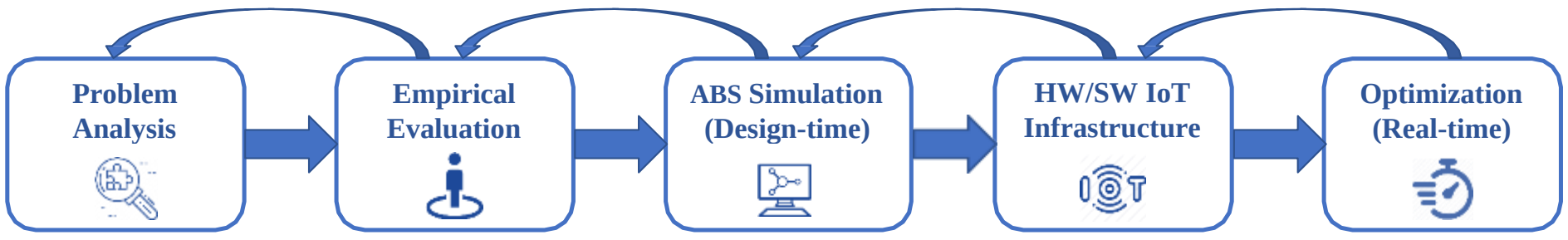

**Figure 7.1** The adopted process.

To address these issues, we extend our previous work ((77), (76) and (6)) by exploring the use of IoT architectures and safety critical systems. Figure 7.1 shows the adopted process to address the aforementioned issues.

In the *problem analysis* phase, we found that static emergency maps are ineffective because they cannot take into account dynamically occurring congestion, obstacles or dangerous areas (70). Furthermore, they disregard individual behaviours (e.g. speed variations and the social attachment that people have to others), and they do not consider certain categories of people (e.g. disabled, etc.). To evaluate such human aspects, in the second phase, we performed *empirical evaluations* with three building evacuation tests. We developed a *simulation/optimization* (Sim/Opt) approach (33) that combines optimization with an agent based social simulation (ABSS) model. This allows us to test more realistic evacuation scenarios and compare them with an optimal approach. ABSS takes into account social attachment (66) of people to others, objects and places, as well as grouping behaviours and the variation in speed of movement, etc.

We previously developed a *optimization algorithm* (netflow) to design the physical building space and help with real-time evacuation. The algorithm decomposes the building space and time into finite elements: unit cells and time slots.

To support the algorithm and to collect data, we designed and implemented a hardware and software *IoT infrastructure*. We installed sensors throughout the selected building, whose data constantly feed into the algorithm to show the best evacuation routes to the occupants. Our IoT system is based on the simulation results performed in the previous phase.

However, the optimization model should be dynamically adapted to human behaviour to give realistic routing recommendations. For example, building occupants may not evacuate immediately; may not follow the recommended optimal paths or may linger before evacuating, collecting their belonging, waiting for colleagues, etc. Therefore we feed the results of an agent based simulation of real-behaviours into our optimization model.

The chapter makes the following contribution of combining the netflow algorithm with ABSS. The approach is tested on various scenarios and compared with the frequently used shortest path algorithm. We use a real-world case study the Alan Turing building in Italy and use the PedSim platform and Cplex solver to simulate our occupants. The novelty of the work is in combining agent based social simulation, that more accurately models human behaviour, with an optimization algorithm fed from an IoT system. Section 7.2 describes previous works, whilst section 7.3 provides the results of the empirical evaluation. The agent based model is presented in section 7.4, and the hardware/software IoT infrastructure in section 7.5. Section 7.6 explains the network optimisation algorithm. Section 7.7 describes the optimisation and agent based simulation scenarios and results. Section 7.8 concludes the chapter.

## 7.2 Literature Review

**ABSS.** The original studies and empirical analysis of pedestrian behaviours were based on photography, time-lapse videos and direct observation (46). The 1990s saw the rise of ABSS (agent based social simulation) to simulate pedestrian behaviours both inside and outside of buildings ((51), (105), (35)). However, unlike our approach these works do not deal with optimality, nor incorporate an IoT approach. In ABSS, an agent is defined as an autonomous software entity that can act upon and perceive its environment (40). When agents are put together they form an artificial society, each perceiving, moving, performing actions, communicating, and transforming the local environment, much like human-beings in real society. In ABSS, the agents typically represent humans or groups of humans. The advantage of an agent-based approach is that nonlinear relationships and heterogeneous behaviours can be easily modelled and understood through the multiple complex interactions. Using Epstein and Axtell's terminology, ABSS is described as a 'bottom-up' approach (36). Or, to use Bonabeau's words, it is the creation of a microscopic model (18). ABSS does not attempt to specify global system behaviour or define a macroscopic model of the situation, but rather focuses on modelling individuals. An effective method used to model pedestrian dynamics in agent based systems is the social force model ((48; 16)). The model is based on physics using a particle system and socio-psychological forces in order to describe human crowd behaviour.

**Sim/Opt.** Concerning research on network optimisation for emergencies, pioneering work was conducted by Choi et al (27), who modeled a building evacuation problem by dynamic flow maximisation where arc capacities may depend on flows in incident arcs. Although dating back to the 1980s and limited to a theoretical analysis, the paper provides a good starting point. Recently, some evacuation planning models have been based on the transshipment problem. For instance, Schloter et al. (94) study classical flow models in order to deal with crowd evacuation planning. Other papers propose a hybrid *Opt/Sim* approach (33). For example, Abdelghany et al. (1) integrates a genetic algorithm with a microscopic pedestrian simulation assignment model. The genetic algorithm looks for an optimal evacuation plan, while simulation guides the search by evaluating the quality of the plans generated. However, the genetic algorithm approach is not guaranteed to find optimal plans.

**IoT.** Recent literature addresses the ability of finding good solutions in the very short time that is needed by a real-time IoT system. Such quick re-computation allows coping with IoT data that dynamically change over time. We addressed IoT systems in our previous work (77). IoT applications typically consist of a set of software components including perception, data processing and storage, and actuation, which are distributed across network(s). In the emergency management, an IoT system should be fault-tolerant (68), performant (6), and energy efficient (76). Unfortunately it is very hard to find papers on IoT systems that have actually been implemented and that give results. Some papers present only simulations assuming ideal human behaviours, or implement an IoT system assuming a very small number of occupants.

The following sections describe the 5 steps (Figure 7.1) that we followed to build such an IoT infrastructure to manage an emergency. Since the introductory sections have already explored the problem, we will start with the second step in the process.

## 7.3 Empirical Evaluation - phase 2

This study considers the evacuation of one floor of Alan Turing building, which is sometimes used for exhibitions. The building floor consists of 29 rooms, 4 main corridors, 4 emergency exits and several IoT sensors and actuators (See Figure 5.1 in Chapter 5). Safety conditions in the building were supervised by the L'Aquila Fire Brigade and Civil Protection department. The complex building structure, as well as the ability to collect data on people during the event, made this an ideal case-study. The results concerning evacuation time in real tests were described in (33); these were used in order to form a baseline for our simulations. The main results of that work giving the global evacuation time of 3 scenarios are shown in Table 7.1.

**Table 7.1** Empirical findings.

| Test # | Date | Started | Finished | # Evacuees | Test Type |
|---|---|---|---|---|---|
| **1** | 22.03.2018 | 10:45 | 10:52 | 225 | Earthquake |
| **2** | 29.05.2018 | 11:37 | 11:43 | 200 | Fire |
| **3** | 07.03.2019 | 11:05 | 11:14 | 380 | Earthquake |

We observed that, the evacuation lasted 9 minutes in the worst case. Earthquake evacuation takes a little longer than fire evacuation, since people first need to find an internal shelter.

From the real-life simulations we noted: *i)* congestion happened especially in corridors and close to the emergency exits; *ii)* some evacuation exits became temporarily blocked because of both congestion and improper design; *iii)* occupants sometimes followed paths other than those specified; *iv)* occupants were uncertain about which path to follow to the safest exit; *v)* evacuation advice and signs were sometimes ignored by the occupants. There was also a lack of situational awareness for occupants and security personnel due to poor communication and lack of up-to-date information.

The aforementioned issues confirmed the inefficiency of static maps, like the one shown in Figure 5.1. In following sections, we describe an ABSS approach to model human behaviour within the building and an IoT infrastructure which is capable of addressing the above mentioned problems.

## 7.4 Agent-based Social Simulation - phase 3

This section describes the agents and the behavioural model for pedestrian movement during the evacuation scenario. The Alan Turing building (6) was used for the simulation. There are two types of agents: *TopologyAgents* and *GameAgents*.

**TopologyAgents** represent the topology of the building such as obstacles, walls, doors, passageways, and emergency exits. These agents are associated with certain forces and traits. For instance, the wall force acts on Game Agents so that they cannot pass through unless a huge amount of GameAgent force (see below) is applied to the walls. Topology agent traits include passageway flow capacity, and total door flow capacity, etc.

**GameAgents** represent occupants and are modelled using the *BDI* (Belief-Desire-Intention) architecture (see below). Agents have traits such as movement speed, perceptive radius, social force (personal and inter-personal radius). GameAgents use their perceptive radius to navigate and they move towards the desired goal unless an event triggers them to act otherwise.

From Figure 7.2, the *GameAgents* typically follow one of two types of behaviours: form a group or proceed alone.

*GameAgents* interact with *topological agents* during an evacuation scenario by flowing through passage ways, moving towards an exit, and passing through doors (Figure 7.3). This interaction between topological elements and Game agents are described in simulated scenarios.

The ***Belief-Desire-Intention*** architecture (88) is used for *GameAgents*. A *Belief* represents the agent's own knowledge of events and locations. A *Desire* represents the motivational state of an agent; things that the agent would like to do. An *Intention* represents the deliberative state of an agent, i.e. a selected desire. Once an intention is chosen, the agent develops a plan to achieve that intention (goal). The agent's decision making and dynamic path routing are influenced by the desire for avoiding congestion and obstacles. This is updated using real geospatial data obtained from the real-time IoT infrastructure.

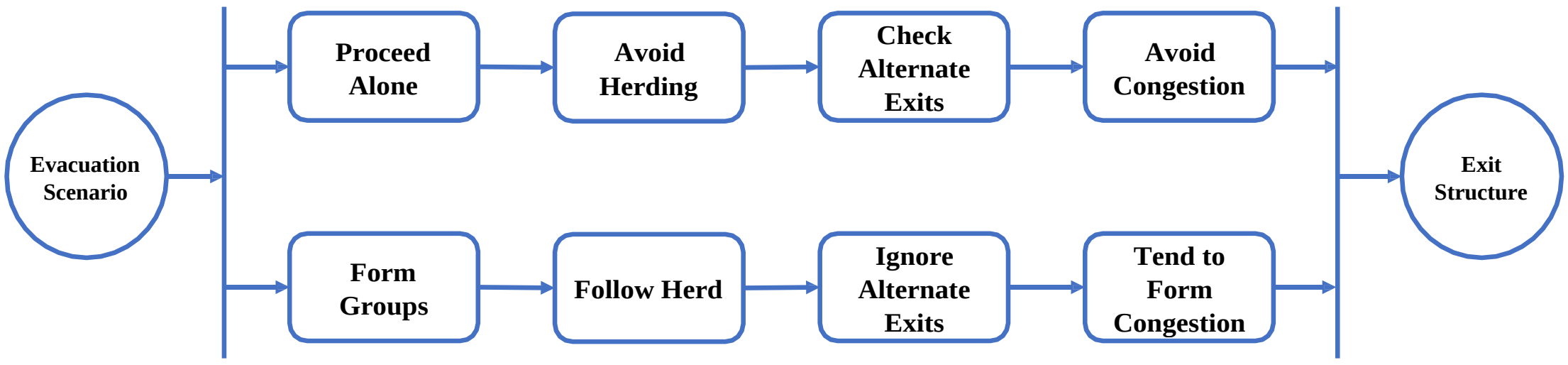

**Figure 7.2** Game agent behaviour.

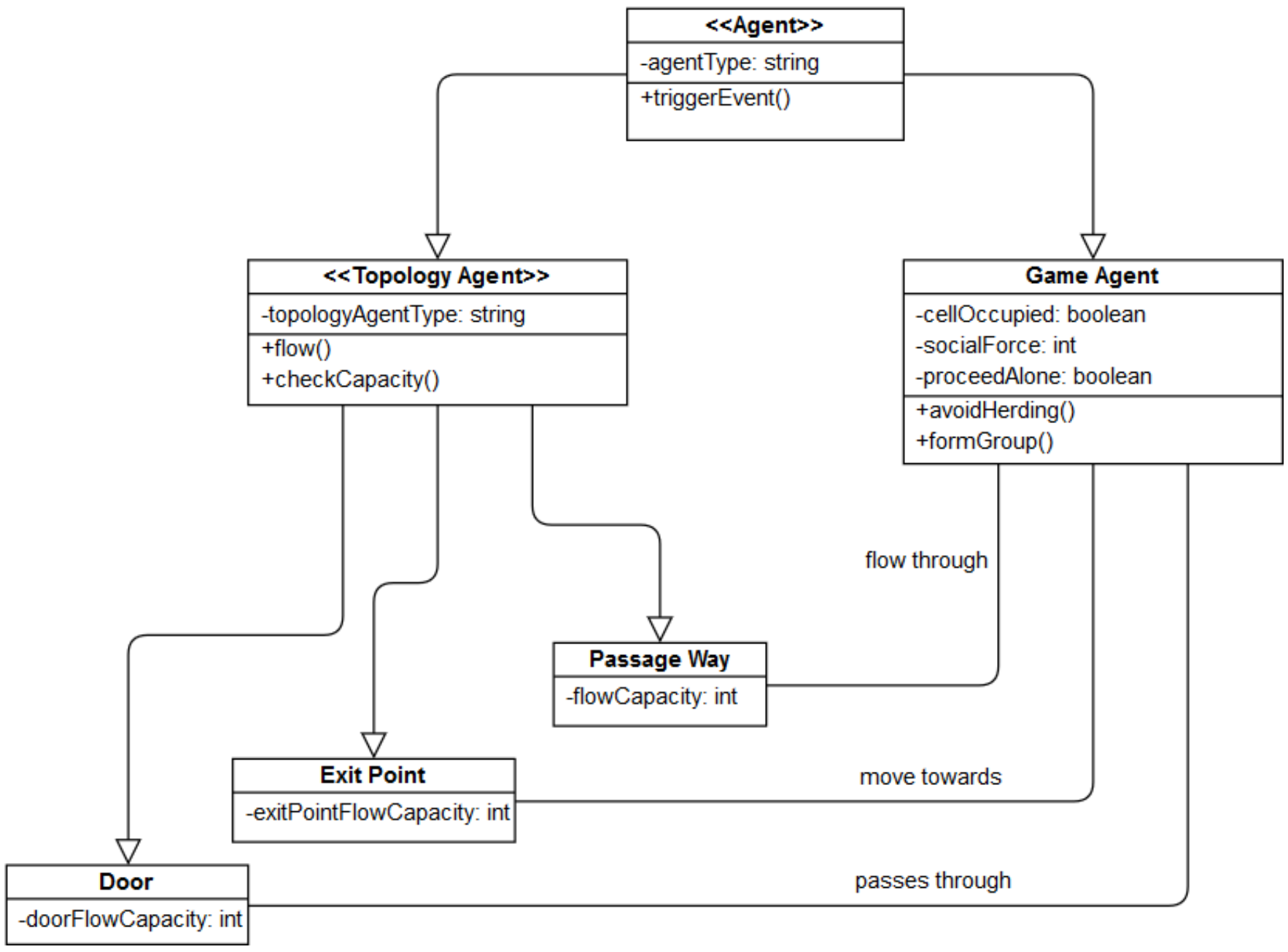

**Figure 7.3** Interaction between Game Agents and Topological Agents

To increase realism we incorporate social grouping. *Herding* is common in emergency situations (52) (86). Thus, the *Game Agents* can form herds, which affect their decisions and movement towards exits. Speed of movement is important, e.g. a man will walk more slowly to match the speed of a woman (101), and groups of individuals typically move more slowly than a single person (93). The slow movement of interacting groups can consequently affect evacuation efficiency (85). To model such behaviours, groups move at a slower speed than individuals. Although realistic, this has far reaching impacts on the overall evacuation time.

## 7.5 Hardware/Software IoT Infrastructure - phase 4

An IoT system is needed for data collection and actuation. However, IoT building designers often face problems with the absence of a reliable and flexible network infrastructure, with installing physical devices in optimal positions, and with legal and privacy issues. The following three issues should be considered before designing the IoT hardware and software:

**1.** *IoT components; their positioning and distribution*. It is crucial to dynamically track how visitors move through a building. The Alan Turing building has CCTV cameras but they were mainly used for security purposes and did not provide any information about visitors' movement, nor their changing numbers. Nevertheless, used the cameras and processed the data using image processing software to provide the people count and flow. We obtained precise data about the people count in each square cell of $3 \times 3m$. We divided the physical surface space into such small cells to be able to implement the network flow routing algorithm.

Additional sensors such as smoke and temperature detectors were installed. These sensors provide an early awareness of the situation in case of an emergency. The position of the IoT sensors highly impacts the accuracy and preciseness of the gathered data. We took advantage of our IoT modelling framework, CAPS (76), to design the physical space and simulate different sensor positions.

**2.** *Software architecture*. In addition to modelling the physical space, our framework is able to create a combined software and hardware view, specifically for surveillance and handling emergency evacuations. Figure 7.4 shows the software architecture. CCTV cameras detect peoples position and movement, which is then fed into the rout-ing algorithm in the controller. The controller decides on the set of actuators based on the situation. In normal situations, the system shows, on a tablet, a 2D-representation of the space, where crowds are located, and how they are moving. In this mode, the optimal flow algorithm is periodically run to estimate the minimum evacuation time required under current conditions. This value can be used to regulate visitor access to the building to comply with safety regulations. If an emergency occurs, in addition to the tablet map, alarm actuators are activated and evacuation signs in each area show the best evacuation routes based on the network model described below.

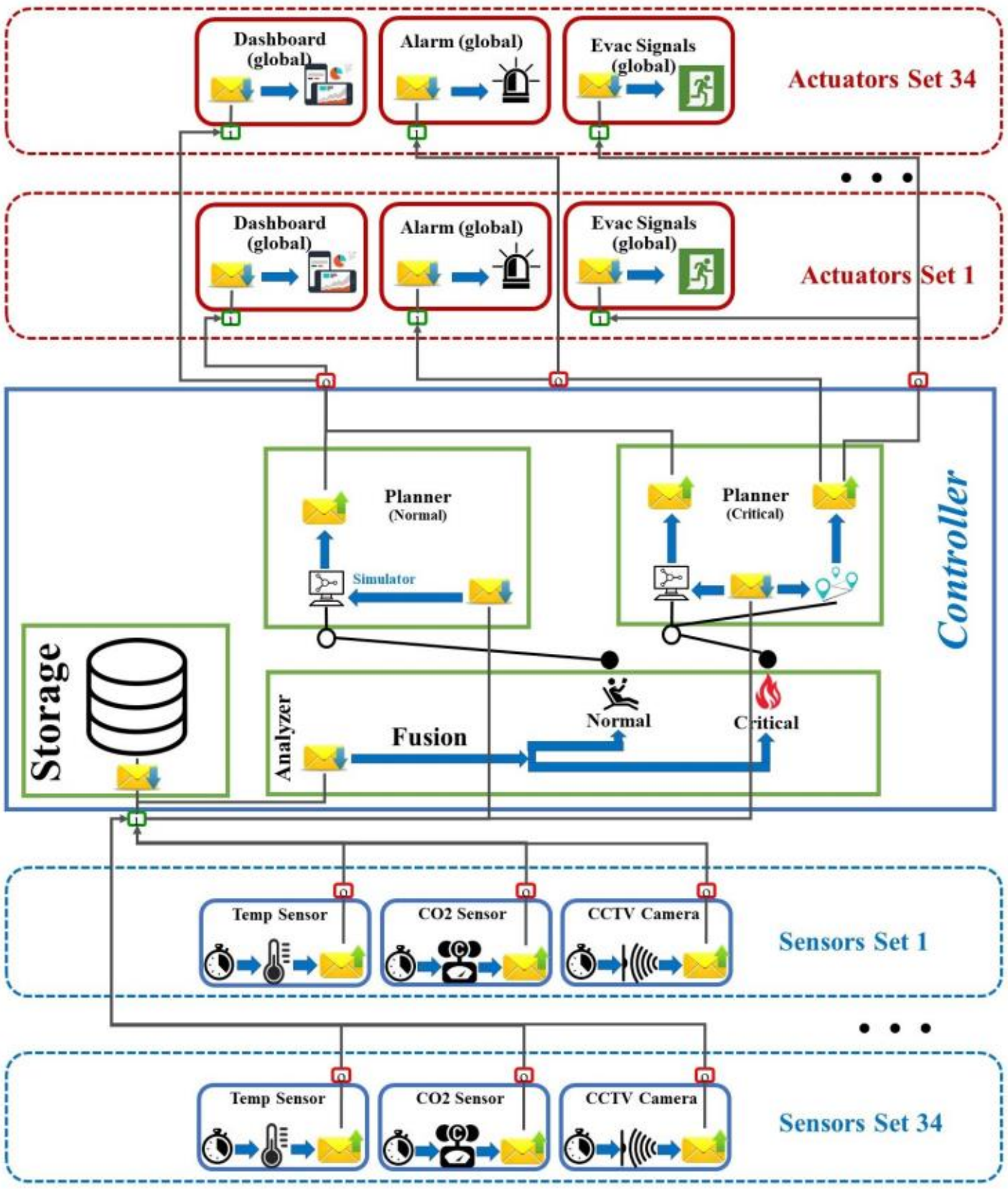

**Figure 7.4** Software architecture.

**3.** *Data and analytics*. We analyze the data on a centralized controller that is in charge of implementing the best crowd routing strategy. The dynamic data that the IoT system feeds into the algorithm are as follows:

- The set of time slots ($T = \{0, 1, \ldots, \tau\}$). These are the monitoring frequency data that the controller needs as input to recalculate the best evacuation routes. We set time slots to be 2.5 seconds. Since the physical space surface is divided into $3 \times 3m$ square cells, the frequency is obtained from the average time a person needs to walk from the centre of one cell to the centre of a neighbouring cell.
- The number of persons that occupy a cell ($i$) at a specific time slot ($t$). This is provided by CCTV cameras and represented in the algorithm by $y_i^t$ denotation.

- The flow of visitors from one cell to a neighbouring cell between two time slots. These are provided by the IoT system and are denoted as $x_{ij}^t$ in the algorithm. This gives the average flow speed in the building.

Some static building data are also fed into the algorithm:

- The maximum number of people that a cell can host at any time ($n_i$); this depends on the cell shape and size. If cells are uniform the capacity is set to a specific number.
- The maximum number of people that, independently of the congestion, can traverse the passageway between two neighbouring cells in a time unit ($c_{ij} = c_{ji}$). Congestion is taken into account in the algorithm formulation phase.

Using the aforementioned data from the IoT system, the following section proposes a network flow algorithm to handle emergency routing.

## 7.6 Network Flow Optimization Algorithm - phase 5

In order to avoid repetition, we removed the algorithm explanations that are already discussed in section 4.2.

## 7.7 Simulation

We now describe the different optimization and ABSS scenarios to test and evaluate the designed IoT system. We are interested in the following research questions.
The first question focuses on the optimization algorithm:

**RQ1**: What is the optimal evacuation time associated with the *Network Flow* and *Shortest Path* algorithms? Which gives a better performance?
The second research question concerns the *Sim/Opt* approach:

**RQ2**: What is the evacuation time under different ABSS scenarios?
The next question investigates the solutions where the evacuation time approaches the optimal value:

**RQ3**: How to optimize the building dimensions in order to reduce the evacuation time?

### 7.7.1 Answer to RQ1: Optimal Evacuation Time

We split each room into unit cells, each behaving as a (virtual) square room that can be traversed in a unit time slot (See (6), page 9). We chose $3 \times 3$ metre cells since this means that only a small area of the floor is not covered with cells and there is a short

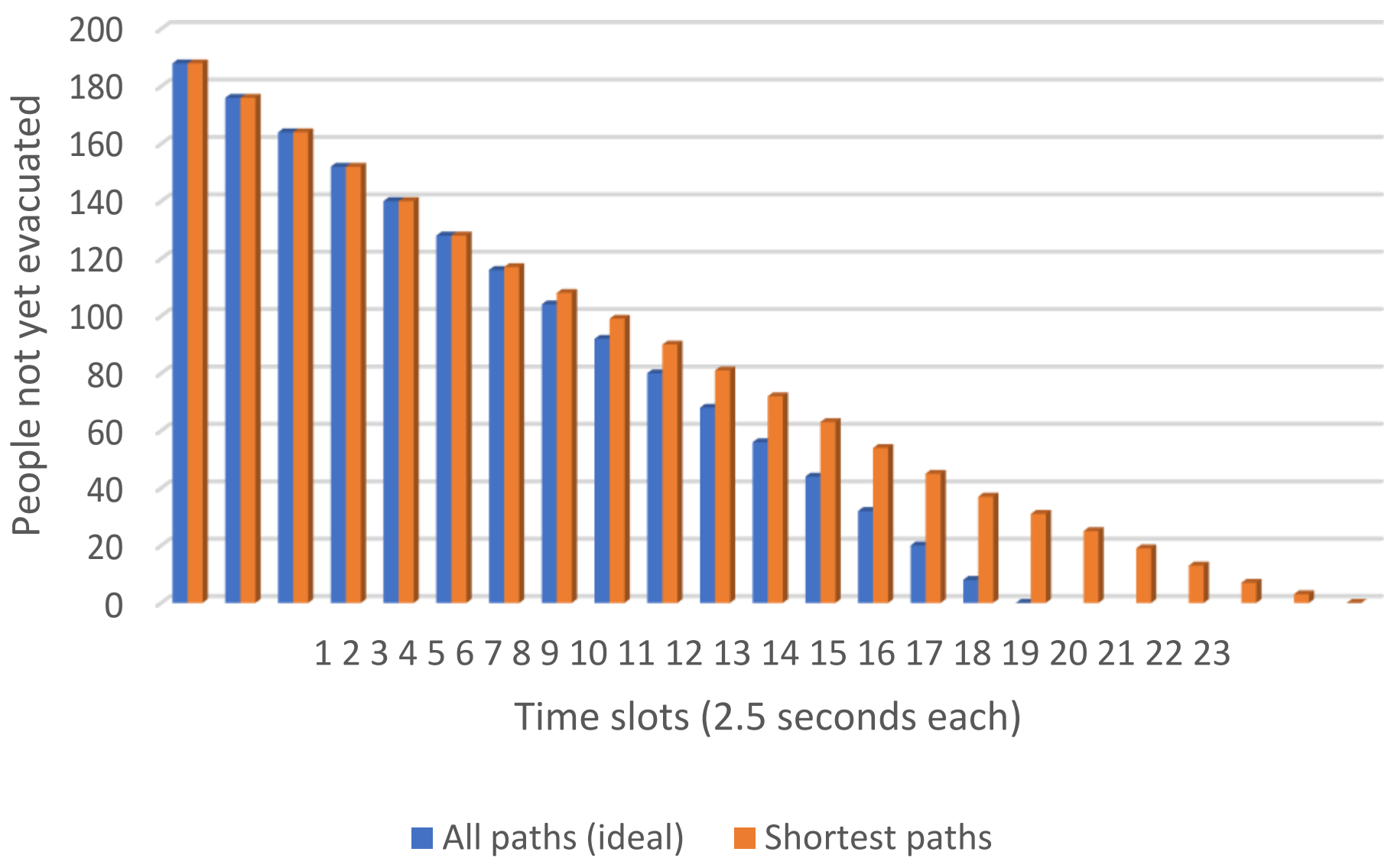

**Figure 7.5** Network flow vs. shortest paths evacuation: Scenarios 2.

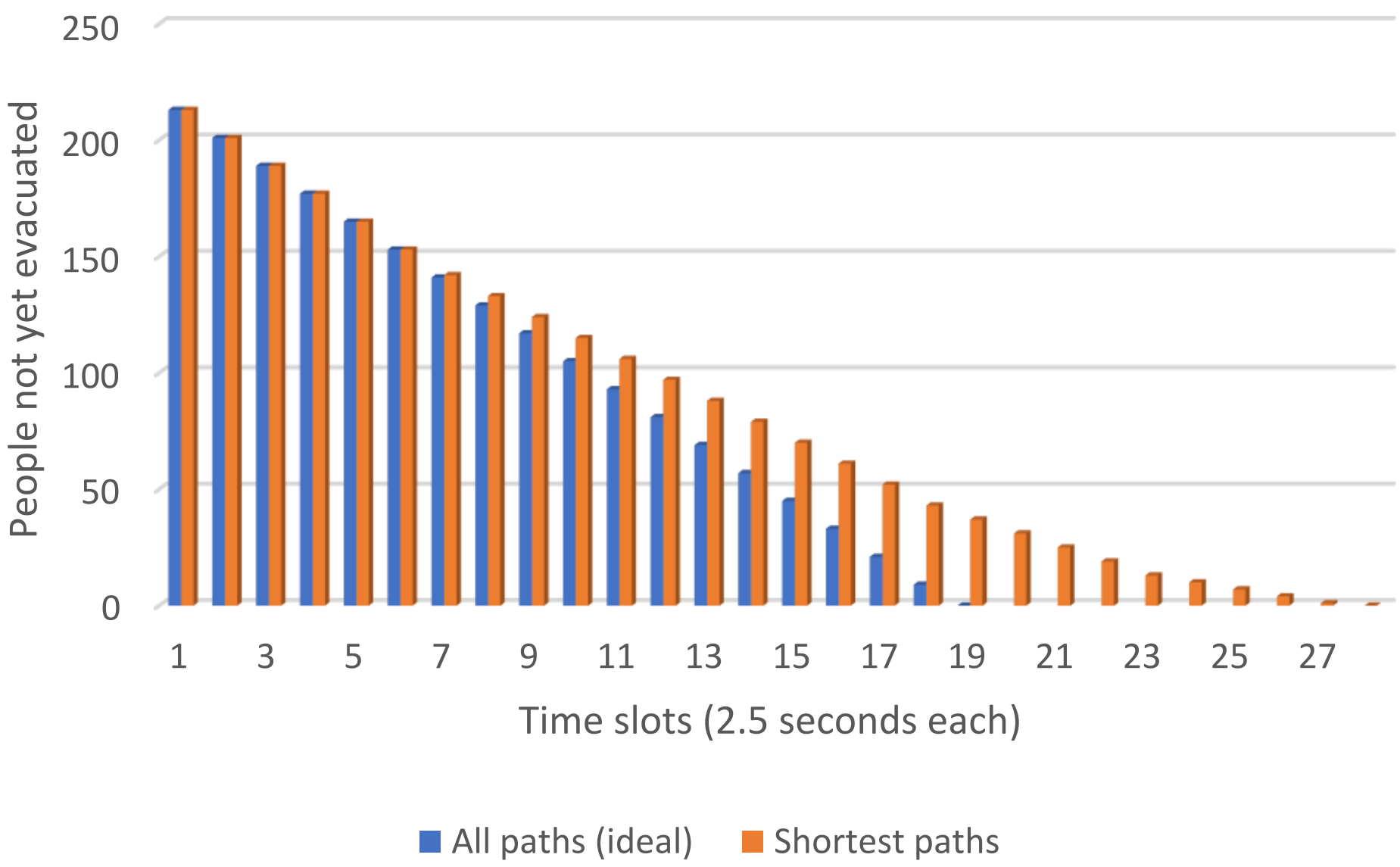

**Figure 7.6** Network flow vs. shortest paths evacuation: Scenarios 1.

CPU time that is needed for real-time applications. The plan embedding resulted in a graph of 144 nodes, including nodes 0, which are safe places.

The simulation code was written in the OPL language and problems were solved by CPLEX version 12.8.0. Experiments were run on a Core i7 2.7GHz computer with 16 GB of RAM under Windows 10 pro 64-bits. We computed the minimum time required for 200, 225 and 380 persons, randomly distributed in the rooms, to reach a safe place. This data is based on the above-mentioned experiments in the Alan Turing building, when the IoT system recorded the simultaneous presence of such populations in the building as a peak value. We solved problem (1-4) for $\tau = 1, 2, \ldots$ until a solution 264 was found. To get a realistic model, parameters such as walking velocity under various conditions, door entrance and room capacities were set based on the literature review: walking velocity = 1.2 $m/s$ (112), door capacity = 1.2 $p/m/s$ (32), cell capacity = 1.25 $p/m^2$ (the UK fire safety).

***Scenario 1***. In the first simulation, we considered an initial occupancy of $N = 200$. This depicts an ideal situation in which agents autonomously choose the best path from all the available building routes. Of course, such an ideal evacuation is perhaps unrealistic and managing one is not easy and probably impractical. In general, evacuation is conducted through predetermined routes. Thus, we suppose that the prescribed evacuation routes are the shortest paths from any cell to the safe place. To evaluate this, we find the sub-graph of $G$ formed by the shortest paths from any cell to 0 (as from static evacuation maps), construct its time-indexed network and solve problems (1)-(4) for increasing $\tau$. From Figure 7.5, evacuating 200 individuals using the proposed netflow algorithm takes 42.5 seconds. Compared to our netflow model, the shortest paths approach increases the optimal evacuation time by 35%. In addition, the CPU time associated with our netflow model is 1.62 seconds (69 % less than shortest paths). This is totally acceptable for real-time applications.

***Scenario 2***. We repeated the simulation increasing the number of people to 225 (based on the real evacuation test). In this case everyone can reach a safe place after 19 time slots, i.e., 47.5 seconds. In a second simulation, we again suppose that the prescribed evacuation routes are the shortest paths from any cell to a safe place. In this situation, evacuating 225 individuals takes 1 minutes and 10 seconds, over 47% more than optimal flows (Figure 7.6).

***Scenario 3***. We then increased the number of people to 380. The netflow algorithm improves the evacuation time by 30% (Figure 7.7). In all scenarios, the netflow algorithm gives a short CPU time so as to be compliant with real-time applications.

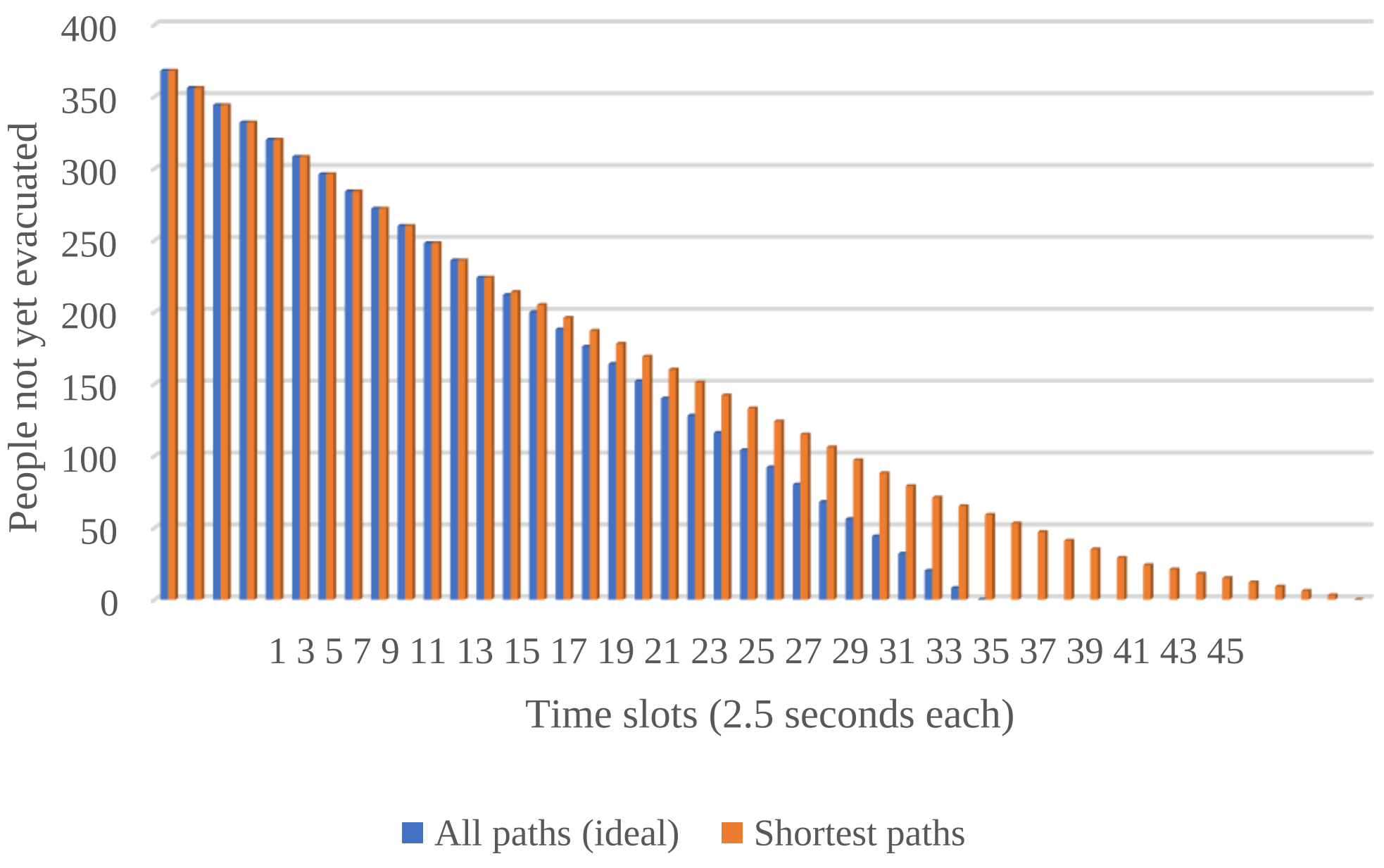

**Figure 7.7** Ideal vs. shortest paths evacuation: Scenario 3.

***Findings***: In the three scenarios comparing netflow and shortest path, there are similar flows for some time (15 seconds in the first and second scenarios and 32,5 seconds in the third). After these times, the shortest path approach experiences congestion and evacuation slows down. As expected, the tail of people still in the building increases with initial occupancy.

### 7.7.2 Answer to RQ2: Agent-based Social Simulations

We consider the Alan Turing building with a real population (*GameAgents*) of: 200, 225 and 380 persons. All agents and scenarios use the following parameters:

- *Walking Velocity* - ranging between $0.7m/s$ to $1.2m/s$, in accordance with the average walking speed in (106; 101).
- *Social Force* - an individual agent's radius is arbitrarily set to 0.2 m, obtained by using the *biacromial diameter* in (98). Thus, agents do not pass through each other and maintain a minimum discernible distance from each other. This also facilitates setting the maximum number of agents per cell, room and passage-flow.
- *Wall Force* - wall force is set 0.1 m, i.e., agents cannot pass beyond 0.1 m from the wall; this prevents agents sticking to walls and passing through obstacles.
- Door Flow Capacity - $1.2p/m/s$, (112).
- Cell capacity - $1.25p/m^2$, (32).

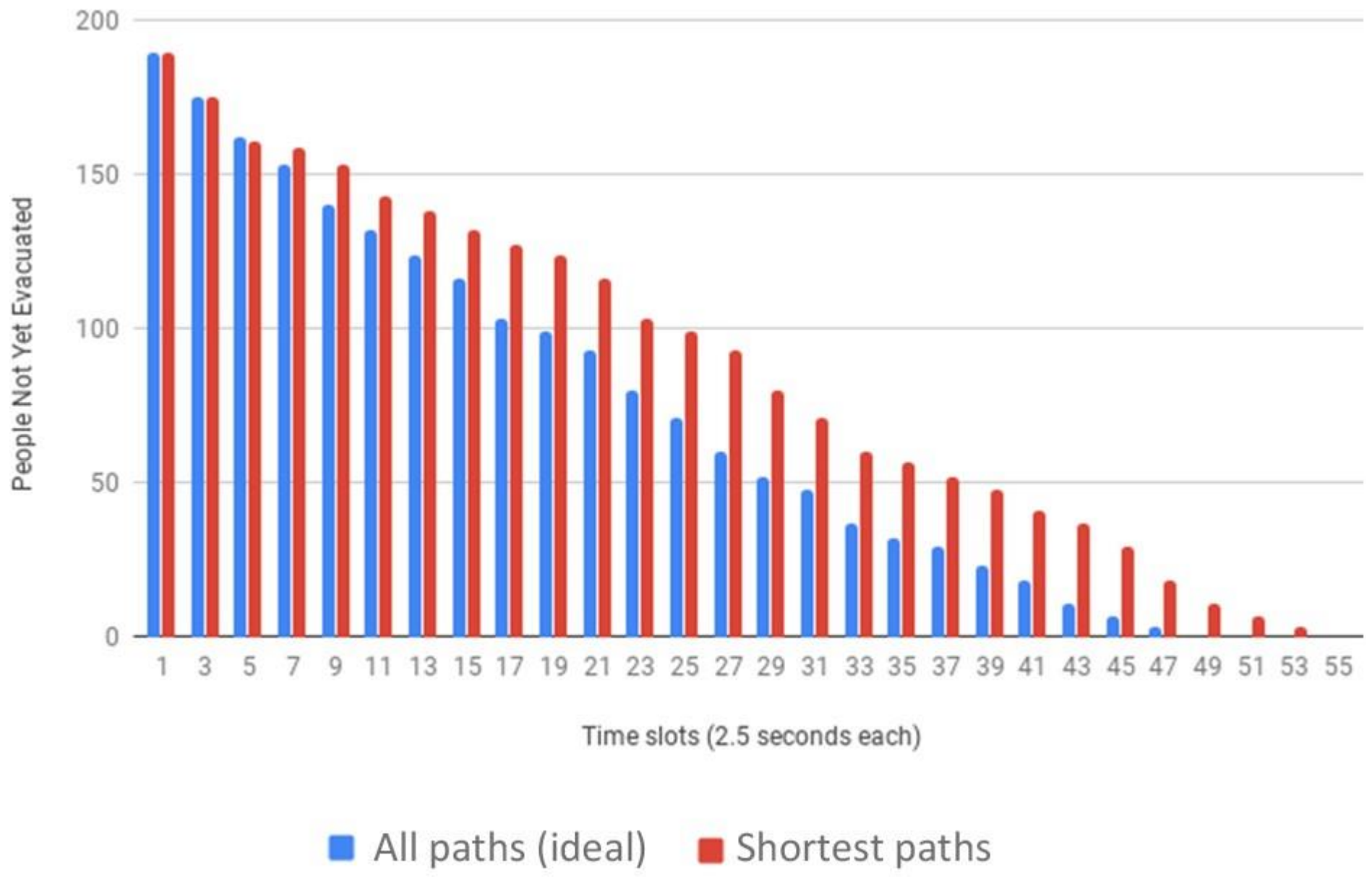

**Figure 7.8** Netflow vs. shortest paths evacuation considering grouping and attachment, Scenarios 1.

All agents use the ***Belief-Desire-Intention*** agent architecture:

- *Belief* - All agents believe that a disaster is unfolding and must somehow escape.
- *Desire* - All agents have the basic desire or *goal* to reach an exit.
- *Intention* - Since the agents *perceive* their surroundings they seek to find the shortest and/or optimal paths to reach the exit (based on the algorithms).

The following simulations were carried out using PedSim Microscopic Simulator on a PC running Ubuntu 18.04, with 8 GB ram, and an i5 processor with 2.5 GHz base clock speed.

***Scenario 1***. In a first set of simulations we simulated both the shortest path and network flow algorithms with varying walking velocities for 200 *GameAgents*. From the upper part of Table 7.2, with a walking velocity to 0.7 m/s, everyone can reach a safe place after 1 minute and 25 seconds using netflow and after 1 minute and 42.5 seconds using shortest paths. Thus, netflow decreases the evacuation time by 17%. Setting the walking velocity to 0.9 m/s and 1.2 m/s, netflow decreases the evacuation time by 11% and 27% respectively.

In a second set of simulations, we consider both shortest path and netflow algorithms under a random assignment of velocities. From the bottom part of Table 7.2, evacuating 200, 225 and 380 *GameAgents* using netflow reduces the evacuation time by over 11%, 9% and 8% respectively.

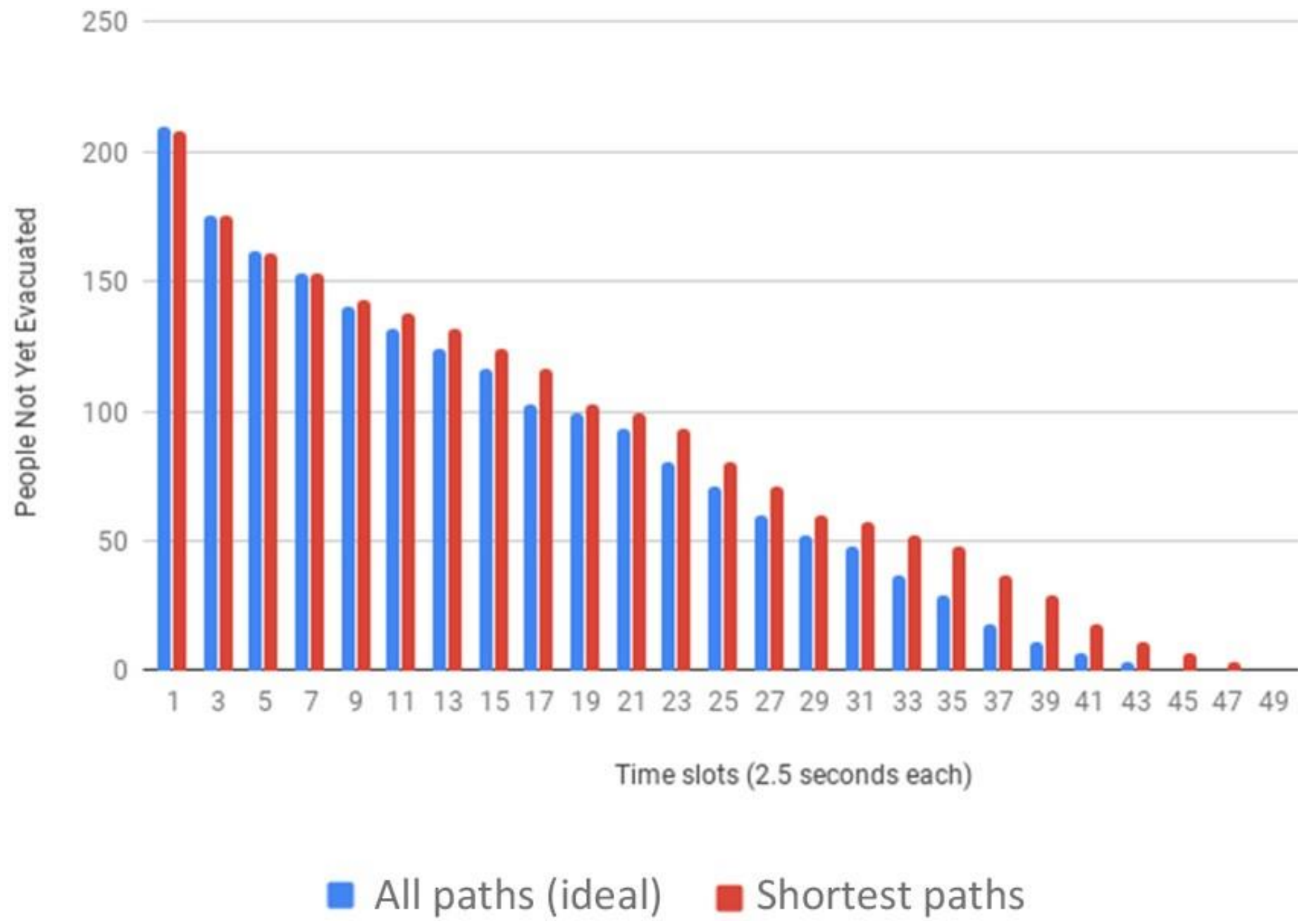

**Figure 7.9** Netflow vs. shortest paths evacuation considering grouping and attachment, Scenarios 2.

**Table 7.2** Evacuation time under uniform (a) and random (b) walking velocity. Time slots: 2.5 seconds each.

| # | $\tau$ *(all paths)* | $\tau$ *(shortest paths)* |
|---|---|---|
| Case $a_1$: *0.7 m/s* | 41 | 34 |
| Case $a_2$: *0.9 m/s* | 35 | 31 |
| Case $a_3$: *1.2 m/s* | 37 | 27 |
| Case $b_1$: *200 ppl* | 53 | 47 |
| Case $b_2$: *225 ppl* | 54 | 49 |
| Case $b_3$: *380 ppl* | 69 | 63 |

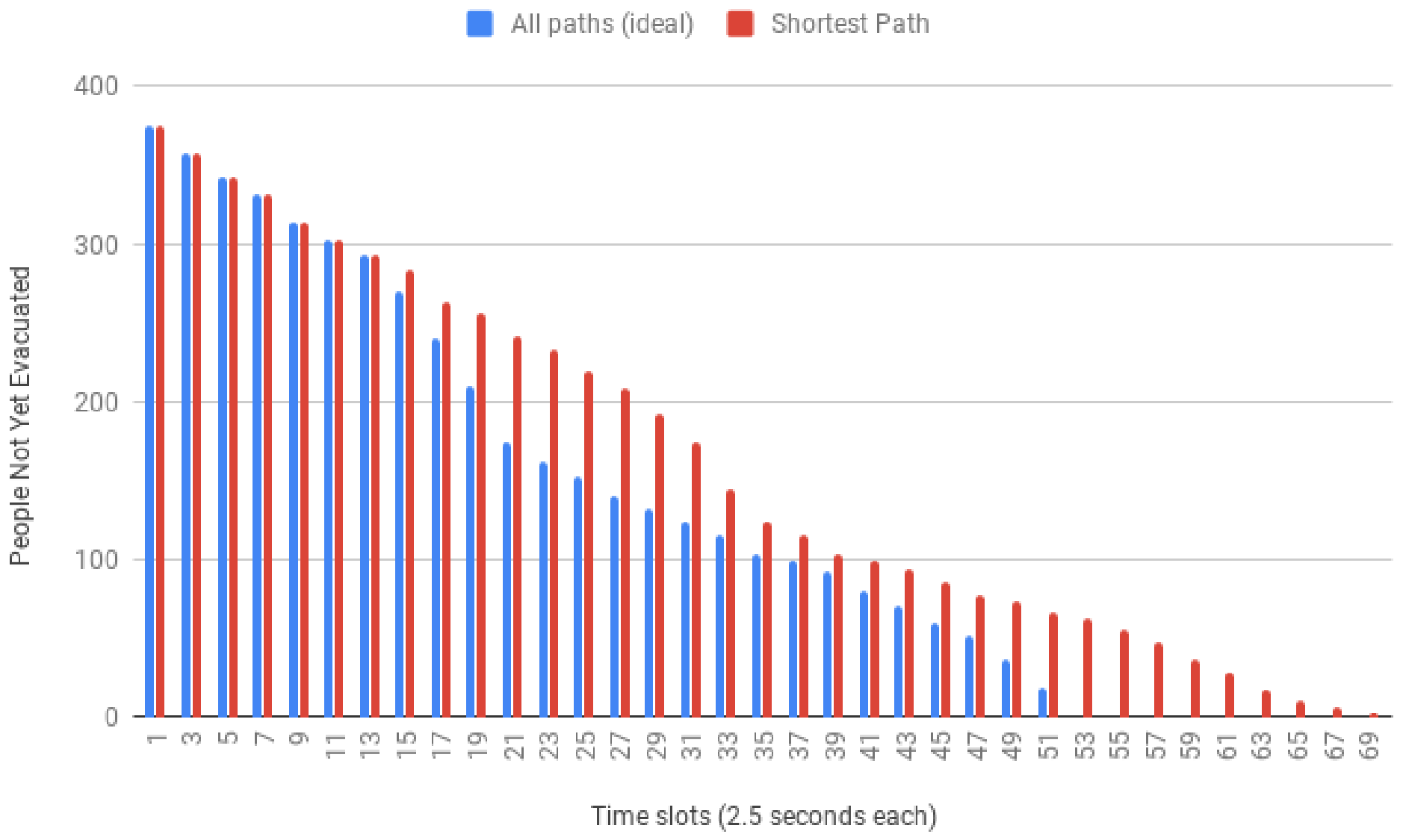

**Figure 7.10** Netflow vs. shortest paths evacuation considering grouping and attachment, Scenario 3.

| |
|---|
| ***Findings***: The network flow algorithm avoids congestion in building bottlenecks, whilst shortest path slows down the evacuation due to its inability to properly manage overcrowding. |

***Scenario 2*** . In a second set of simulations, we simulate social attachment and groupings ((13; 14)). A group of agents is a single immutable entity that move together. We consider random groups consisting of 3 to 7 agents. This gives an interesting scenario as congestion at exits become more pronounced. In this set of simulations, agents walking velocity randomly varies between 0.7 m/s and 1.2 m/s. From Wagnild et al. (106), walking velocity highly depends on the company and the speed of the slowest person in the group.

From Figure 7.8, evacuating 200 agents takes 2 minutes and 12.5 seconds and 1 minute and 57.5 seconds for shortest paths and netflow respectively.

Figure 7.9 shows the results of evacuating 225 agents. The evacuation surprisingly takes less time than the case with 200 agents (1 minute and 57.5 seconds with shortest paths and 1 minute and 47.5 seconds with netflow). Although other simulation runs with the same settings gave us expected results, i.e., higher evacuation time for higher number of agents, it was interesting to include this particular set of results to show that randomized grouping and attachment constraints may increase or decrease congestion and evacuation time. This is why the above *variation* of results occurred.

In a third set of simulations, the agent population was 380 (Figure 7.10); evacuation time increased to 2 minutes 52.5 seconds with shortest path and 2 minutes 7.5 seconds with netflow.

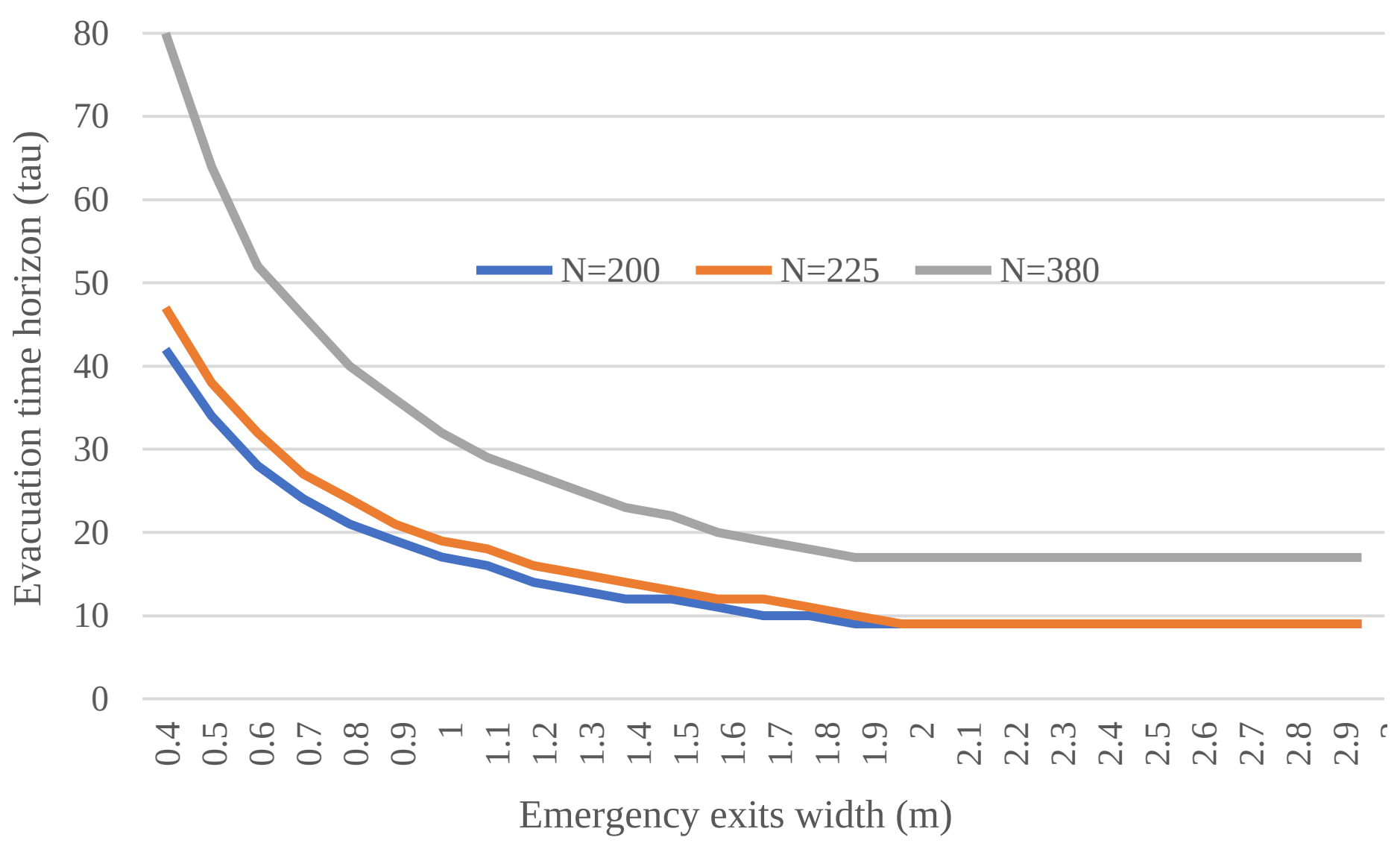

**Figure 7.11** Evacuation time variations with respect to the width of emergency exits.

| |
|---|
| ***Findings***: Grouping and attachment slows down evacuation, compared with scenario 1. Evacuation time increases with the number of agents because socially attached agents will not leave the building without their colleagues. |

### 7.7.3 Answer to RQ3: Optimization of Building Dimensions

We made a small change to the building structure to observe the impact on evacuation time. The pure optimization approach was used to optimize emergency exits and *Sim/Opt* was used to assess the impact of adding a second door to each room.

***Emergency exits***. Considering the same occupancy as above (i.e. 200, 225 and 380), we increased and decreased the width of emergency exits. From Figure 7.11, evacuation obviously takes longer by decreasing the emergency exit width. The interesting point is that the evacuation time horizon sharply slopes downward with wider exits, *but up to a certain dimension: i.e. 2 metres for all three cases*. For exits wider than this the evacuation time remains constant because it is not a function of congestion at exits, but depends on traveling time and congestion at internal doors. In the next subsection, we use our ABM Sim/Opt approach to further optimize internal door design and make the building optimally evacuable.

***Internal doors***. A wider door may not necessarily lead to a proportional increase in agent flow. Helbing et al (47) obtained the same result when they proposed that two close doors are more efficient than a single door of double width. This set of simulations aims at understanding the effects of congestion when there are close double doors, rather than a single large width door. For a proper comparison, we considered random walking velocity assignment and random grouping.

**Table 7.3** Evacuation time for strategic double door placement. Time slots: 2.5 seconds each.

| # | $\tau$ *(shortest paths)* | $\tau$ *(all paths)* |
|---|---|---|
| Case 1: *200 ppl* | 41 | 27 |
| Case 2: *225 ppl* | 45 | 37 |
| Case 3: *380 ppl* | 69 | 49 |

***Findings***: Double doors significantly impact the total evacuation time, which is shorter when compared to section 7.7.2, scenario 2. The 1 m doors are adjacent. When $N = 200$, there is a 34.14% decrease in evacuation time. For $N = 225$ and $N = 380$ the decrease is 13.95% and 3.95% respectively. Although evacuation time significantly decreases, the results seem to wane as the total number of people is increased. This is because only internal doors are doubled and not exits.

## 7.8 Conclusion

The netflow driven micro agent simulation, optimized with the plausible constraints, gives a realistic approach to evacuation compared to the shortest path approach. Based on the results, we can design buildings and evacuation systems that consider crowds. Our system is generic and can be applied to different buildings if interior plans are available. Since the IoT sensors are battery driven they remain active during a power cut. Our case study building has a back-up generator in case of a power failure so that the computer running the algorithm would continue to function. Where a back-up generator is not available, the algorithm could be distributed on a remote machine. The IoT system can count the number of people in a building and detect their location. This data is dynamically fed into the simulation tool for evaluating real-time evacuation.

# CONCLUSION

This thesis investigated on the following main question: *How we optimally model, monitor, analyze, simulate, and architect IoT-based indoor environments to best handle emergency evacuation whenever needed?* We found out that evacuation maps that are generally designed by civil protection operators provide static plans that impose more danger to building occupants. Thus, we designed IoT architectures with algorithmic core which enables quick evacuation of people from a risky area. We improved the systems by testing its quality attributes (such as performance and energy efficiency) and choosing the most efficient architectures for the evacuation infrastructure. We enhanced our IoT systems by considering real human behavior, such as grouping and attachment.

Based on these results, we can design topologies and evacuation systems that are better suited to accommodate the required crowd of pedestrians. As mentioned, the work employs a software architecture that helps improve and optimize evacuation time. From the topology data obtained, it is pipelined into a simulation environment (CPLEX/PedSim in our case). Once this data and the distribution of people is fed into the system, the algorithm for evacuation provides a generic initial pathway to the agents and as time progresses, provides an alternate/optimal paths for various agents and suggest the paths suitable to each individual. This is achieved by dividing the floor plan first into blocks and subsequently into number of cells. The cells may have different orientations-horizontal and vertical. The paths are not allowed to cut through walls. The walls can be in any place even inside the rooms. Safe passage through the doors are then computed connecting to safe areas. This modular nature of the algorithm makes it easy to scale for future variations and building architectures, and add more constraints by add-on modules. The agent-based simulation tool we used (PedSim) can itself be upgraded to more generic path-finding algorithm. The following are the advantages that are due to the present simulation tool and the employed algorithm:

*i)* all possible trajectories are computed;

*ii)* the optimal paths to safe areas are selected from the simulated egress pathways;

*iii)* occupants are regulated as per the constraints to reduce congestion;

*iv)* erratic and random motions are reduced;

*v)* the evacuation plan is made available to decision-makers and occupants;

*vi)* simulation is carried out at design time, although monitoring the status of the evacuation is done in real time using Internet of Things systems;

*vii)* an optimal evacuation procedure model can be evolved from the resulting simulations.

In our system, the internet of things infrastructure helps counting the number of persons in each block and detecting their location (cell numbers). This can further be dynamically fed into the simulation tool for real-time optimization of exit paths. The present system is currently being applied to various real indoor/outdoor areas under VASARI and Venice city Italian projects.